\documentclass[useAMS,usegraphicx,usenatbib]{mn2e}
\usepackage{aas_macros}
\usepackage[a4paper,centering, totalwidth=520pt, totalheight=700pt]{geometry}
\usepackage[fleqn]{amsmath} 
\usepackage{graphicx}
\usepackage{amssymb}
\usepackage{enumerate}
\usepackage{color}
\usepackage{hhline}
\usepackage{caption}
\usepackage{multirow}
\usepackage[normalem]{ulem}
\usepackage{longtable}
\usepackage{afterpage}

\usepackage{fixltx2e}[1999/12/01]

\usepackage[hyperindex]{hyperref}
\hypersetup{
breaklinks = {false},
colorlinks = {false},
linkcolor={black},
pdfpagemode = {None}, 
pdfborder = {0 0 1},
pdftitle = {ELDAR, a new method to identify AGN in multi-filter surveys: the ALHAMBRA
test-case},
pdfsubject = {Identification of AGN},
pdfauthor = {J. Chaves-Montero et al.},
}

\title[Identifying AGN in multi-filter surveys with ELDAR]
{ELDAR, a new method to identify AGN in multi-filter surveys: the ALHAMBRA
test-case}
\begin{document}

\author[Chaves-Montero et al.]{
\parbox[h]{\textwidth}
{Jon\'as Chaves-Montero$^1$\textsuperscript{\thanks{\href{mailto:jchaves@cefca.es}{jchaves@cefca.es}}},
Silvia Bonoli$^1$, 
Mara Salvato$^2$,
Natascha Greisel$^1$,
Luis A. D\'iaz-Garc\'ia$^1$,
Carlos L\'opez-Sanjuan$^1$,
Kerttu Viironen$^1$,
Alberto Fern\'andez-Soto$^{3,4}$,
Mirjana Povi\'c$^{5,6}$,
Bego\~na Ascaso$^{7}$,
Pablo Arnalte-Mur$^{8,9}$,
Josefa Masegosa$^6$,
Israel Matute$^{10}$,
Isabel M\'arquez$^6$,
A. Javier Cenarro$^1$,
L. Raul Abramo$^{11}$,
Alessandro Ederoclite$^1$, \&
Emilio J. Alfaro$^{12}$.}
\vspace*{12pt}
\\\hspace*{3pt}({\it Affiliations can be found after the references})
}
\maketitle

\date{\today}

\begin{abstract}
We present {\small ELDAR}, a new method that exploits the potential of medium-
and narrow-band filter surveys to securely identify active galactic nuclei (AGN)
and determine their redshifts. Our methodology improves on traditional
approaches by looking for AGN emission lines expected to be identified against
the continuum, thanks to the width of the filters. To assess its performance, we
apply {\small ELDAR} to the data of the ALHAMBRA survey, which covered an
effective area of $2.38\,{\rm deg}^2$ with 20 contiguous medium-band optical
filters down to F814W~$\simeq 24.5$. Using two different configurations of
{\small ELDAR} in which we require the detection of at least 2 and 3 emission
lines, respectively, we extract two catalogues of type-I AGN. The first is
composed of 585 sources ($79\,\%$ of them spectroscopically-unknown) down to
F814W~$=22.5$ at $z_{\rm phot}>1$, which corresponds to a surface density of
$209\,{\rm deg}^{-2}$. In the second, the 494 selected sources ($83\,\%$ of them
spectroscopically-unknown) reach F814W~$=23$ at $z_{\rm phot}>1.5$, for a
corresponding number density of $176\,{\rm deg}^{-2}$. Then, using samples of
spectroscopically-known AGN in the ALHAMBRA fields, for the two catalogues we
estimate a completeness of $73\,\%$ and $67\,\%$, and a redshift precision of
$1.01\,\%$ and $0.86\,\%$ (with outliers fractions of $8.1\,\%$ and $5.8\,\%$).
At $z>2$, where our selection performs best, we reach $85\,\%$ and $77\,\%$
completeness and we find no contamination from galaxies.
\end{abstract}

\begin{keywords}
galaxies: active -- galaxies: distances and redshifts -- methods: data analysis
-- techniques: photometric -- quasars: emission lines -- surveys
\end{keywords}


\section{Introduction}

Active galactic nuclei (AGN) are among the brightest objects in the Universe.
They are powered by the accretion of matter onto a supermassive black hole
(SMBH): as the gas approaches the SMBH, its temperature rises and starts to emit
radiation across the entire electromagnetic spectrum. Nevertheless, AGN not only
show a continuum emission from the gas in the accretion disk, but they also
exhibit multiple emission lines from the X-ray to the infrared spectral range.
In turn, the emission lines may be broad or narrow, depending on the orientation
of the AGN with respect to the observer and the obscuring material
\citep[according to the AGN unification scheme,] [] {antonucci93, urry95}. AGN
with broad emission lines are referred to as ``type-{\small I}'', while AGN with
just narrow emission lines as ``type-{\small II}''. We will employ this
observational classification along the rest of the paper.

For their many applications in different fields of astrophysics, from
high-energy physics to cosmology, a complete census of AGN is fundamental. AGN
are studied in the context of galaxy evolution models
\citep[e.g.,][]{heckman14}, as there are evidences of tight correlations between
SMBH and galaxy properties \citep[e.g.,][]{kormendy95, gebhardt00}, although a
causal origin of these correlations is not universally accepted
\citep[e.g.,][]{peng07, jahnke11}. In addition, thanks to their large
luminosities, the optically brightest type-{\small I} AGN (commonly referred to
as quasars) allow us to trace the matter distribution since early times
\citep[currently, the most distant spectroscopically-confirmed quasar is at
$z=7.1$, see][]{mortlock11}. Quasars can also be used to constrain cosmology:
\citet{busca13} successfully detected Baryonic Acoustic Oscillations (BAO) in
the Ly~$\alpha$ forest, and future galaxy surveys plan to measure the BAO
feature through the clustering of quasars \citep[e.g., the Extended Baryon
Oscillation Spectroscopic Survey is expected to reach a $1.6\,\%$ precision
measuring spherically averaged BAO with quasars, see][]{dawson16, zhao16}.
Finally, they have even been proposed as standard candles \citep{wang14,
watson11, risaliti16}.

There are many techniques for detecting AGN, such as traditional colour-colour
selections \citep[e.g.,] [] {matthews63}, optical variability \citep[e.g.,] []
{schmidt10}, and the combination of optical data and observations in radio
\citep[e.g.,][] {white00}, X-ray \citep[e.g.,][] {barger03, brusa03}, and/or
infrared \citep[e.g.,] [] {lacy04}. The strengths and weaknesses of these
methods are different. For instance, X-ray selection allows to be complete,
missing only the most obscured sources, at the cost of being very time
consuming. On the other hand, broad-band photometric surveys are less
time-expensive but they are biased towards unobscured type-{\small I} AGN, and
they include a significant contamination from stars and galaxies.

The emergence of multi-filter surveys, such as the Classifying Objects by
Medium-Band Observations - a spectrophotometric 17-filter survey -
\citep[COMBO-17,] [] {wolf04, wolf08}, the Cosmic Evolution Survey
\citep[COSMOS,] [] {Scoville07}, the Advance Large Homogeneous Area Medium Band
Redshift Astronomical survey \citep[ALHAMBRA,] [] {moles08}, the Survey for
High-$z$ Absorption Red and Dead Sources \citep[SHARDS,] []{perez13}, the
Physics of the Accelerating Universe Survey \citep[PAUS,] [] {marti14}, and the
upcoming Javalambre Physics of the Accelerating Universe Astrophysical Survey
\citep[J-PAS,] [] {benitez14}, open the possibility of exploring new methods for
detecting AGN. Multi-band photometric data, in fact, combine the strengths of
broad-band photometric and spectroscopic surveys, resulting in a low-resolution
spectra for every pixel of the sky observed, e.g. the ALHAMBRA survey has
spectral resolution of $R=310$. The aim of this work is precisely to produce a
new pipeline to identify AGN and to compute their redshifts using just data from
multi-filter surveys. We take advantage of the low-resolution spectroscopic
nature of these data in order to identify strong spectral features typical of
active galaxies.

We test our new algorithm, dubbed as Emission Line Detector of Astrophysical
Radiators ({\small ELDAR}), by applying it to the data from the ALHAMBRA survey
\citep{moles08, molino14}. This survey is an optimal test-case for {\small
ELDAR} because it observed $\simeq 4\,{\rm deg}^2$ using 20 contiguous
medium-band filters (Full Width Half Maximum FWHM~$\simeq300\,\rm{\AA}$) in the
optical range and 3 broad-band filters ($J$, $H$, and $K_s$) in the infrared. In
addition, \citet{matute12} showed that it is possible to compute precise
redshifts ($\sigma_{\rm NMAD}\simeq1\,\%$) for spectroscopically-known quasars
using just ALHAMBRA data. Here, we extract two catalogues of type-{\small I} AGN
using two different {\small ELDAR} configurations, the first maximising
completeness and the second minimising contamination. Then, we analyse the main
properties of these catalogues and we estimate their completeness, redshift
precision, and galaxy contamination by applying the same {\small ELDAR}
configurations to samples of spectroscopically-known type-{\small I} AGN and
galaxies within the ALHAMBRA fields.

This paper is structured as follows. In \S\ref{sec:methodology} we introduce
{\small ELDAR} and in \S\ref{sec:ELDAHL} we tune our method to detect
type-{\small I} AGN in ALHAMBRA. In \S\ref{sec:results} we extract two
catalogues of type-{\small I} AGN and we characterise their properties. In
\S\ref{sec:widths} we discuss the potential of our methodology for surveys with
narrower bands and in \S\ref{sec:conclusions} we summarise our conclusions.

Throughout this paper the optical and near-IR magnitudes are in the AB system,
we always use the spectral flux density per unit wavelength, and we assume a six
parameter $\Lambda$CDM cosmology with $H_0=67.8\,\text{km s}^{-1}
\text{Mpc}^{-1}$, $\Omega_\Lambda=0.692$, and $\Omega_{\rm M}=0.308$
\citep{planck16a}.


\section{ELDAR algorithm}
\label{sec:methodology}

The new methodology to detect AGN in multi-filter surveys that we introduce in
this paper, {\small ELDAR}, consists of two main steps: i) {\it
template-fitting}, that aims at pre-selecting AGN candidates and at obtaining a
redshift probability distribution function (PDZ) for each of them, and ii) {\it
spectro-photometric confirmation}, whose objective is to securely confirm the
previous candidates by detecting typical AGN emission lines in the multi-band
photometric data and to refine the photo-$z$ estimation.

In what follows, we describe in more detail the two steps of {\small ELDAR}.


\subsection{Template-fitting step} 
\label{sec:lephare}

The main purpouse of this first step is to pre-select AGN candidates and to
obtain a PDZ for each of them. While any template-fitting code may be used for
this pre-selection phase, in this work we adopt the code PHotometric Analysis
for Redshift Estimate ({\small LePHARE}) \citep{arnouts99}. {\small LePHARE} is
a template-fitting code extensively used to compute photo-$z$s for galaxies and
AGN \citep[e.g.,] [] {ilbert09, salvato09, salvato11, fotopoulou12, matute12}.
Here we provide a general discussion on how to correctly configure {\small
LePHARE}, as the templates and parameters of the code have to be carefully
chosen and optimised to detect AGN depending on the characteristics of the
survey to be analysed. In addition, in \S\ref{sec:ELDARALH} we provide the
specific configuration of {\small LePHARE} for the case of the ALHAMBRA survey.

\begin{itemize}

\item {\it Template selection.} {\small LePHARE} classifies each source and
computes its redshift depending on the Spectral Energy Distribution (SED) of the
template that produces the best-fit to the photometric data, where a template is
a theoretical or empirical curve that describes the flux of different
astronomical objects as a function of $\lambda$. The library of templates to be
used in {\small LePHARE} has to be meticulously chosen, especially when working
with AGN \citep{hsu14}: while it should be comprehensive enough to include the
broad variety of SEDs of the types of sources that are sought, the number of
templates should not be too large to avoid degeneracies.

The templates are divided into two main categories in {\small LePHARE}: stellar
and extragalactic. The first includes the SEDs of stars, while the second
presents the SEDs of extragalactic objects at rest-frame, which are shifted in
redshift during the fitting procedure. To build our stellar library we include
254 stellar templates from the publicly available distribution of {\small
LePHARE}. They are divided into 131 templates of normal stellar spectral types
and luminosity classes at solar abundance, metal-poor F-K dwarfs, and G-K giants
\citep{pickles98}; 4 templates of white dwarfs \citep{bohlin95}; 100 templates
of low mass stars \citep{chabrier00}; and 19 templates of sub-dwarfs
\citep{bixler91}. We include all of them to cover as many stellar types as
possible and thus, to avoid the classification of stars as AGN.

For the extragalactic library, we only include templates of active galaxies, as
these are the sources we are targeting. With this approach, we ensure that no
AGN are wrongly classified as `normal' galaxies (i.e. galaxies whose SEDs are
not dominated by nuclear activity), while all normal galaxies will be discarded
by the spectro-photometric confirmation step (see \S\ref{sec:lscode}). The AGN
templates to be included in the extragalactic library are survey specific, as
the AGN types that can be unambiguously confirmed in a given survey depend on
its characteristics, e.g., its depth, area, and the width of its photometric
bands. In particular, the width of the survey bands determines the approximate
minimum Equivalent Width (EW) of the emission lines that can be detected by
{\small ELDAR} (see \S\ref{sec:ELDARALH}). As the EWs of AGN emission lines
depend on the type of active galaxy, we should only include templates of AGN
with emission lines strong enough to be detected by our method.

\item{\it Redshift range and precision.} The extragalactic templates included in
the {\small LePHARE} library are located at rest-frame. During the fitting
procedure, {\small LePHARE} creates a grid of templates displaced in redshift,
where the redshift step and maximum redshift are defined by the user. As
\citet{benitez09b} observed, the size of the redshift step should depend on the
number of filters available and the overlap between them. As for the maximum
redshift, we set it to the redshift above which no strong spectral features 
appear to within the survey wavelength coverage.

Effectively, the PDZ generated by {\small LePHARE} is defined as:

\begin{equation}
{\rm \small PDZ}(z)=\frac{G(z)}{G(z_{\rm best})},
\end{equation}

\noindent where $G(z)=\exp[-\chi^2_{\rm min}(z)/2]$, $\chi^2_{\rm min}(z)$ is
the $\chi^2$ resulting from the template that best fits the data at redshift
$z$, and $z_{\rm best}$ is the redshift at which the data is best-fitted. With
this definition, the PDZ is not properly a probability density function, and to
generate one for each object the PDZ of the previous expression should be
normalised by its integral.

\item {\it Dust attenuation.} The extinction law of AGN varies as a function of
redshift \citep[e.g.,][]{gallerani10}, reflecting different mechanisms for dust
production and/or destruction. A correct modelling of the effect of dust is
required because it absorbs UV and optical light, which then re-emits in the
infrared modifying the SEDs of AGN. In {\small LePHARE} we employ the Milky Way
\citep{allen76}, Small Magellanic Cloud \citep{prevot84}, Large Magellanic Cloud
\citep{fitzpatrick86}, and starburst \citep{calzetti00} extinction laws, which
are shown in fig.~7 of \citet{bolzonella00}.

The dust attenuation ($A_V$) of an active galaxy depends on its orientation with
respect to the observer and it is defined as

\begin{equation}
A_V = R_V \times E(B-V),
\end{equation}

\noindent where $E(B-V)$ is the colour excess and $R_V$ is a coefficient that
depends on the extinction law. We introduce colour excesses from $0$ to $0.10$
in steps of $0.02$, from $0.10$ to $0.30$ in steps of $0.04$, and from $0.30$ to
$1.00$ in steps of $0.10$. We include colour excesses as high as 1 to account
for very extinct AGN. We set finer steps for low colour excesses because
some AGN templates are empirical, and thus they already include some extinction.

\item {\it Luminosity prior.} Setting luminosity priors is important to avoid
unrealistic solutions \citep{salvato09}, and they should be chosen depending on
the type of objects that we want to target. Quasars, for example, are
traditionally defined as objects with $M_B \le -23$ \citep[e.g.,] []
{Osterbrock91}, and setting $M_B=-23$ as upper limit ensures that {\small
LePHARE} rejects low redshift (low-$z$) incorrect solutions.

\end{itemize}


\subsection{Spectro-photometric confirmation step} \label{sec:lscode}

Objects with strong emission lines, such as type-{\small I} AGN, are
particularly suited to be detected in surveys with contiguous medium- and/or
narrow-band filters. This is because emission lines with a large EW completely
dominate the bands in which they fall. Consequently, these bands appear as clear
`peaks' in the multi-band data. The height of these peaks with respect to the
continuum emission depends on i) the EW of the line, ii) the width of the band
where the emission line falls, and iii) the shape of the continuum. Assuming
that the AGN continuum emission is flat in the bands adjacent to the band where
the line falls, {\small ELDAR} is able to detect lines with EW greater than

\begin{equation}
\label{eq3:ew}
{\rm EW}_{\rm min} = \frac{B_{\rm FWHM}}{(1+z)B_{\rm SNR}}\sigma_{\rm line},
\end{equation}

\noindent where $B_{\rm FWHM}$ is the FWHM of the band where the line falls,
$B_{\rm SNR}$ is the Signal-to-Noise Ratio (SNR) in this band, $z$ is the
redshift of the source, and $\sigma_{\rm line}$ is a parameter that denotes the
confidence with which we want to confirm lines, e.g. $\sigma_{\rm line}=1$ means
a $1\,\sigma$ detection. Therefore, the detection of emission lines depends on
the intrinsic properties of each source (e.g. their $z$ and EW) and on the
characteristics of the survey to be analysed (e.g. the bandwidth of the bands
and their depth.

However, Eq.~\ref{eq3:ew} just sets an approximate value for ${\rm EW}_{\rm
min}$. This is because the assumption of a flat continuum is not usually correct
for AGN, especially at $z<2.5$ where the slope of the AGN continuum is very
steep and blue. Moreover, the value of ${\rm EW}_{\rm min}$ is also an lower
limit for emission lines that fall in between two bands or are broader than the
survey bands.

With these caveats in mind, the objective of this second step of {\small ELDAR}
is to search for typical AGN emission lines in the multi-band photometry of the
sources that we want to classify. We improve on the ability of template fitting
codes in unambiguously confirm emission line objects, as they do not include
special weights for the bands where emission lines fall and, as the number of
bands dominated by the continuum emission is always greater than the number of
bands dominated by emission lines, they are not specifically designed for
detecting these objects. 

The detection of AGN emission lines allows not only to confirm sources as active
galaxies but also to reject stars and galaxies assigned to AGN templates in the
first step of {\small ELDAR}. Moreover, it provides a method to discriminate
between different redshift solutions given by the PDZ. Operationally, the
confirmation step works as follows:

\begin{enumerate}

\item We start by selecting, for each source, the redshifts at which the SED is
best-fitted by an extragalactic template ($\chi^2_{\rm AGN}<\chi^2_{\rm star}$)
and the value of the PDZ is greater than 0.5. We set a lower limit in the PDZ in
order to include the information provided by {\small LePHARE} from the fitting
of the SED. We check the dependence of the results on different PDZ lower limits
in Appendix B. For each of these possible redshift solutions, $z_{\rm guess}$,
we perform the steps that follow.

\item According to each $z_{\rm guess}$, we calculate which AGN emission lines
with EW greater than ${\rm EW}_{\rm min}$ are expected to lie within the
wavelength coverage of the survey, and in which band they should fall. We then
confirm the detection of a line if:

\begin{equation}
\label{eq3:linedetection}
F_{\rm cen} > \left\{
\begin{aligned}
&F_{\rm blue} +\sigma_{\rm line}\,S_{\rm cen},\\
&F_{\rm red}  +\sigma_{\rm line}\,S_{\rm cen},\\
&F_{\rm blue} +\sigma_{\rm line}\,S_{\rm blue},\\
&F_{\rm red}  +\sigma_{\rm line}\,S_{\rm red},\\
\end{aligned}
\right.
\end{equation}

\noindent where $F_{\rm cen}$ is the flux in the band where the line should fall
according to $z_{\rm guess}$, $F_{\rm blue}$ ($F_{\rm red}$) is the flux in the
band bluewards (redwards) of the band where the line should fall, and $S_{\rm
cen}$, $S_{\rm blue}$, and $S_{\rm red}$ are their errors. By construction, we
are unable to confirm lines that fall either in the first or in the last band of
the filter system.

\begin{figure}
\begin{center}
\includegraphics[width=0.45\textwidth]{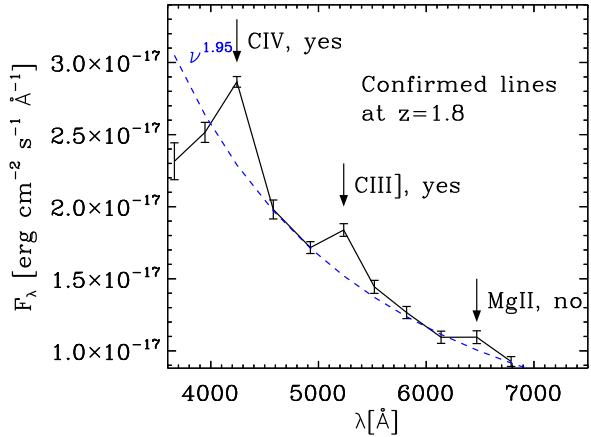}
\end{center}
\caption{\label{fig:linedetect} Multi-band ALHAMBRA photometry of a
spectroscopically-known type-{\small I} AGN at $z=1.8$. At this redshift, the
lines C~{\small IV}, C~{\small III}], and Mg~{\small II} fall within the
ALHAMBRA medium-band wavelength range. {\small ELDAR} confirms C~{\small IV} and
C~{\small III}] with more than $1\,\sigma$ confidence in the 3rd and 6th band,
respectively. On the other hand, Mg~{\small II} is not confirmed because the
flux in the 9th band, where this line should fall according to $z_{\rm spec}$,
does not fulfil all the requirements of Eq.~\ref{eq3:linedetection}. The blue
dashed line shows a power law to guide the eye on the AGN continuum.}
\end{figure}

\begin{figure}
\begin{center}
\includegraphics[width=0.45\textwidth]{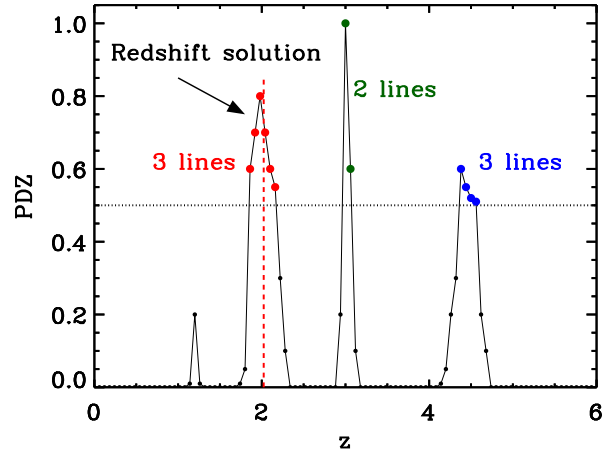}
\end{center}
\caption{\label{fig:explain} Illustrative example of a PDZ in which we include
information about the number of AGN emission lines detected by {\small ELDAR}.
The black small dots indicate redshift solutions with PDZ~$<0.5$, the green dots
solutions with PDZ~$>0.5$ for which {\small ELDAR} detects 2 AGN emission lines,
and the red and blue dots solutions with PDZ~$>0.5$ for which {\small ELDAR}
detects 3 AGN emission lines. The red dashed line shows the final redshift
solution for the source, $z_{\rm phot}$. See the text for further information
about how $z_{\rm phot}$ is computed.}
\end{figure}

\hspace*{18pt}In Fig.~\ref{fig:linedetect} we show a spectroscopically-known
type-{\small I} AGN at $z_{\rm spec}=1.8$ observed by the ALHAMBRA survey (we
will present the characteristics of the survey in \S\ref{sec:alhambra}). We show
arrows pointing to the bands where C~{\small IV}, C~{\small III}], and
Mg~{\small II} should fall according to $z_{\rm spec}$. The blue dashed line
indicates a power law that guides the eye on the AGN continuum emission and it
allows us to easily see the flux excess in the bands where the AGN emission
lines fall. For this source, C~{\small IV} and C~{\small III}] are detected by
{\small ELDAR} while Mg~{\small II} is not confirmed because it does not fulfil
all the requirements of Eq.~\ref{eq3:linedetection}.

\hspace*{18pt}There are some redshift intervals for which two different emission
lines may fall in consecutive bands, and thus the line detection is not secure.
However, the typical separation between the strongest AGN emission lines
(EW~$>8$) with rest-frame central wavelength $\lambda_{\rm c}<4\,000\,{\rm \AA}$
is large enough for these lines to never fall in consecutive bands in surveys
with filters narrower than FWHM~$\sim400\,{\rm \AA}$. In any case, if lines with
different EW fall in consecutive bands, the line with the largest EW can still
be confirmed.

\hspace*{18pt}In surveys with no contiguous bands another complication might
arise at redshifts in which AGN emission lines fall between two bands, as the
flux of the line gets dispersed. However, in most cases the greatest part of the
line falls in one band and just its tail in other/s. In this case, the line is
detected in the band where the greatest part of its flux falls. We further
explore this issue in \S\ref{sec:widths}.

\hspace*{18pt}To account for redshift errors and physical processes that may
displace emission lines from the band where they should fall, such as line
shifts and anisotropic profiles \citep[see ][]{vandenberk01}, we search for
emission lines not only in the band where they should fall according to $z_{\rm
guess}$, but also in the two adjacent bands.

\item We confirm as AGN the sources for which we detect at least $\mathcal{N}$
emission lines at the expected redshift, where $\mathcal{N}$ is chosen depending
on the number of lines that the survey filter system allows to detect, as well
as on a compromise between the completeness and the level of galaxy and star
contamination that we want to achieve. Obviously, the contamination from
galaxies and stars decreases by increasing $\mathcal{N}$ (see
\S\ref{sec:degeneracy} for a discussion about potential contaminants for the
ALHAMBRA survey).

\item Once a source is confirmed as AGN, we check at which $z_{\rm guess}$ the
largest number of lines is detected, rejecting the other values. If we end up
with a single $z_{\rm guess}$, we accept it as the final photo-$z$ solution,
$z_{\rm phot}$. Otherwise, we group contiguous $z_{\rm guess}$ into intervals,
and we look for the interval with the greatest average PDZ. In this case, we
then compute the final redshift solution as

\begin{equation}
\label{eq3:zres}
z_{\rm phot} = \frac{\sum_i^n\,z_{{\rm guess},i}
\,{\rm \small PDZ}(z_{{\rm guess},i})}
{\sum_i^n\,{\rm \small PDZ}(z_{{\rm guess},i})},
\end{equation}

\noindent where the summation goes through the $n$ values of $z_{\rm guess}$
in the selected interval.

\end{enumerate}

In Fig.~\ref{fig:explain} we show an illustrative example of this procedure. We
start by selecting $z_{\rm guess}$, i.e. the redshifts at which the SED of the
object is best-fitted by an AGN template and the value of the PDZ is greater
than 0.5. These redshift solutions are the red, green, and blue points. Then, we
pick the $z_{\rm guess}$ for which the largest number of lines is detected (in
this example, the red and blue dots). After that, we group the red points into
one interval and the blue ones into another. Later, we reject the blue-points
interval because the mean PDZ of the red-points interval is greater. Finally, we
compute $z_{\rm phot}$ with the red-points interval using Eq.~\ref{eq3:zres}.

The above steps define the backbone of the spectro-photometric confirmation.
Additional criteria can be added to refine the procedure. For instance, as the
Ly~$\alpha$ line is the strongest AGN emission line in the UV, in the present
work we require i) the Ly~$\alpha$ line to be detected in sources with redshift
solutions for which this line should fall within the survey wavelength coverage,
and ii) the flux in the band where it falls to be at least $75\,\%$ of the
maximum flux in any of the other bands. Even if the Ly~$\alpha$ line is the
strongest in the UV, we set a $75\,\%$ limit to account for the possibility of
the line falling in between two bands and/or other emission lines surpassing its
flux. With this condition we want to reject cold stars whose continuum emission
may be confused with the Lyman-break of high-$z$ AGN. We explore the dependence
of the results on this criterion in Appendix B.


\begin{figure}
\begin{center}
\includegraphics[width=0.45\textwidth]{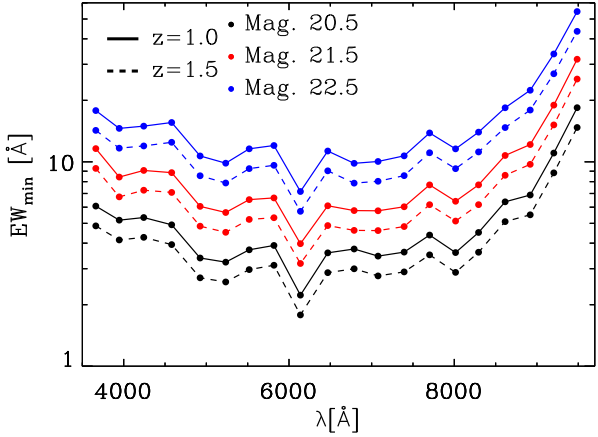}
\end{center}
\caption{\label{fig:eqw} Minimum EW of the emission lines that can be detected
in each ALHAMBRA optical band for $\sigma_{\rm line}=1$, as a function of the
magnitude of the band and the redshift of the source.}
\end{figure}

\begin{figure}
\begin{center}
\includegraphics[width=0.45\textwidth]{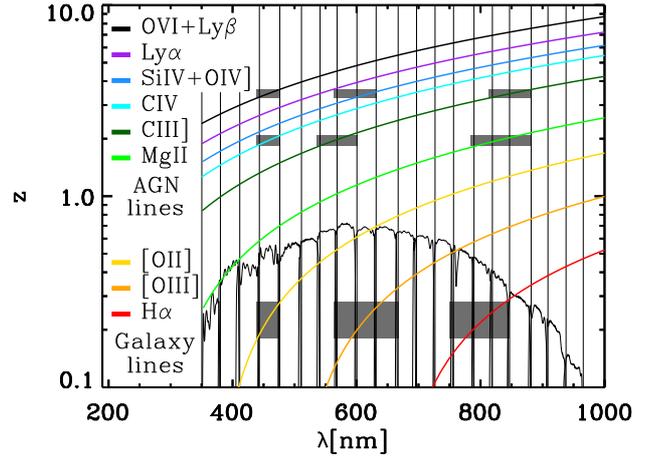}
\end{center}
\caption{\label{fig:diff-lines} Evolution of the central wavelength of AGN and
galaxy emission lines as a function of redshift. We also display the
transmission curves of the ALHAMBRA medium-band filters, which allow us to see
in which band the emission lines are located as a function of $z$. The grey
areas highlight the redshift intervals for which there is a degeneracy among the
triplet of galaxy emission lines \{[O~{\small II}], [O~{\small III}],
H~$\alpha$\}, and the triplets of AGN emission lines \{C~{\small IV}, C~{\small
III}], Mg~{\small II}\} and \{O~{\small VI}+Ly~$\beta$, Si~{\small
IV}]+O~{\small IV}], C~{\small III}]\}.}
\end{figure}

\section{Applying ELDAR to ALHAMBRA data} 
\label{sec:ELDAHL}

In the previous section we introduced {\small ELDAR}, our new procedure to
detect AGN. Here, we introduce the ALHAMBRA survey, we discuss some effects that
may affect the quality of {\small ELDAR}'s results, and we show how we have
optimised our method for analysing the ALHAMBRA data. In \S\ref{sec:results} we
will blindly apply the configurations introduced in this section to the ALHAMBRA
data in order to extract a new catalogue of type-{\small I} AGN.

\subsection{The ALHAMBRA survey}
\label{sec:alhambra}

ALHAMBRA\footnote{\href{http://www.alhambrasurvey.com/?lang=en}
{http://www.alhambrasurvey.com}} is a medium-band photometric survey that
observed $\simeq4\,\text{deg}^2$ of the sky distributed over 8 non-overlapping
fields. These fields were selected to be in common with other surveys, such as
the Deep Extragalactic Evolutionary Probe 2 (DEEP2), the Sloan Digital Sky
Survey (SDSS), COSMOS, the Hubble Deep Field North (HDF-N), the Deep Groth Strip
Survey (GROTH), and the European Large Area ISO Survey (ELAIS). The ALHAMBRA
filter system consists of 20 contiguous medium-band filters of width
$\simeq300\,\rm{\AA}$, which cover the optical range from $3\,500$ to
$9\,700\,\rm{\AA}$, and the 3 broad-band infrared filters $J$, $H$, and $K_s$.
The magnitude limit ($5\,\sigma$, 3'') is $\simeq23.7$ for the blue optical
filters, $\simeq22.2$ for the red optical filters, and $\simeq22$ for the
infrared filters \citep{apariciovillegas10}. Due to the width of its filters and
the contiguous wavelength coverage from the near UV to the near-infrared, the
ALHAMBRA survey is an optimal test-case for {\small ELDAR}.

The last public data release of ALHAMBRA is introduced in \citet[M14
hereafter]{molino14}. It covered an area of $\simeq3\,\text{deg}^2$ over 7
fields, detecting $438\,356$ sources brighter than 24.5 mag in the synthetic
detection band, F814W. This band was generated by combining the 9 reddest
ALHAMBRA bands to mimic the {\it Hubble Space Telescope} (HST) - Advanced Camera
for Surveys (ACS) F814W band.

The ALHAMBRA filter system produces precise redshift estimates for blue and red
galaxies, as shown by M14. Specifically, M14 found a redshift precision of
$\sigma_z\simeq1\,\%$ for spectroscopically-known galaxies with F814W~$<22.5$
within the ALHAMBRA fields. Moreover, in a first attempt to characterise the
ability of ALHAMBRA to produce precise photo-$z$s for type-{\small I} AGN,
\citet{matute12} applied {\small LePHARE} to a sample of 170
spectroscopically-known type-{\small I} AGN within the ALHAMBRA fields, finding
a redshift precision of the same order as for galaxies.

As we stated in the previous section, the properties of the survey filter system
are essential to determine i) the approximate minimum EW of the emission lines
that can be detected and ii) the redshift precision that can be achieved. In
Fig.~\ref{fig:eqw} we show an estimation of the minimum equivalent width of an
emission line that can be detected in each ALHAMBRA medium-band (as defined in
Eq.~\ref{eq3:ew}), as a function of the redshift of the source, its magnitude,
and using $\sigma_{\rm line}=1$. By definition, the value of ${\rm EW}_{\rm
min}$ decreases for bright sources (higher SNR) and at high-$z$. In addition,
the value of ${\rm EW}_{\rm min}$ grows for the reddest bands. This is because
the efficiency of the ALHAMBRA CCDs significantly decreases for
$\lambda>9\,000\rm{\AA}$ (see the overall transmission of the ALHAMBRA filter
system in Fig.~\ref{fig:diff-lines}).

\begin{table}
\begin{center}
\caption{\label{tab:lines} Emission lines employed to confirm type-{\small I}
AGN in ALHAMBRA. At least 2 and 3 emission lines must be detected to validate
objects using the {\small ELDAR}'s 2- and 3-lines mode, respectively (see
\S\ref{sec:ELDARALH}).}
\begin{tabular}{ccc}
\hline
Line&$\lambda_{\rm c}(\rm{\AA})$&$\left<{\rm EW}(\rm{\AA})\right>$\\
\hline
O~{\small VI}+Ly~$\beta$ &1030&15.6$\pm$0.3\\
Ly~$\alpha$     &1216&91.8$\pm$0.7\\
Si~{\small IV}+O~{\small IV}]&1397&8.13$\pm$0.09\\
C~{\small IV}   &1549&23.8$\pm$0.1\\
C~{\small III}] &1909&21.2$\pm$0.1\\
Mg~{\small II}  &2799&32.3$\pm$0.1\\
\hline
\end{tabular}
\end{center}
\textbf{Notes}. The values of the central wavelengths and EWs are computed using
184 quasars observed by the HST \citep[emission lines with $\lambda_{\rm
c}<1300\,{\rm \AA}$,] [] {telfer02} and $2\,000$ quasars observed by SDSS
\citep[emission lines with $\lambda_{\rm c}>1300\,{\rm \AA}$,] [] {vandenberk01}.
\end{table}

In Table~\ref{tab:lines} we list the AGN emission lines that are potentiality
detectable with at least $1\,\sigma$ precision for i) sources with magnitude
$\lesssim21.5$ in the band where the line falls, and ii) an observed central
wavelength, $\lambda_{\rm obs}=\lambda_c(1+z)$, smaller than $9\,000\,{\rm
\AA}$. In addition, as the ALHAMBRA bands are contiguous, these lines can be
detected for the entire redshift interval where they fall within the ALHAMBRA
optical coverage. We do not look for emission lines with $\lambda_{\rm
c}>3\,000\,{\rm \AA}$, such as [O~{\small II}], H$\beta$, [O~{\small III}], or
H~$\alpha$, because these lines also appear in star-forming galaxies. Whereas it
is possible to use them to discriminate between type-{\small I} AGN and
star-forming galaxies as the lines of type-{\small I} AGN are much broader, the
low spectral resolution of ALHAMBRA prevent us to employ them (we expect this to
be possible in surveys with narrower bands). Therefore, given the lines that we
can use to detect AGN and their strengths, we will be able to securely identify
type-{\small I} AGN at $z>1$ (unobscured broad emission line AGN with no or very
little contribution from the host). As a consequence, we focus on the detection
of type-{\small I} AGN in this work, and we tune {\small ELDAR} accordingly.

\begin{figure}
\begin{center}
\includegraphics[width=0.45\textwidth]{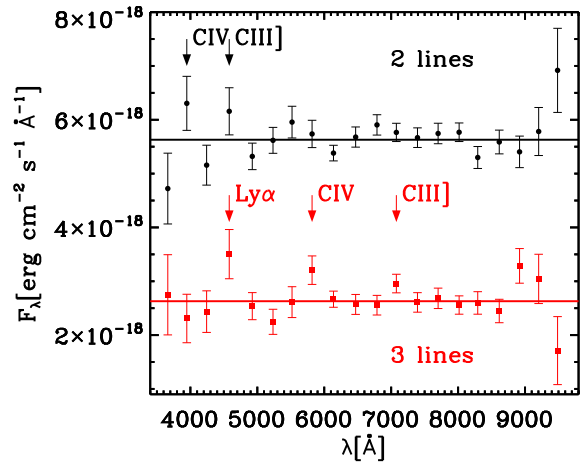}
\end{center}
\caption{\label{fig:exspu} Two ALHAMBRA mock realisations of a source with a
flat SED and F814W~$=21.5$. The black points show fluxes measured in each
ALHAMBRA medium-band in the first realisation and the red squares in the second.
The measured fluxes are not on the top of the solid lines, which indicate the
underlying SED of the mock source, due to photometric errors. The arrows point
to bands where {\small ELDAR} detects emission lines with $1\,\sigma$
significance, and the names indicate with which AGN emission lines these
spurious lines could be confused. The fluxes of the first (second) realisation
are displaced $+(-)1.5\times10^{-18}\,{\rm erg\, cm^{-2}s^{-1}\AA^{-1}}$ for
visual purposes. We show these two examples because these objects could be
confirmed as AGN by {\small ELDAR} due to the presence of spurious lines.}
\end{figure}

\subsection{Effects that may reduce the redshift precision and increase the 
contamination from galaxies}
\label{sec:degeneracy}

Before optimising {\small ELDAR} for detecting type-{\small I} AGN in the
ALHAMBRA survey, we will explore three effects that may decrease the quality of
the {\small ELDAR}'s results: i) confusion between pairs/triplets of AGN and
galaxy emission lines, which increases contamination from galaxies; ii)
confusion between different pairs/triplets of AGN emission lines, which reduces
the redshift precision; and iii) detection of spurious lines, which may reduce
the redshift precision and introduce galaxy contamination. There are examples of
all of these issues in Appendix A.

The confusion between different pairs/triplets of emission lines arises due to
the limited spectral resolution of multi-filter surveys. The misidentification
appears at redshifts where the relative difference between the central
wavelengths of different pairs/triplets of emission lines is the same, and thus
they fall in the same survey bands. The number and width of these redshift
intervals depend on the width of the survey bands, where the narrower the
bands the smaller the incidence. In Fig.~\ref{fig:diff-lines} we display the
observed central wavelength of several AGN and galaxy emission lines as a
function of $z$. Moreover, we plot the transmission curves of the ALHAMBRA
medium-band filters. They guide the eye to see the band where different emission
lines fall as a function of $z$. The grey areas highlight the redshift interval
for which a triplet of galaxy emission lines can be confused with triplets of
AGN emission lines. This is at $z\simeq0.2$ where the galaxy emission lines
\{[O~{\small II}], [O~{\small III}], H~$\alpha$\} can be confused with the
AGN lines \{C~{\small IV}, C~{\small III}], Mg~{\small II}\} at $z\simeq2$
and \{O~{\small VI}+Ly~$\beta$, Si~{\small IV}+O~{\small IV}], C~{\small
III}]\} at $z\simeq3.5$. 

The incidence of line misidentification for pairs of galaxies and AGN emission
lines is much higher than for triplets, causing low-$z$ low-$z$ star-forming
galaxies to be confused with high-$z$ type-{\small I} AGN. This is important
because the number density of star-forming galaxies is much greater than the
number density of type-{\small I} AGN. In addition, misidentification of AGN
emission lines may lead to catastrophic redshift solutions. We study this in
detail in \S\ref{sec:precision}.

Finally, the presence of spurious lines in the multi-band data can be a possible
source of mis-classification. We define a spurious line as a line detected by
{\small ELDAR} in a band where no emission lines should fall according to
$z_{\rm spec}$. They mostly appear due to photometric errors, and their
incidence depends on the criterion chosen to confirm emission lines,
$\sigma_{\rm line}$, where the smaller its value the higher the frequency.
The also may appear due to the blending of two sources or stellar spikes. 

To get a rough estimation of the impact of spurious lines in ALHAMBRA, we
consider the case of a mock source with a flat SED. Then, we compute the
magnitude and uncertainty in each ALHAMBRA medium-band, where the uncertainties
are computed using ALHAMBRA empirical errors\footnote{We use all the ALHAMBRA
objects with good photometry and F814W~$>24.5$ to compute empirical error curves
for each ALHAMBRA band as a function of the magnitude in the band.}. After that,
we perturb the magnitude in each band $10^5$ times using Gaussian distributions
with width equal to the $1\,\sigma$ uncertainty in the band, generating $10^5$
random realisations of the mock source. In Fig.~\ref{fig:exspu} we show two of
these realisations. In the first one, {\small ELDAR} detects spurious lines in
the 2nd and 4th bands, the same bands where \{C~{\small IV}, C~{\small
III}]\} fall at $z=1.48$. In the second, {\small ELDAR} confirms spurious lines
in the 4th, 8th, and 12th bands, the same bands where \{Ly~$\alpha$, C~{\small
IV}, C~{\small III}]\} fall at $z=2.76$. Therefore, these objects could be
wrongly classified as type-{\small I} AGN by certain configurations of {\small
ELDAR}. Moreover, the incidence of spurious confirmations is even higher for
objects with real emission lines.

The number of sources wrongly confirmed as type-{\small I} AGN due to spurious
lines depends on $\sigma_{\rm line}$ and $\mathcal{N}$, where the higher their
values the lower the contamination. Therefore, it is very important to take this
into account before choosing the value of $\mathcal{N}$ and $\sigma_{\rm line}$.
In addition, another effect that increases the number of spurious lines is a bad
calibration of the zeropoints of the survey bands; however, this is not an issue
for us because the values ALHAMBRA zeropoints are very robust (for a detailed
discussion see M14).

\subsection{Specific configuration of ELDAR for the ALHAMBRA survey}
\label{sec:ELDARALH}

Here we configure {\small ELDAR} to identify type-{\small I} AGN in the ALHAMBRA
survey. In order to do this, we start by optimising {\small LePHARE}, and then
we tune the configuration of the spectro-photometric step to extract two samples
of type-{\small I} AGN, where the first prioritises completeness and the second
a reduced galaxy contamination.

Given the width of the ALHAMBRA bands, the only type of AGN that we can securely
detect are the ones with broad emission lines, i.e. type-{\small I} AGN.
Consequently, we will only introduce templates describing the SED of these
objects in the extragalactic library of {\small LePHARE}. Specifically, we
select the empirical templates of quasars and AGN used in \citet{salvato09,
salvato11} and the synthetic templates of quasars included in the {\small
LePHARE} distribution. The resulting library encompasses 49 templates, where 31
of them are synthetic templates that employ different power laws for the AGN
continuum and EWs for the emission lines. From this list, we select the
templates that give the best results in terms of completeness and redshift
precision for a sample of spectroscopically-known AGN within the ALHAMBRA
fields, which we call {\small AGN-S}.

The AGN-S sample is obtained by performing a crossmatch between the
spectroscopically confirmed point-like type-{\small I} AGN (sources with Q or A
flags) from the Million Quasar Catalogue\footnote{
\href{http://heasarc.gsfc.nasa.gov/w3browse/all/milliquas.html}
{http://heasarc.gsfc.nasa.gov/w3browse/all/milliquas.html}} \citep[MQC,]
[references within] {flesch15} and the ALHAMBRA sources with F814W~$<23$. The
MQC is a largely complete compendium of AGN from the literature through 21 June
2016. We do the match for objects separated by less than 2 arcsec and, in the
two cases where we find two ALHAMBRA sources within 2 arcsec of the MQC object,
we visually confirm the match by looking at the ALHAMBRA photometry (in both
cases we validate the match with blue objects that clearly exhibit broad
emission lines). In addition, we also perform a crossmatch between the ALHAMBRA
sources with F814W~$<23$ and the 637 type-{\small I} AGN from the COSMOS-Legacy
X-ray catalog (C-COSMOS) \citep[][]{civano16, marchesi16} with an optical
counterpart and spectroscopic redshift, following the same matching procedure as
for the MQC. We end up with a total of 295 sources for the AGN-S sample.

\begin{table}
\begin{center}
\caption{\label{tab:temp11} Extragalactic templates that we introduce in
{\small LePHARE}.}
\begin{tabular}{rll}
\hline
Index&Template&Class\\
\hline
1&I22491\_70\_TQSO1\_30 			& Quasar $30\,\%$ + Gal. $70\,\%$[1]\\
2&I22491\_60\_TQSO1\_40 			& Quasar $40\,\%$ + Gal. $60\,\%$[1]\\
3&I22491\_50\_TQSO1\_50 			& Quasar $50\,\%$ + Gal. $50\,\%$[1]\\
4&I22491\_40\_TQSO1\_60 			& Quasar $60\,\%$ + Gal. $40\,\%$[1]\\
5&pl\_I22491\_30\_TQSO1\_70 	& Quasar $70\,\%$ + Gal. $30\,\%$[1]\\
6&pl\_I22491\_20\_TQSO1\_80 	& Quasar $80\,\%$ + Gal. $20\,\%$[1]\\
7&pl\_QSO\_DR2\_029\_t0  		  & Quasar low lum.[1]\\
8&pl\_QSOH 									  & Quasar high lum.[1]\\
9&pl\_TQSO1 									& Quasar high IR lum.[1]\\
10&qso-0.2\_84								& Quasar synthetic[2]\\
11&QSO\_VVDS							    & Quasar[3]\\
12&QSO\_SDSS							    & Quasar[4]\\
\hline
\end{tabular}
\end{center}
\textbf{References}. [1] \citet{salvato09}, [2] {\small LePHARE} distribution,
[3] VVDS composite \citep{gavignaud06}, and [4] SDSS composite
\citep{vandenberk01}. Templates starting with ${\rm pl}\_$ are extended into the
UV using a power law \citep[see][]{salvato09}.
\end{table}

\begin{figure}
\begin{center}
\includegraphics[width=0.45\textwidth]{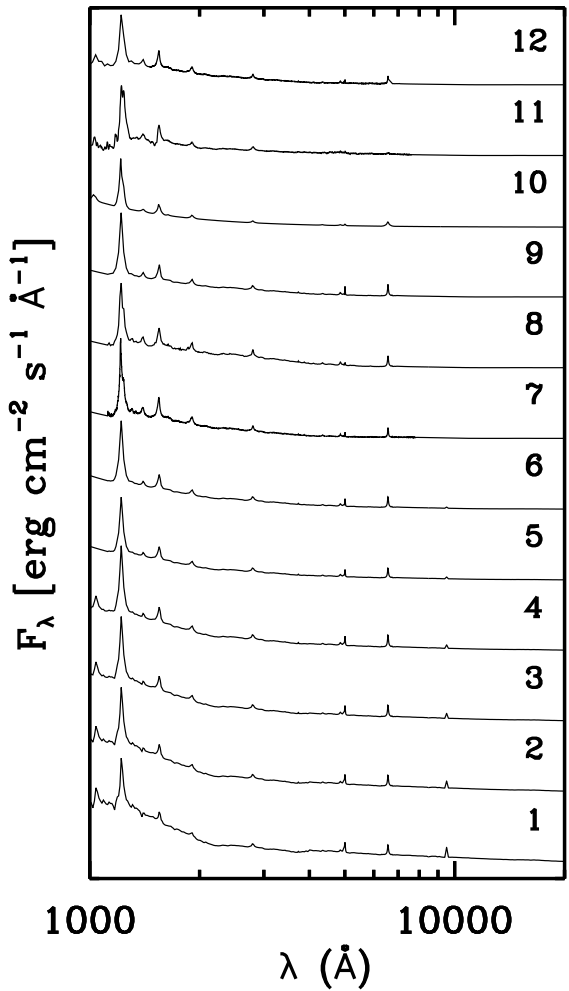}
\end{center}
\caption{\label{fig:temp} Extragalactic templates that we introduce in {\small
LePHARE}. They are sorted in the same order as in Table~\ref{tab:temp11} and
the fluxes are expressed per unit wavelength.}
\end{figure}

Then, to select the final list of templates:

\begin{itemize}

\item We start by running {\small LePHARE} on the AGN-S sample, and then
we reject all the templates that are not assigned to any source at its
spectroscopic redshift.

\item We compute the redshift precision for the AGN-S sample (using the mode of
the PDZ produced by {\small LePHARE}) by employing the remaining templates but
one at a time, and we reject the templates that do not degrade the redshift
precision.

\end{itemize}

We end up with the 12 templates listed in Table~\ref{tab:temp11} and plotted in
Fig.~\ref{fig:temp}. The templates 1-6 are from \citet{salvato09} and show the
SEDs of starburst galaxies and type-{\small I} AGN in different proportions; the
templates 7-9 are also from \citet{salvato09} and present the SEDs of pure
type-{\small I} AGN; the template 10 is from the {\small LePHARE} distribution
and describes the SED of a synthetic quasar; and the templates 11-12 are quasar
composite templates, the first from the VIMOS-VLT Deep Survey \citep[VVDS,
][]{gavignaud06} and the second from the SDSS survey \citep{vandenberk01}.

All these templates are at rest-frame. In order to compute precise redshifts for
type-{\small I} AGN, we have to define the redshift interval and step for
displacement (see discussion in \S\ref{sec:lephare}). We set the maximum
redshift to be $z=6$, as at $z>6$ most of the AGN emission lines of
Table~\ref{tab:lines} are outside the ALHAMBRA optical coverage,
making impossible for {\small ELDAR} to confirm any source. As for the redshift
step, we set it to be $\Delta z=0.01$, which is approximately the redshift
precision that can be achieved using ALHAMBRA data. We have checked that a
finer redshift step does not produce a higher redshift precision for the AGN-S
sample.

We impose a flat prior on the absolute magnitude in the ALHAMBRA band F830W of
$-30<M_{\rm F830W}<-20$, which is a luminosity prior appropriate for our search
of type-{\small I} AGN. The prior is set in the F830W band because it is the
medium-band whose central wavelength is the closest to the one of the ALHAMBRA
synthetic detection band, F814W.

After tuning {\small LePHARE}, we need to define the configuration of the
spectro-photometric confirmation step. We have to select $\mathcal{N}$ and
$\sigma_{\rm line}$, whose values depend on the levels of galaxy contamination
and completeness that we want to achieve. In the present analysis we decide to
extract two different samples of type-{\small I} AGN by defining two different
{\small ELDAR} configurations, the first prioritising completeness and the
second a small galaxy contamination. The specific characteristics of these
configurations are the following:

\begin{itemize}

\item {\bf 2-lines mode}: We require $\mathcal{N}=2$, $\sigma_{\rm line}=1.5$,
and F814W~$=22.5$ as limiting magnitude. The first requirement sets the minimum
redshift for confirming sources to $z=1$, as it is the minimum redshift for
which two AGN emission lines of Table~\ref{tab:lines} fall within the ALHAMBRA
optical coverage.

\item {\bf 3-lines mode}: We demand $\mathcal{N}=3$, $\sigma_{\rm line}=0.75$,
and F814W~$=23$ as limiting magnitude. The requirement of detecting at least
three AGN emission lines fixes the minimum redshift to $z=1.5$. It also enables
the possibility of confirming fainter sources and lines with lower contrast, as
a higher value of $\mathcal{N}$ reduces the galaxy contamination (see Appendix
B). Nevertheless, we relax this condition to $\mathcal{N}=2$ only for sources at
$z_{\rm phot}>5$ to increase the completeness, as the total number of emission
lines within the ALHAMBRA medium-band wavelength coverage at $5<z<5.6$ is 3
\{O~{\small VI}+Ly~$\beta$, Ly~$\alpha$, Si~{\small IV}+O~{\small IV}]\} and
at $5.6<z<6$ is just 2 \{O~{\small VI}+Ly~$\beta$, Ly~$\alpha$\}.

\end{itemize}

The previous {\small ELDAR} configurations are selected to minimise the fraction
of false detections while pushing the completeness and magnitude limit. For the
2-lines mode 2e select select a greater value of $\sigma_{\rm line}$ than for
the 3-lines mode to reduce the contamination from galaxies due to spurious
lines. In Appendix B we explore the completeness, redshift precision, and galaxy
contamination in the case of different values of $\mathcal{N}$, $\sigma_{\rm
line}$, and F814W magnitude cuts.

For objects with the Lyman-break within the ALHAMBRA medium-band wavelength
coverage, we set the additional requirement that these objects cannot
have a $3\,\sigma$ flux detection in more than one band with a central
wavelength smaller than the Lyman-break ($912\,\rm{\AA}$) at rest-frame. We
allow flux detection in one band because of metal lines with $\lambda_{\rm
c}<912\,\rm{\AA}$, such as NeVIII and MgX. This criterion aims at rejecting
low-$z$ galaxies for which the $4000\,\rm{\AA}$ break is confused with the
Lyman-break.

Finally, as low-$z$ galaxies have extended Point Spread Function (PSF) whereas
type-{\small I} AGN at $z>1$ are point-like, we do not apply {\small ELDAR} to
sources with extended morphology. To characterise the morphology, we employ the
Stellarity parameter of SExtractor \citep{bertin96}, which is 1 for point-like
sources and 0 for extended ones, and we do not run {\small ELDAR} on sources
with Stellarity~$<0.2$. We do not select a higher cut-off because in ground base
surveys, if data obtained with bad seeing are stacked together, the PSF gets
smeared \citep[see][for a demonstration with AGN]{hsu14}. However, if the value
of Stellarity is smaller than $0.2$, the probability of the source to be
point-like is very low for ALHAMBRA sources with F814W~$<23$ (see M14). We
explore further contamination from low-$z$ galaxies in
\S\ref{sec:contamination}.

The same steps followed here to tune {\small ELDAR} for the ALHAMBRA survey can
be used to adjust the {\small ELDAR} configuration for surveys with different
filter systems and depths.

\begin{figure*}
\begin{center}
\includegraphics[width=0.325\textwidth]{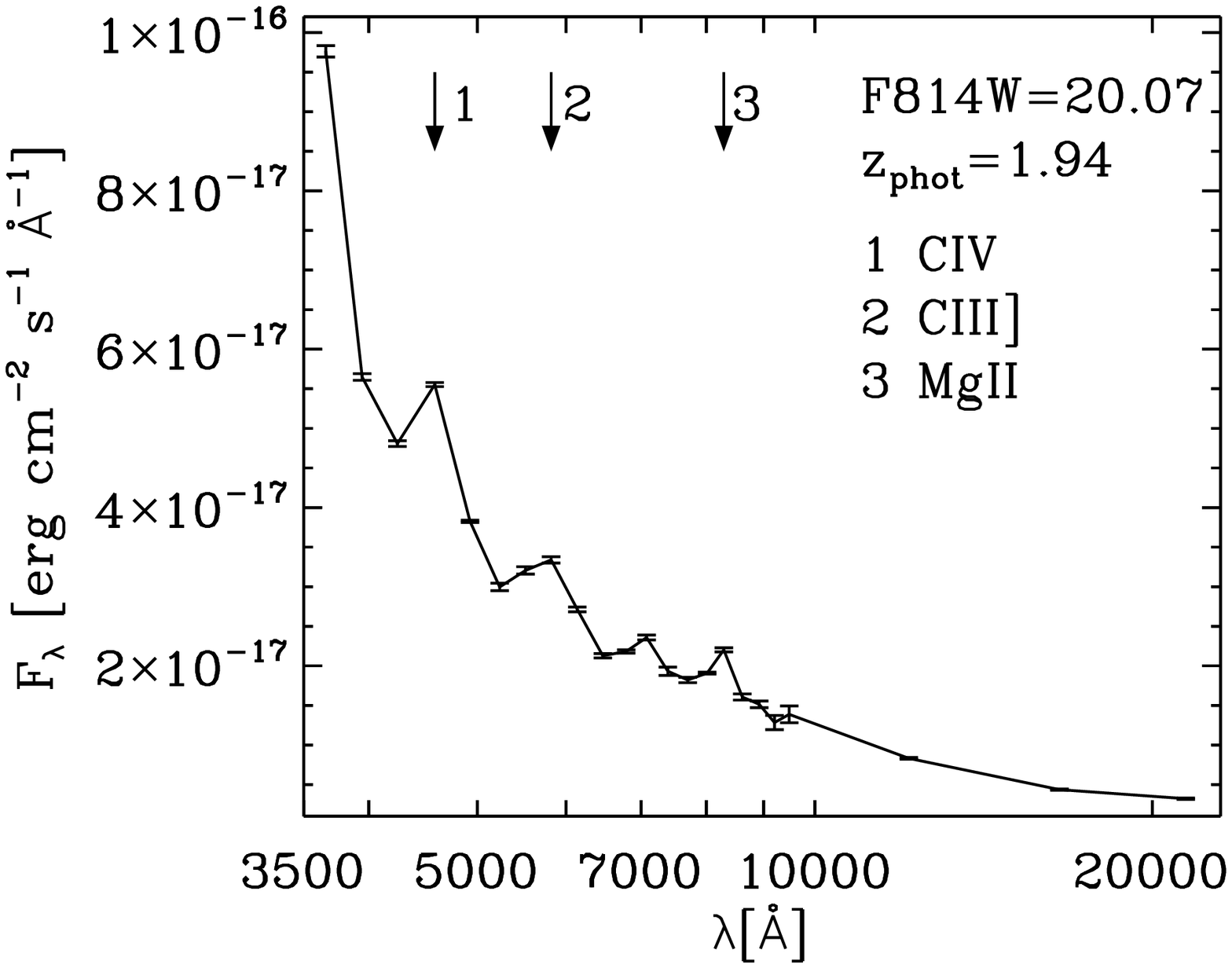}\includegraphics[width=0.325\textwidth]{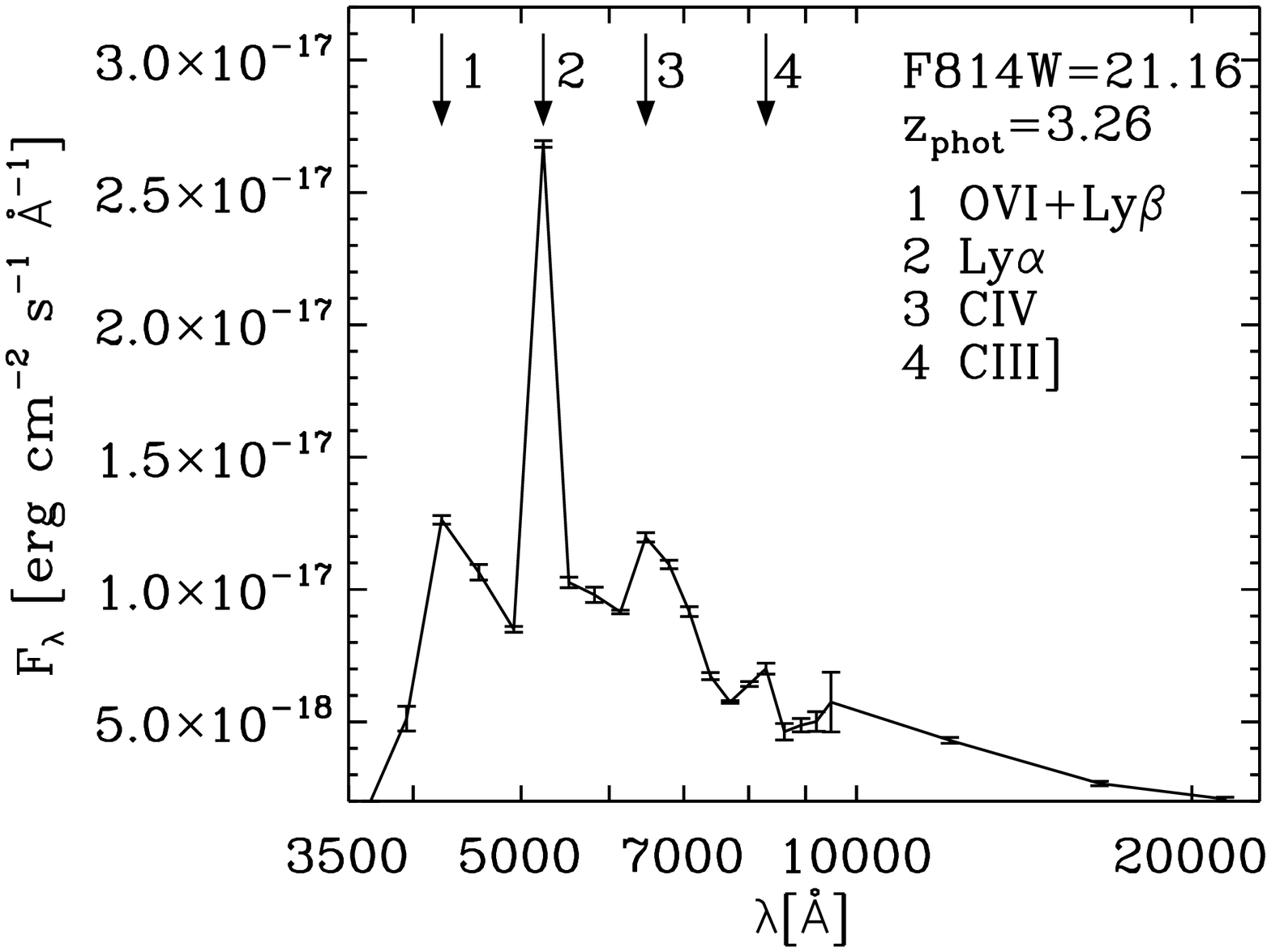}\includegraphics[width=0.325\textwidth]{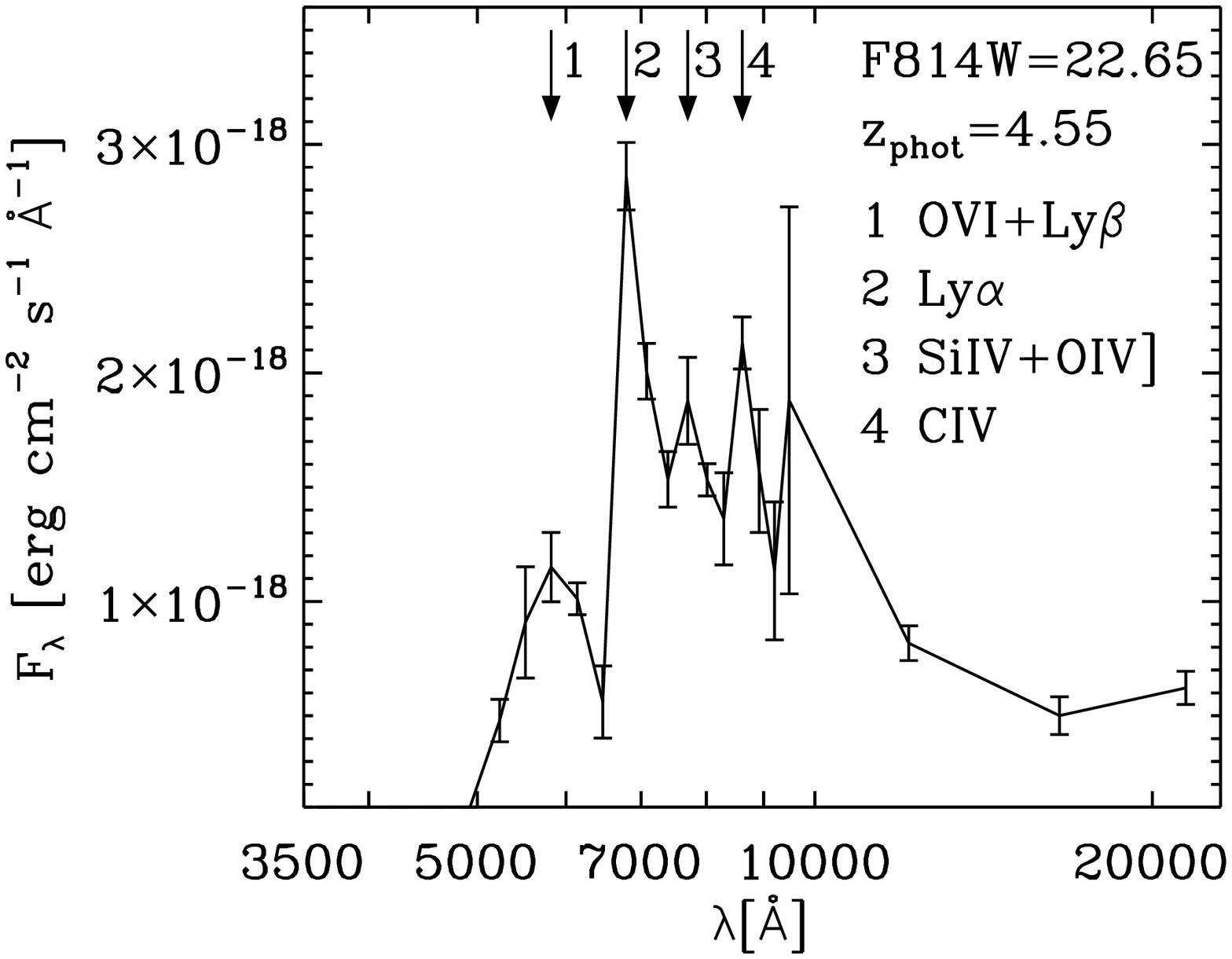}
\end{center}
\caption{\label{fig:ex_text1} SEDs of type-{\small I} AGN detected by {\small
ELDAR} using ALHAMBRA data. None of them are spectroscopically-known. We can
clearly see some peaks in the SEDs of these objects, where they correspond to
typical AGN emission lines. The arrows point to the bands where AGN emission
lines are confirmed by {\small ELDAR}. {\bf Left panel:} object of both the
ALH2L and ALH3L catalogues at $z_{\rm phot}=1.935$ with IDs ALH2L346 and
ALH3L186, respectively. {\small ELDAR} detects three AGN emission lines for this
object. {\bf Central panel:} object of the ALH2L and ALH3L catalogues at $z_{\rm
phot}=3.258$ with IDs ALH2L560 and ALH3L450, respectively. {\small ELDAR}
detects four AGN emission lines for this object. {\bf Right panel:} object of
the ALH3L catalogue at $z_{\rm phot}=4.549$ with ID ALH3L490. {\small ELDAR}
detects four AGN emission lines for this object. It is not confirmed by the
2-lines mode of {\small ELDAR} because this object is fainter than
F814W~$22.5$ (magnitude limit of the 2-lines mode).}
\end{figure*}

\subsection{Summary of the ELDAR configuration for the ALHAMBRA survey}
\label{sec:ELDsum}

In \S\ref{sec:methodology} we described the main characteristics of {\small
ELDAR} and in \S\ref{sec:ELDARALH} we tuned our methodology to identify
type-{\small I} AGN using ALHAMBRA data. In what follows, we summarise how
{\small ELDAR} works and its main properties for this specific survey:

\begin{itemize}

\item To extract a catalogue of type-{\small I} AGN from the ALHAMBRA data we
run {\small LePHARE} over all non-extended sources of the ALHAMBRA survey
(Stellarity~$>0.2$) using templates describing the SEDs of stars and
type-{\small I} AGN. We reject the objects best-fitted by stellar templates.

\item After that, {\small ELDAR} looks for the AGN emission lines gathered in
Table~\ref{tab:lines} at the redshifts in which the value of the PDZ is greater
than 0.5. Later, it confirms the AGN emission lines detected with $\sigma_{\rm
line}=1.5$ for the 2-lines mode and with $\sigma_{\rm line}=0.75$ for the
3-lines mode. These requirements set a minimum redshift for detecting
type-{\small I} AGN of $z_{\rm min}=1$ for the first mode and $z_{\rm min}=1.5$
for the second.

\item Next, the 2-lines mode of {\small ELDAR} confirms as type-{\small I} AGN
the sources with F814W~$<22.5$ and at least two detected AGN emission lines, and
the 3-lines mode validates the objects with F814W~$<23$ and at least three
detected AGN emission lines. Additionaly, we require for both modes the
detection of Ly~$\alpha$ for objects at $z_{\rm phot}>2$ and that the flux in
the band where Ly~$\alpha$ falls has to be greater than the $75\,\%$ of the
maximum flux in any of the other band. We also demand no flux detection in more
than one band whose central wavelength is smaller than the Lyman-break at
rest-frame.

\item Finally, both modes compute the redshift of the confirmed sources using
Eq.~\ref{eq3:zres} (see the mock example in Fig.~\ref{fig:explain}).

\end{itemize}



\section{The ALHAMBRA ALH2L and ALH3L catalogues}
\label{sec:results}

To determine the effectiveness of {\small ELDAR} in detecting type-{\small I}
AGN, in this section we apply the 2- and 3-lines modes of {\small ELDAR} to the
ALHAMBRA data. We will end up with two type-{\small I} AGN samples, the ALH2L an
ALH3L catalogues, respectively. We will present their properties and discuss
their quality in terms of redshift precision, completeness, and contamination
from galaxies and stars.

We start by selecting the ALHAMBRA sources to be analysed. From the $446\,361$
sources of the M14 catalogue with good photometry (Satur\_Flag and
DupliDet\_Flag equal to zero), we pick the $41\,367$ no extended objects
(Stellarity~$>0.2$) with F814W~$<23$. We then run {\small LePHARE} on these
sources, rejecting the $20\,580$ objects best-fitted by stellar templates
($\chi^2_{\rm star}<\chi^2_{\rm AGN}$). The number of stars that we find is
approximately the same as the number of stars detected in ALHAMBRA using a
combination of the apparent geometry of the sources, their F814W magnitudes, and
optical and near infrared colours (see M14). After that, we apply the
spectro-photometric confirmation step to the remaining sources. For the 2-lines
mode of {\small ELDAR} we end up with 585 type-{\small I} AGN with $z>1$ and
F814W~$<22.5$ (ALH2L catalogue) and for the 3-lines mode with 494 type-{\small
I} AGN with $z>1.5$ and F814W~$<23$ (ALH3L catalogue). They have 316 sources in
common and it is worth to notice that 461 and 408 sources of the ALH2L and ALH3L
catalogues, respectively, are not spectroscopically-known. Both catalogues are
publicly available and they are detailed in Appendix C.

In Fig.~\ref{fig:ex_text1} we display the SEDs of three
spectroscopically-unknown sources that {\small ELDAR} confirms as type-{\small
I} AGN. In the figure the arrows point to the bands where {\small ELDAR} detects
AGN emission lines. In the left panel we show a type-{\small I} AGN at $z_{\rm
phot}=1.94$ that belongs to both the ALH2L and ALH3L catalogues. For this object
the 2- and 3-lines modes of {\small ELDAR} detect the lines C~{\small IV},
C~{\small III}], and Mg~{\small II}. They do so despite the blue and steep
continuum of $z\sim2$ type-{\small I} AGN (we remind the reader that our
methodology assumes a flat continuum). The template that best-fits this source
is the number 10 (qso-0.2\_84 template) including a very low colour excess
($E(B-V)=0.02$). In the central panel we present an object of both the ALH2L and
ALH3L catalogues at $z_{\rm phot}=3.26$ for which {\small ELDAR} detects the
complex O~{\small VI} + Ly~$\beta$ and the lines Ly~$\alpha$, C~{\small IV}, and
C~{\small III}]. At $z\sim3$ the AGN continuum is flatter than at $z\sim2$, and
thus the detection of AGN emission lines is more straightforward at this
redshift. This object is also best-fitted by the template number 10, in this
case without any extinction. In the right panel we display the SED of an object
of the ALH3L catalogue at high-$z$ ($z_{\rm phot}=4.55$) for which {\small
ELDAR} detects the complexes O~{\small VI} + Ly~$\beta$ and Si~{\small IV} +
O~{\small IV}], and the lines Ly~$\alpha$ and C~{\small IV}. It is not included
in the ALH2L catalogue because its magnitude, F814W~$=22.65$, is dimmer than the
magnitude limit for this catalogue, set at F814W~$=22.5$. This object is
best-fitted by the template number 10 with a very low colour excess
($E(B-V)=0.04$). Moreover, it is one of the eight objects of the ALH3L catalogue
at $z_{\rm phot}>4$, where just the one at the highest redshift ($z_{\rm
phot}=5.41$) has been spectroscopically confirmed \citep[at $z_{\rm
spec}=5.41$,] [] {matute13}.

\begin{figure}
\begin{center}
\includegraphics[width=0.45\textwidth]{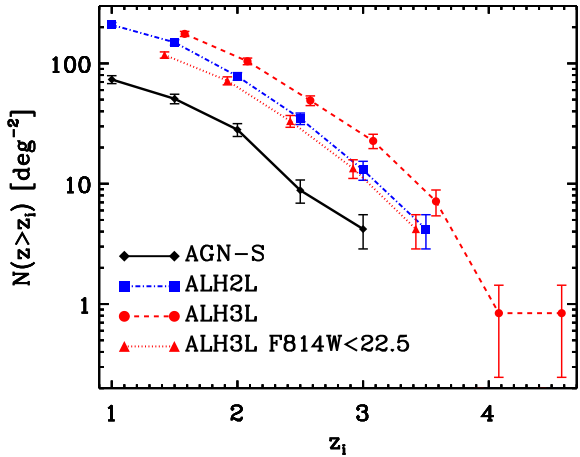} 
\end{center}
\caption{\label{fig:nden} Number density of type-{\small I} AGN at $z>z_i$. The
black solid line indicates the results for the AGN-S sample, which includes all
the spectroscopically-known type-{\small I} AGN within the ALHAMBRA fields, the
blue dot-dashed and red dashed lines for the ALH2L and ALH3L catalogues,
respectively, and the red dotted line for the objects of the ALH3L catalogue
with F814W~$<22.5$. The error bars denote Poisson errors.}
\end{figure}

\begin{figure}
\begin{center}
\includegraphics[width=0.45\textwidth]{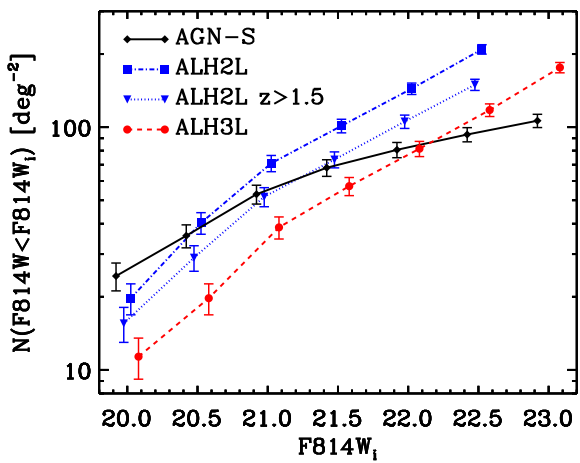}
\end{center}
\caption{\label{fig:photo-z} Number density of type-{\small I} AGN brighter than
F814W$_i$. The colour coding is the same as in Fig.~\ref{fig:nden} for the AGN-S
sample and the ALH2L and ALH3 catalogues. The blue dotted line indicates the
results for the sources of the ALH2L catalogue with $z>1.5$. If we consider just
objects with $z>1.5$ and F814W~$<22.5$, the ALH2L catalogue contains $30\,\%$
more sources than the ALH3L catalogue.}
\end{figure}

\begin{figure}
\begin{center}
\includegraphics[width=0.45\textwidth]{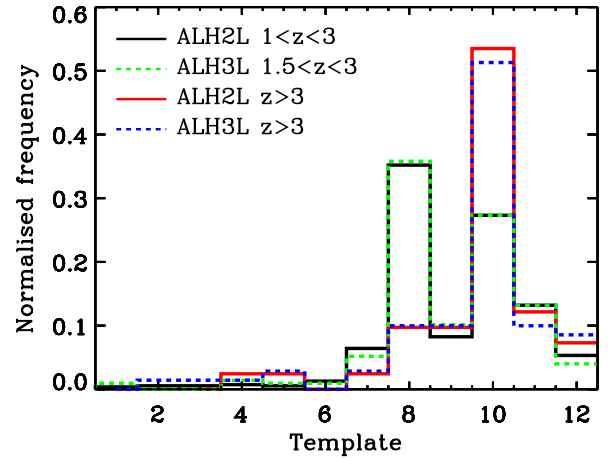}
\end{center}
\caption{\label{fig:qsotemp} Distribution of the best-fitting templates for the
ALH2L and ALH3L catalogues. The numbers in the x-axis correspond to the numbers
in Table~\ref{tab:temp11} and Fig.~\ref{fig:temp}. In general, type-{\small I}
AGN at $z<3$ prefer the template 8 and at $z>3$ the template 10.}
\end{figure}

\begin{figure*}
\begin{center}
\includegraphics[width=0.975\textwidth]{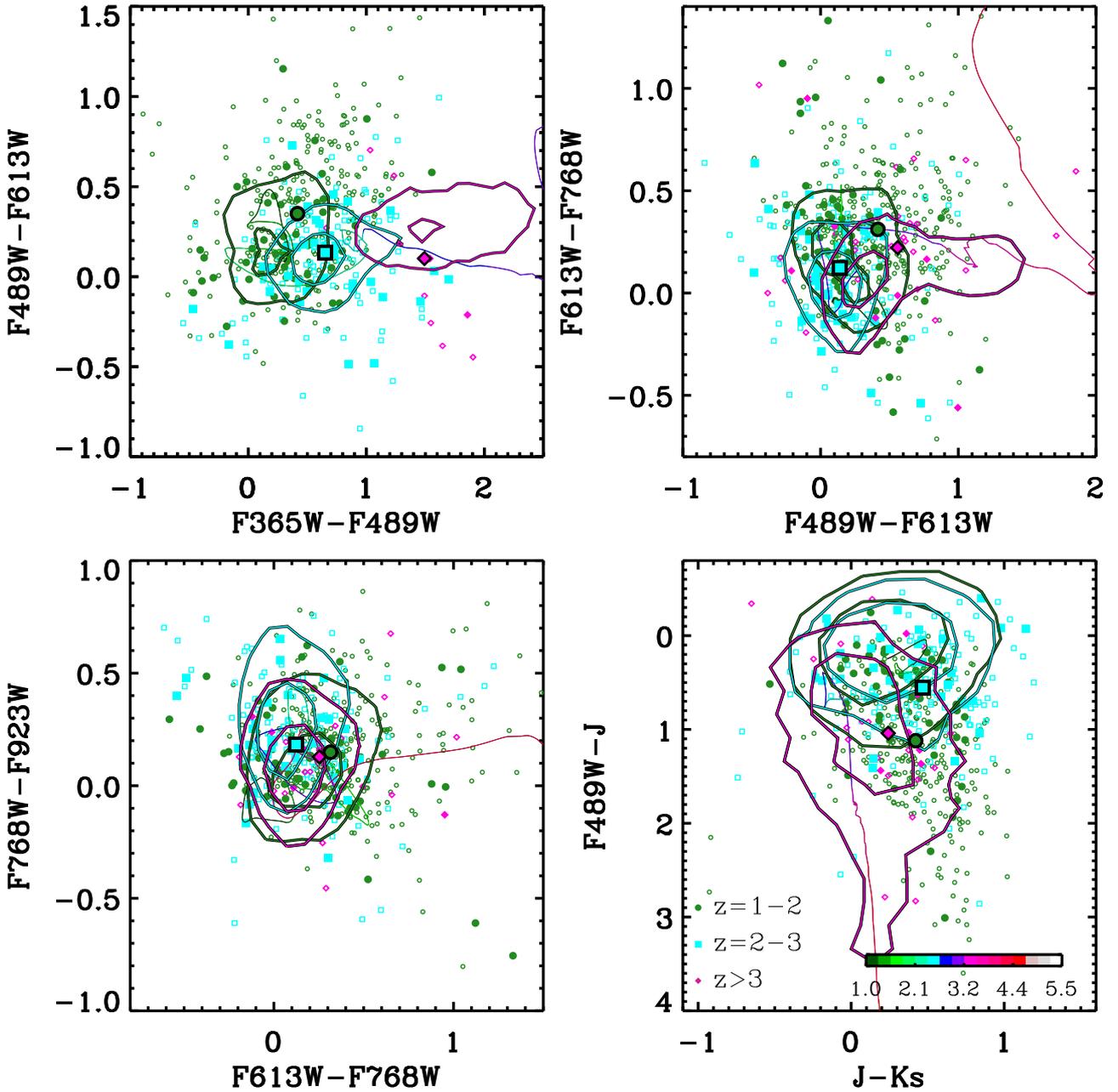}
\end{center}
\caption{\label{fig:colors} Colour-colour diagrams for the ALH2L catalogue. Only
objects with photometric errors smaller than 0.2 mag in the bands shown are used
to generate each panel. The colour of the symbols and lines indicates the
redshift of the sources, as stated in the legend. Filled (open) symbols denote
ALH2L objects that are (not) in common with the AGN-S sample, and big symbols
indicate the median colours for all the objects at a certain redshift. Contours
outline the colour loci of quasars from the SDSS-DR12 Quasar catalogue
(top-left, top-right, and bottom-left panels) and the SDSS-DR6 Quasar catalogue
with a counterpart in UKIDSS-LAS (bottom right panel), where the inner contours
encloses the $0.5\,\%$ of the sample and the outer contour the $3\,\%$. From the
top-left to bottom-right, for the SDSS quasars we show $g-r$ vs $u-g$, $r-i$ vs
$g-r$, $i-z$ vs $r-i$, $g-J$ vs $J-K_s$. Narrow lines display the
evolution of the colours of the template pl\_QSOH as a function of $z$.}
\end{figure*}

\begin{figure*}
\begin{center}
\includegraphics[width=0.975\textwidth]{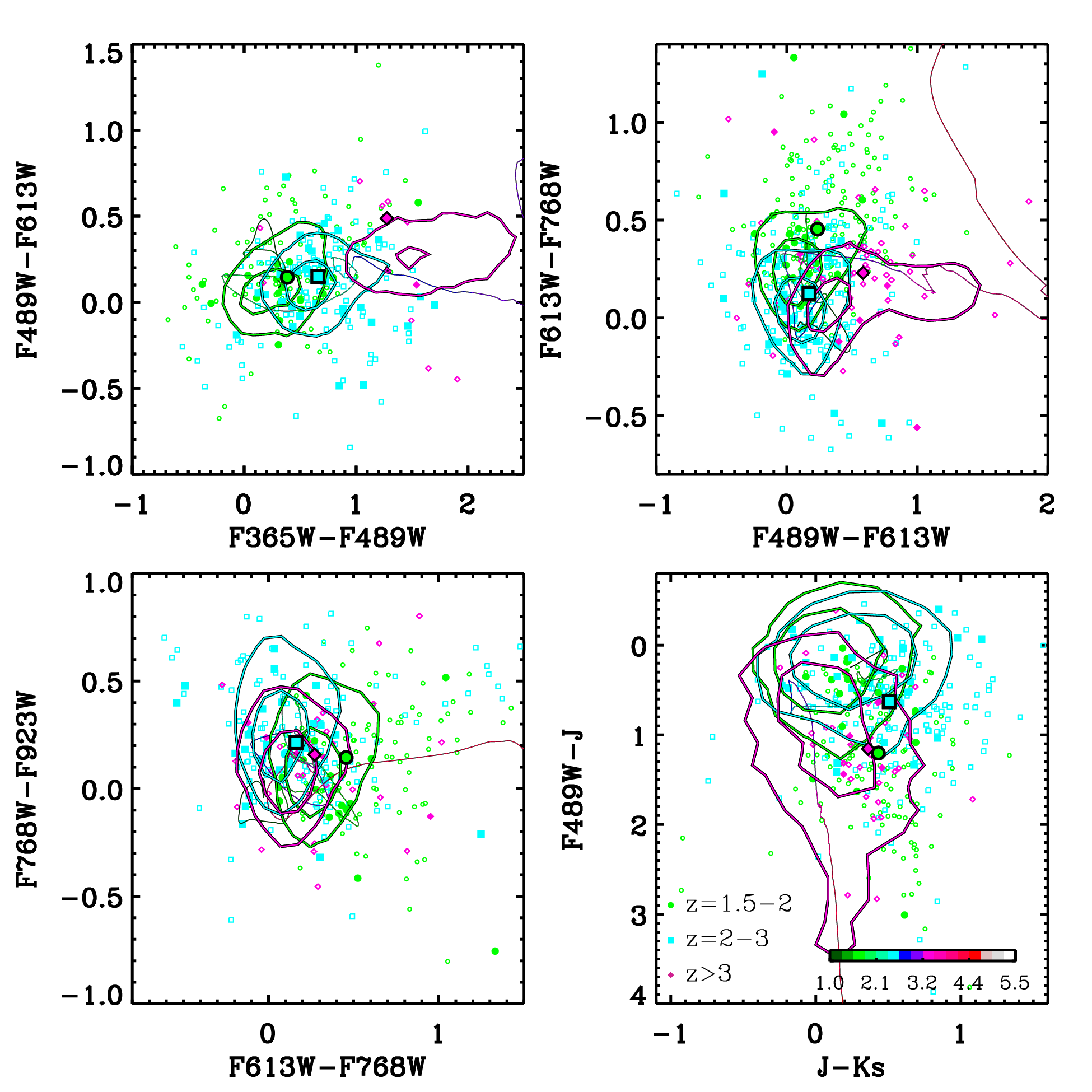}
\end{center}
\caption{\label{fig:colors1} Colour-colour diagrams for the ALH3L catalogue. We
employ the same colour coding as in Fig.~\ref{fig:colors}.}
\end{figure*}

\subsection{Properties of the ALH2L and ALH3L catalogues}

In this section we show the magnitudes, redshifts, best-fitting templates, and
colours of the objects of the ALH2L and ALH3L catalogues.

To compute the number density of the ALH2L and ALH3L catalogues we need the
effective area of the ALHAMBRA survey. To obtain it, we employ a mask generated
by \citet{arnalte14} that excludes low exposure time areas, obvious defects in
the images, and circular regions around saturated stars. After applying this
mask, the effective area of the ALHAMBRA survey is $2.38\,\text{deg}^2$. We
apply the same mask to the ALH2L and ALH3L catalogues, finding 498 and 419
objects within the mask, respectively, which correspond to a surface number
density of $\simeq209\,\text{deg}^{-2}$ and $\simeq176\,\text{deg}^{-2}$. In
Fig.~\ref{fig:nden} we display the number density of both catalogues as a
function of redshift. The blue dot-dashed and red dashed lines indicate the
number density for the ALH2L and ALH3L catalogues, respectively, and the black
line does so for the AGN-S sample, which includes all the spectroscopically-known
type-{\small I} AGN within the ALHAMBRA fields. As there are not obvious gaps in
the redshift distribution of the ALH2L and ALH3L catalogues, we conclude that
{\small ELDAR} uniformly identifies type-{\small I} AGN as a function of
redshift. This is thanks to the continuity of the ALHAMBRA medium-bands.
Non-contiguous bands would introduce gaps in the redshift distribution due to
emission lines falling in between them.

In Fig.~\ref{fig:nden} we also show the number density for the objects of the
ALH3L catalogue with F814W~$<22.5$. As we can see, the 2- and 3-lines modes of
{\small ELDAR} approximately detect the same number of type-{\small I} AGN at
$z>2$ with F814W~$<22.5$. The main strength of the first is that it allows to
detect more objects than the second at $z<2$, whereas the best virtue of the
3-lines mode is that it allows to robustly confirm type-{\small I} AGN at lower
SNR.

In Fig.~\ref{fig:photo-z} we display the number density of type-{\small I} AGN
for the AGN-S sample and for the ALH2L and ALH3L catalogues as a function of the
F814W magnitude. The number of sources detected by {\small ELDAR} grows like a
power law up to the magnitude limit of the catalogues (F814W~$=22.5$ and
F814W~$=23$ for the ALH2L and ALH3L catalogues, respectively); however, for the
AGN-S sample it increases more slowly at F814W~$>21$. Consenquently, given that
that the contamination from galaxies and stars for the ALH2L and ALH3L
catalogues is low (see \S\ref{sec:quality}), the approach of {\small ELDAR} is
the most effective way of detecting faint type-{\small I} AGN in multi-filter
surveys.

In Fig.~\ref{fig:photo-z} we can also see that for sources at $z>1.5$ and with
F814W~$<22.5$, the number of objects in the ALH2L catalogue is $30\,\%$ greater
than in the ALH3L catalogue. We will discuss the completeness of both catalogues
in \S\ref{sec:quality}.

In Fig.~\ref{fig:qsotemp} we display the best-fitting template solution for the
ALH2L and ALH3L catalogues as a function of $z$. In general, the distribution of
templates for both catalogues is very similar, where the $35\,\%$ and $25\,\%$
of the sources at $z<3$ prefer the templates 8 and 10, and the $50\,\%$ of the
objects at $z>3$ the template 10. The template 8 presents the SED of a high
luminosity quasar and the number 10 depicts the SED of a synthetic quasar whose
continuum emission follows a power law. We also find that the $85\,\%$ of the
sources of the ALH2L and ALH3L catalogues are fitted by templates with low
extinction ($E(B-V)<0.2$), in agreement with the fact we are targeting
unobscured type-{\small I} AGN.

In an attempt to investigate whether the sources that {\small ELDAR} confirms as
type-{\small I} are the same kind of objects as the AGN that spectroscopic
surveys confirm, which are usually preselected using colour-colour diagrams, in
Fig.~\ref{fig:colors} and \ref{fig:colors1} we display four colour-colour
diagrams for the sources of the ALH2L and ALH3L catalogues, respectively, and
SDSS quasars. For the SDSS quasars we show their broad-band SDSS colours, while
for our ALHAMBRA objects we use the medium-bands colours closest to each
broad-band SDSS colour\footnote{We compute the ALHAMBRA colours with the
medium-bands whose central wavelength is the closest to the one of SDSS bands.
The correspondence is $u$ and F365W, $g$ and F489W, $r$ and F613W, $i$ and
F768W, and $z$ and F923W.}. In the figures, the symbols indicate the colours of
individual ALHAMBRA sources and the contours denote the colour loci of
spectroscopically confirmed quasars from the SDSS-DR12 Quasar catalogue
\citep{paris17} (top-left, top-right, and bottom-left panels) and the SDSS-DR6
Quasar catalogue with counterparts in the United Kingdom Infrared Telescope
Infrared Deep Sky Survey Large Area Survey (UKIDSS-LAS) \citep{peth11} (bottom
right panel). Narrow-lines show the colours of the pl\_QSOH template as a
function of $z$. The average colours of the ALH2L and ALH3L samples are
consistent with the colours of quasars observed after broad-band target
selection. The larger colour distribution of the ALH2L and ALH3L samples
(partially due to the fact that medium-bands are more sensitive to spectral
features) indicates that our method is able to select objects with broader
colour ranges. On the other hand, at $z<2$ the median colours of the objects of
the ALH2L and ALH3L catalogues are displaced with respect to the centre of the
SDSS contours. This is because SDSS does not systematically target quasars at
$z<2.15$.

We conclude that {\small ELDAR} is not only able to select and characterise the
typical quasars selected by broad-band surveys, but it has the potential of
detecting a broader range of quasar types.

\begin{figure}
\begin{center}
\includegraphics[width=0.45\textwidth]{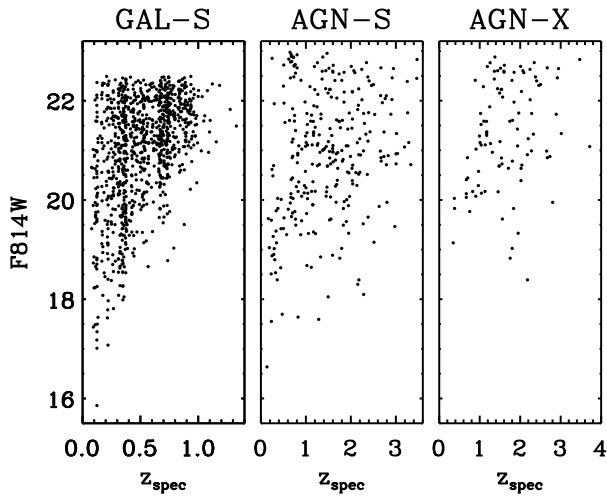}
\end{center}
\caption{\label{fig:sampprop} Redshift and F814W magnitude distribution for the
GAL-S, AGN-S and AGN-X samples. We employ these samples to asses, respectively,
the galaxy contamination, redshift precision, and completeness of the ALH2L and
ALH3L catalogues.}
\end{figure}

\subsection{Quality of the ALH2L and ALH3L catalogues}
\label{sec:quality}

In order to asses the quality of the ALH2L and ALH3L catalogues, we need samples
of spectroscopically-known type-{\small I} AGN and galaxies within the ALHAMBRA
fields. We will employ the AGN-S sample (see \S\ref{sec:ELDARALH}) and two new
samples: the first consists of X-ray selected type-{\small I} AGN in the
ALHAMBRA COSMOS field (a sub-sample of the AGN-S sample). The second includes
galaxies within the same ALHAMBRA field. We name them AGN-X and GAL-S,
respectively.

We separate X-ray selected AGN to generate the AGN-X sample because X-ray
selection produces complete samples of type-{\small I} AGN \citep{brandt05} and
thus, we can use the AGN-X sample to robustly estimate the completeness of the
ALH2L and ALH3L catalogues. In addition, X-ray selected AGN catalogues have a
low contamination from galaxies and stars \citep{lehmer12, stern12}. This
catalogue is composed of 105 sources with F814W~$<23$.

To obtain the GAL-S sample we match the objects from the DR2 of the zCOSMOS
$10$k-bright spectroscopic sample \citep[zCOSMOS hereafter]{lilly09} with secure
redshift (flags 3.x and 4.x) and the ALHAMBRA sources with F814W~$<23$. zCOSMOS
includes randomly selected galaxies with F814W~$<22.5$ at $z_{\rm spec}<1.5$ in
the COSMOS field, where the sampling rate is $\simeq 0.35$ in the area in common
with the ALHAMBRA survey. Following the same procedure as for the AGN-S sample
to do the match, we find a total of 1051 sources.

In Fig.~\ref{fig:sampprop} we display the magnitude and redshift distribution
for the objects of the GAL-S, AGN-S, and AGN-X samples. In the following
sections we will employ them to explore, respectively, the galaxy contamination,
redshift precision, and completeness produced by the 2- and 3-lines modes of
{\small ELDAR}. The results of this analysis are summarised in
Table~\ref{tab:accuracy}.

\begin{figure}
\begin{center}
\includegraphics[width=0.45\textwidth]{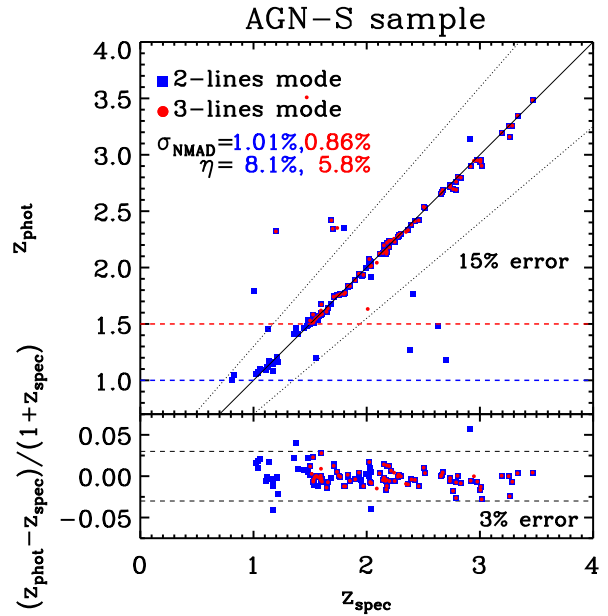}
\end{center}
\caption{\label{fig:qsopres} Comparison between photometric and spectroscopic
redshifts for the AGN-S sample using the 2- and 3-lines modes of {\small ELDAR}.
Blue squares and red circles denote objects identified by the 2- and 3-lines
modes. The solid line indicates the $1:1$ relation, the dotted lines the
threshold between good redshift solutions and outliers, and the blue (red)
dashed line the redshift cut-off for the 2-lines (3-lines) mode. The normalised
median absolute deviation, $\sigma_{\rm NMAD}$, and the fraction of outliers,
$\eta$, are in blue and red for the 2- and 3-lines modes, respectivey. The
bottom panel shows a measurement of the photo-$z$ accuracy for each source.
Dashed lines indicate $3\,\%$ errors.}
\end{figure}

\subsubsection{Redshift precision} \label{sec:precision}

We define the fraction of redshift outliers in a sample, $\eta$, as the
percentage of objects with catastrophic redshift solutions for which $|z_{\rm
spec}-z_{\rm phot}|> 0.15\,(1+z_{\rm spec})$. We estimate this fraction for the
ALH2L and ALH3L catalogues by applying the 2- and 3-lines modes of {\small
ELDAR} to the AGN-S sample, respectively. We find that the fraction of outliers
is a bit larger for the ALH2L catalogue, $\eta=8.1\,\%$, than for the ALH3L
catalogue, $\eta=5.8\,\%$. This is because the larger the number of lines
required to confirm an object, the lower is the probability for this object to
be an outlier. These outliers are caused by a degeneracy between pairs of AGN
emission lines, such as \{C~{\small III}], Mg~{\small II}]\} at $z=1.2$ and
\{Ly~$\alpha$, C~{\small III}]\} at $z=2.3$, and \{C~{\small IV}, C~{\small
III}]\} at $z=1.7$ and \{Ly~$\alpha$, C~{\small IV}\} at $z=2.4$. We show the
ALHAMBRA photometric data of some of these outliers in Appendix A.

To compute the redshift precision for the catalogues, we employ the normalised
median absolute deviation, $\sigma_{\rm NMAD}$, defined by \cite{hoaglin83} as

\begin{equation}
\sigma_{\rm NMAD}=1.48 \; \text{median}
\left(\frac{|z_{\rm phot}-z_{\rm spec}|}{1+z_{\rm spec}}\right).
\end{equation}

\noindent We use $\sigma_{\rm NMAD}$ because it is designed to be less sensitive
to redshift outliers than the standard deviation of photometric and
spectroscopic redshifts. In a distribution without redshift outliers they would
have the same value. Applying the 2- and 3-lines modes to the AGN-S sample, we
obtain a redshift precision of $\sigma_{\rm NMAD}=1.01\,\%$ and $\sigma_{\rm
NMAD}=0.86\,\%$, respectively. Therefore, the precision reached for type-{\small
I} AGN using the 3-lines mode of {\small ELDAR} is even greater that the one
achieved for galaxies and type-{\small I} AGN in other ALHAMBRA studies, see M14
and \citet{matute12}, respectively.

\begin{table}
\begin{center}
\caption{\label{tab:accuracy} Results for the AGN-S, AGN-X, and GAL-S samples
using the {\small ELDAR}'s 2- and 3-lines modes.}
\begin{tabular}{ccccc}
\hline
Sample&Mode&Compl. $(\%)$&$\sigma_{\rm NMAD}(\%)$&$\eta(\%)$\\ \hline
\multirow{2}{*}{AGN-S}&2-lines&71.7&{\bf1.01}&{\bf8.1}\\
                      &3-lines&65.2&{\bf0.86}&{\bf5.8}\\ \hline
\multirow{2}{*}{AGN-X}&2-lines&{\bf73.3}&1.15&6.8\\
                      &3-lines&{\bf66.7}&0.91&0.0\\ \hline \hline
Sample&Mode&\multicolumn{3}{c}{Galaxies confirmed as AGN}\\ \hline
\multirow{2}{*}{GAL-S}&2-lines&\multicolumn{3}{c}{4 ({\bf31\,\%})}\\
                      &3-lines&\multicolumn{3}{c}{1 ({\bf9\,\%})}\\ \hline
\end{tabular}
\end{center}
{\bf Notes.} Bold numbers indicate the estimated redshift precision,
completeness, and galaxy contamination for the ALH2L and ALH3L catalogues. The
galaxy contamination is extrapolated from the results for the GAL-S sample
assuming that the ALHAMBRA COSMOS field is representative for all the ALHAMBRA
fields.
\end{table}

In Fig.~\ref{fig:qsopres} we show the comparison between the spectroscopic and
photometric redshifts of the sources of the AGN-S sample, where the photo-$z$s
are computed using the 2- and 3-lines modes of {\small ELDAR}. The two modes
produce precise results ($\sigma_{\rm NMAD}\leq1\,\%$) with a fraction of
outlier smaller than $10\,\%$. The results are particularly good at $z>2$, where
we do not find any outlier for the 3-lines mode. In the bottom panel of
Fig.~\ref{fig:qsopres} we display relative precision of the photo-$z$s produced
by {\small ELDAR}. We find that it is greater than $3\,\%$ for $86\,\%$ and
$93\,\%$ of the sources using the 2- and 3-lines mode, respectively, which shows
that {\small ELDAR} produces accurate photo-$z$s for most of the sources.

In Table~\ref{tab:comparison} we gather the redshift precision and outlier
fraction for several X-rays selected samples (references in the caption). Most
of the sources of the \citet{cardamone10, luo10, hsu14} samples are AGN whose
SED is dominated by the host galaxy, and thus the photo-$z$ of these objects are
straightforward to compute because the $4\,000\,{\rm \AA}$ break is visible. On
the other hand, the \citet{salvato09, salvato11, fotopoulou12, matute12} samples
mostly contains type-{\small I} AGN. All these surveys, but the Lockman Hole
area, which only has broad-band filters, have broad-, medium-, and narrow-band
filters. As a consequence, the Lockman Hole sample is the one with the lowest
redshift precision and the highest fraction of outliers. In this work, using the
AGN-S sample, we obtain the best results in terms of redshift precision, which
is because of the contiguous coverage of the optical range by the 20 medium-band
filters of ALHAMBRA. Although the fraction of outliers that we obtain is not the
lowest one, we want to highlight that the AGN-S sample is not X-ray selected. If
we apply our methodology just to the AGN-X sample, we find no outliers using the
3-lines mode of {\small ELDAR}.

In Appendix~\ref{app:C} we study the redshift precision as a function of
magnitude, redshift, and the value of the {\small ELDAR} free parameters.

\begin{table}
\begin{center}
\caption{\label{tab:comparison} Redshift precision and fraction of outliers for
different AGN/quasar catalogues.}
\begin{tabular}{cccccc}
\hline
Ref.&Bands&Depth&$\sigma_{\rm NMAD}(\%)$&$\eta(\%)$\\\hline
$(a)$&30&$i^*_{AB}<22.5$& 1.2 & 6.3\\
$(b)$&32&$R<26$         & 1.2 &12.0\\
$(c)$&42&$R<26$         & 5.9 & 8.6\\
$(d)$&31&$i^*_{AB}<22.5$& 1.1 & 5.1\\
$(e)$&21&$R_c<22.5$     & 8.4 &21.4\\
$(f)$&23&$m_{678}<23.5$ & 0.9 &12.3\\
$(g)$&50&$R<23$         & 1.1 & 4.2\\
$(h)$&23&F814W~$<22.5$  & 1.01& 8.1\\
$(i)$&23&F814W~$<23$    & 0.86& 5.8\\
\hline
\end{tabular}
\end{center}
{\bf Notes.}
$(a)$ XMM-Newton-COSMOS \protect\citep[QSOV sample,][]{salvato09}. 
$(b)$ The Multiwavelength Survey by Yale-Chile
\protect\citep[X-ray sources,][]{cardamone10}. 
$(c)$ {\it Chandra} Deep Field-South \protect\citep[X-ray sources,][]{luo10}.
$(d)$ XMM-Newton- and {\it Chandra}-COSMOS \protect\citep[QSOV sample,][]{salvato11}. 
$(e)$ Lockman Hole area \protect\citep[QSOV sample,][]{fotopoulou12} 
$(f)$ ALHAMBRA \protect\citep[][]{matute12}. 
$(g)$ Extended {\it Chandra} Deep Field South \protect\citep[X-ray sources,][]{hsu14}.
$(h)$ ALH2L catalogue (this work).
$(i)$ ALH3L catalogue (this work).
\end{table}


\subsubsection{Contamination from galaxies and stars} 
\label{sec:contamination}

Because of their large number density and emission lines, star-forming galaxies
are potentially the largest sample of objects that may be incorrectly classified
as type-{\small I} AGN by {\small ELDAR}. This is because most stellar types do
not have broad emission lines like type-{\small I} AGN. We will estimate the
galaxy contamination in the ALH2L and ALH3L catalogues by applying the 2- and
3-lines modes of {\small ELDAR} to the GAL-S sample, where this sample allows us
to estimate the galaxy contamination up to F814W~$=22.5$.

\begin{figure}
\begin{center}
\includegraphics[width=0.45\textwidth]{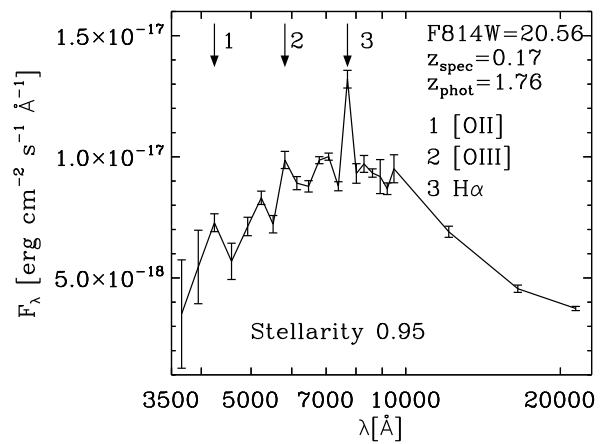}
\end{center}
\caption{\label{fig:text_fig4} Object of the GAL-S sample at $z_{\rm spec}=0.17$
that it is classified as type-{\small I} AGN by the 2- and 3-lines modes of
{\small ELDAR} at $z_{\rm phot}=1.76$. This source is confirmed by {\small
ELDAR} because there is a degeneracy between the triplet \{[O~{\small II}],
[O~{\small III}], H~$\alpha$\} at $z=0.17$ and the triplet \{C~{\small IV},
C~{\small III}], Mg~{\small II}\} at $z=1.76$.}
\end{figure}

After applying the 2- and 3-lines modes of {\small ELDAR} to the $1\,051$
galaxies of the {\small GAL-S} sample, we end up with a total of 4 and 1 objects
wrongly classified as type-{\small I} AGN, respectively. All of them show clear
emission lines, have values of Stellarity~$>0.6$, and are at $z_{\rm
spec}<0.35$. In addition, we have visually inspected their spectra to confirm
that they are low-$z$ star-forming galaxies. In Fig.~\ref{fig:text_fig4} we show
the only galaxy of the GAL-S sample that it is wrongly classified as
type-{\small I} AGN by both the 2- and 3-lines modes. It is a point-like object
(Stellarity~$=0.95$) at $z_{\rm spec}=0.17$. This source is confirmed by our
methodology because there is a degeneracy between the triplet \{[O~{\small II}],
[O~{\small III}], H~$\alpha$\} at $z=0.17$ and the triplet \{C~{\small IV},
C~{\small III}], Mg~{\small II}\} at $z=1.76$. The other galaxies that are
wrongly classified at type-{\small I} AGN by the 2-lines mode are objects for
which there is a degeneracy between pairs of galaxy and AGN emission lines. None
of them is confirmed due to spurious lines. This source of contamination is
avoided because of the optimally selected value of $\sigma_{\rm line}$ for the
2- and 3-lines modes.

The effective area of the ALHAMBRA COSMOS field is $0.203\,\text{deg}^2$, which
is $8.5\,\%$ of the total effective area of the ALHAMBRA survey,
$2.38\,\text{deg}^2$. To compute the galaxy contamination for the ALH2L and
ALH3L catalogues, we will assume that the ALHAMBRA COSMOS field is
representative for the rest of the ALHAMBRA fields. As the sampling rate for
zCOSMOS is $\simeq 0.35$ within the ALHAMBRA COSMOS field and $87\,\%$ of the
galaxies at $z_{\rm spec}<0.35$ has secure redshifts, we estimate a galaxy
contamination of 154 objects for the ALH2L catalogue and 38 for the ALH3L
catalogue. This corresponds to a galaxy contamination of $31\,\%$ for the first
and $9\,\%$ for the second. On the other hand, we expect the galaxy
contamination at $z>2$ to be zero because {\small ELDAR} assigns photo-$z$
smaller than $z_{\rm phot}=2$ to the 4 galaxies wrongly classified as
type-{\small I} AGN.

In Appendix~\ref{app:C} we study the galaxy contamination as a function of
magnitude, redshift, and the value of the {\small ELDAR} free parameters.

We do not explore the contamination from stars because we already reject all the
sources best-fitted by stellar templates and because normal stellar types do not
show emission lines with large EWs. It is possible that stellar types with a
very blue SED, e.g. O, A, and B, could be best-fitted by AGN templates; however,
they would be rejected during the spectro-photometric step because they do not
present emission lines with EWs large enough to be detected in ALHAMBRA. Another
source of contamination could be Wolf-Rayet stars since they present broad
emission lines of ionised helium, carbon, and nitrogen. Nevertheless, the
predicted total number of Wolf-Rayet stars in our region of the galaxy is
smaller than $1\,600$ \citep{hucht01}, and thus this kind of stars cannot be an
important source of contamination.


\begin{figure}
\begin{center}
\includegraphics[width=0.45\textwidth]{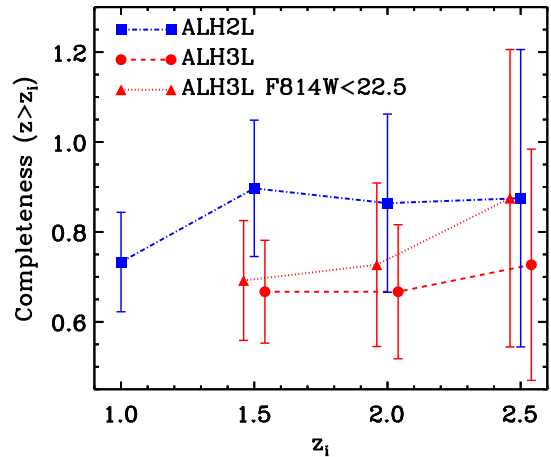}
\end{center}
\caption{\label{fig:AGNcomp} Completeness at $z>z_i$ for objects identified by
{\small ELDAR}. The completeness is defined as the percentage of
spectroscopically-known type-{\small I} AGN successfully confirmed by our method
and it is estimated using just the objects of the AGN-X sample. For sources with
$z>1.5$ the completeness is $\sim90\,\%$ for the ALH2L catalogue.}
\end{figure}

\begin{figure*}
\begin{center}
\includegraphics[width=0.45\textwidth]{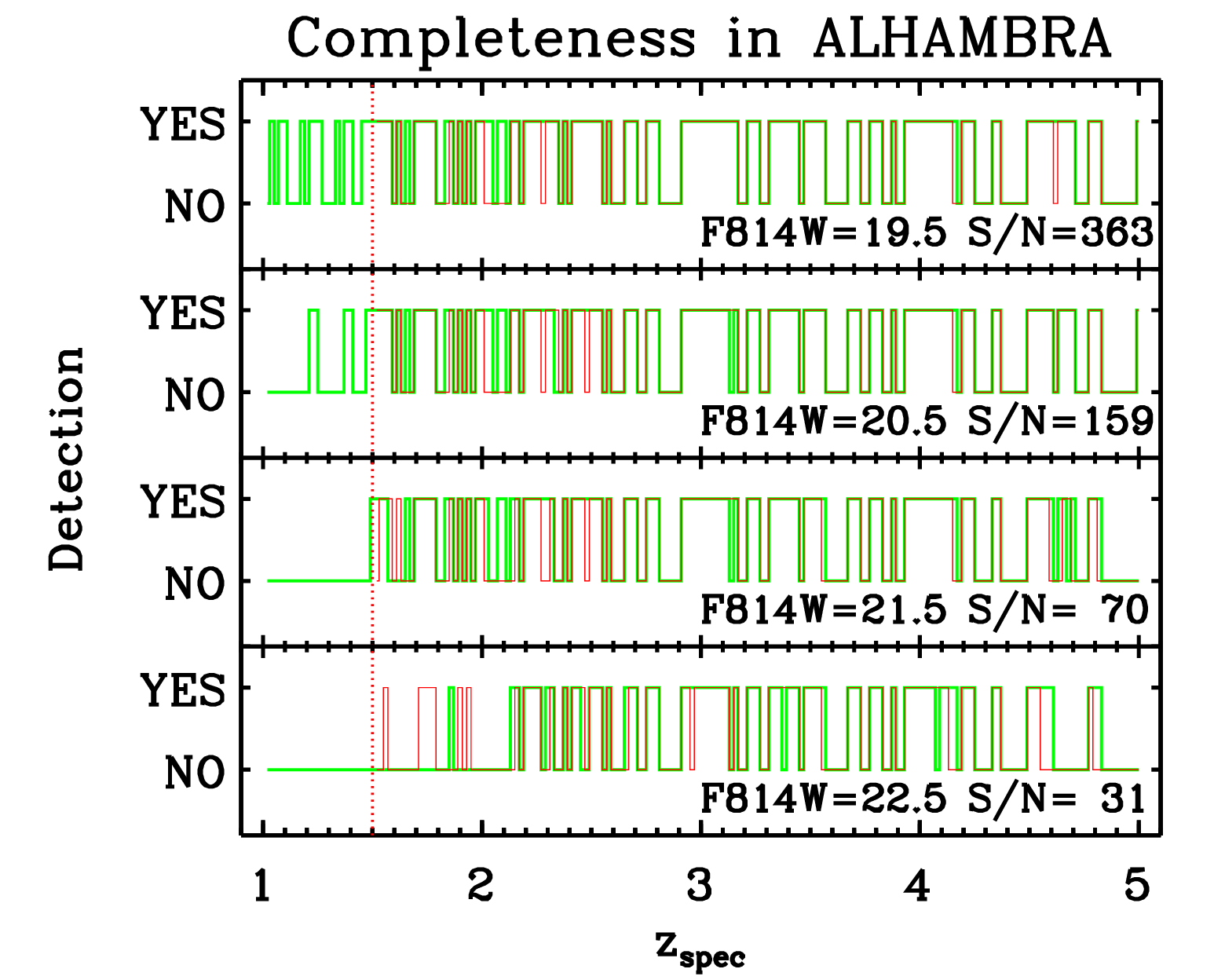} \includegraphics[width=0.45\textwidth]{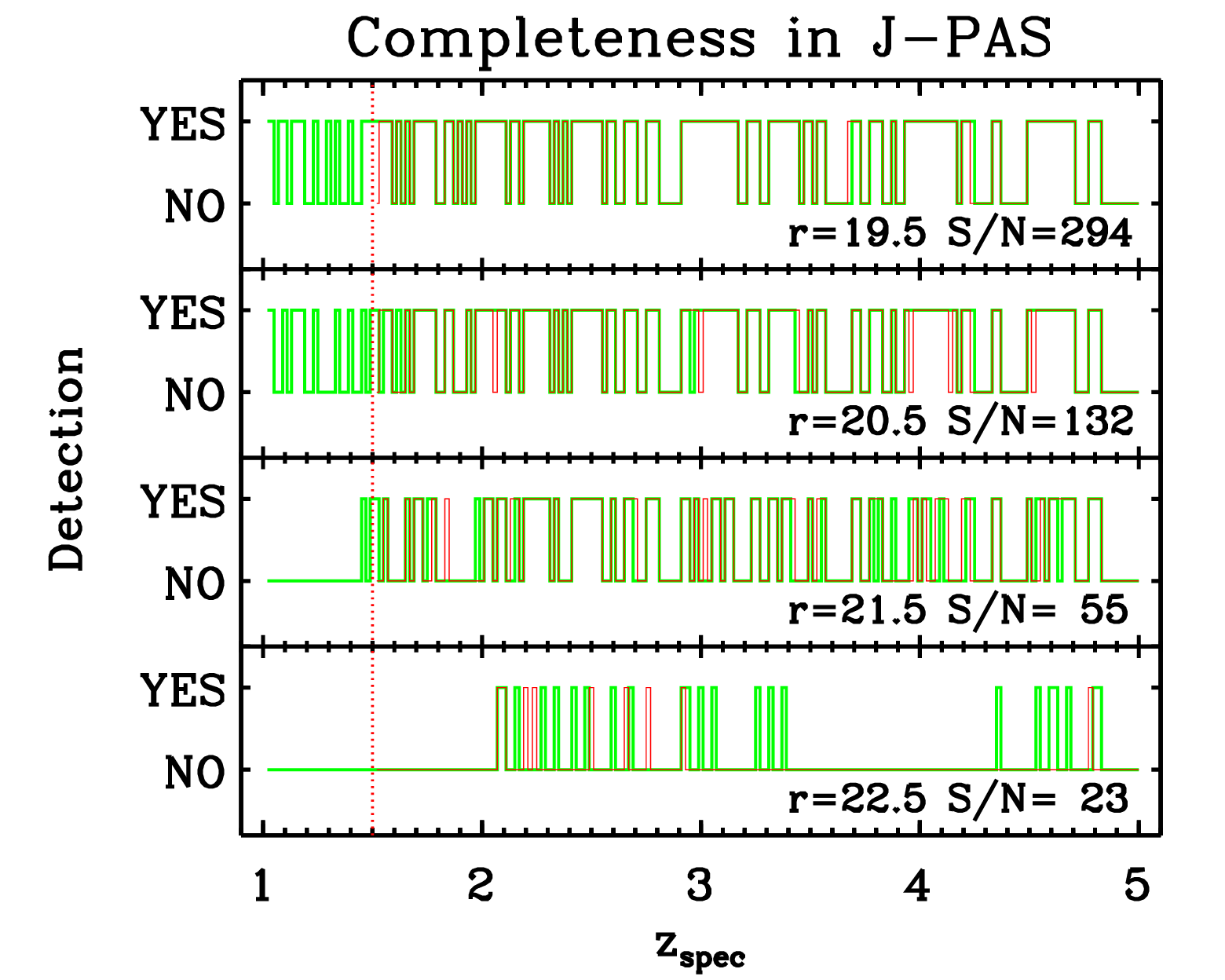}
\end{center}
\caption{
\label{fig:AGNcomp2} Detection of type-{\small I} AGN in ALHAMBRA and J-PAS
using the 2- and 3-lines methods of {\small ELDAR} as a function of the
magnitude in the detection band and the redshift of the mock source. The results
are generated by convolving the synthetic template qso-0.2\_84 with the ALHAMBRA
and J-PAS filter systems. Green and red lines show the results for the 2-lines
and 3-lines modes, respectively. There are bright objects which are not detected
due to emission lines falling in between two bands. The number of confirmations
for the faintest sources is smaller because the SNR is not large enough to
detect all the AGN emission lines that {\small ELDAR} looks for.}
\end{figure*}

\subsubsection{Completeness} \label{sec:completeness}

To estimate the completeness of the ALH2L and ALH3L catalogues, we apply the 2-
and 3-lines modes of {\small ELDAR} to the AGN-X sample. We employ this sample
because, as we explained before, X-ray selection produces largely complete
samples of type-{\small I} AGN. We find a completeness of $73\,\%$ for the first
(44 objects) and $67\,\%$ for the second (34 sources). Of the objects that the
2-lines mode does not classify as type-{\small I} AGN, $88\,\%$ of them have
PDZ$(z_{\rm spec})<0.5$. We check that we do not obtain PDZ$(z_{\rm spec})>0.5$
for them including in {\small LePHARE} all the AGN templates from
\citet{salvato09, salvato11} and from the {\small LePHARE} distribution. For the
3-lines mode we find that $60\,\%$ of the objects not confirmed as type-{\small
I} AGN have PDZ~$(z_{\rm spec})<0.5$. The rest of them are rejected because
{\small ELDAR} does not detect at least 3 AGN emission lines in their
photometry. It is the consequence of objects for which some of their emission
lines have a EW smaller than values listed in Table~\ref{tab:lines}, and thus
the ALHAMBRA bands are not narrow enough to confirm them. We also check that no
source of the AGN-X sample is best-fitted by a stellar template in the first
step of {\small ELDAR}.

In Fig.~\ref{fig:AGNcomp} we display the completeness of the ALH2L and ALH3L
catalogues as a function of $z$. For the ALH2L catalogue the completeness grows
from $z=1$ ($\simeq70\,\%$) to $z=1.5$ ($\simeq90\,\%$), and for higher
redshifts it remains fairly constant. For the ALH3L catalogue the completeness
is largely independent of $z$ and it is $\simeq70\,\%$. For objects of the ALH3L
catalogue with F814W~$<22.5$, we can see that at $z=2.5$ the completeness is the
same as for sources of the ALH2L catalogue at the same redshift. This confirms
that the main strength of the 2-lines mode is to detect type-{\small I} AGN at
low redshift (see Fig.~\ref{fig:nden}).

We do not show the completeness at $z>2.5$ because there are only two objects in
the AGN-X sample at higher redshifts. In Appendix~\ref{app:C} we use the AGN-S
sample to study the completeness as a function of magnitude, redshift, and the
value of the {\small ELDAR} free parameters. We employ the AGN-S because
contains more objects that the AGN-X sample at high-$z$. We find that the
completeness increases to $85\,\%$ and $77\,\%$ for objects of the ALH2L and
ALH3L catalogues at $z>2$, respectively.

\section{Forecasts for narrow band surveys} 
\label{sec:widths}

In this section we forecast the performance of {\small ELDAR} in surveys with
narrower bands than the ALHAMBRA survey, as our method can be applied to any
survey in which the bands are narrow enough to isolate AGN emission lines from
the continuum.

There are several surveys that incorporate contiguous bands narrower than the
ALHAMBRA bands, such as SHARDS (25 bands of FWHM~$\simeq170\,\rm{\AA}$), PAUS
(40 bands of FWHM~$\simeq130\,\rm{\AA}$), and the upcoming J-PAS (54 bands of
FWHM~$\simeq140 \,\rm{\AA}$). As the data from all of these surveys is not
publicly available yet, we decided to forecast the completeness and redshift
precision for J-PAS because it has the greatest number of bands, and thus we
expect to find the largest differences between the results for ALHAMBRA and for
this survey.

To estimate the performance of {\small ELDAR} detecting type-{\small I} AGN in
J-PAS and to make a fair comparison with ALHAMBRA, we generate AGN-mock data for
the ALHAMBRA and J-PAS filter systems. In order to do that, we convolve the
template qso-0.2\_84 with both filter systems and we shift it in redshift
between $z=1$ and $z=5$ using a redshift step of $\Delta z=0.02$. Then, we
create 4 mock sources at each redshift imposing a magnitude of 19.5, 20.5, 21.5,
and 22.5 in the detection band of ALHAMBRA and J-PAS, which are the F814W band
for the first and the $r$ band for the second. We note that these magnitudes
correspond to different SNR in the medium-/narrow-bands of these surveys, given
their different magnitude limits. Next, we compute the error in each band using
a empirical relation for ALHAMBRA mock data and the J-PAS exposure time
calculator for J-PAS mock data (J. Varela, private communication). Finally, we
apply the 2- and 3-lines modes of {\small ELDAR} to both samples, where the only
modification that we include in {\small ELDAR} for J-PAS data is that we change
the redshift step employed in {\small LePHARE} from 0.01 to 0.001. This is done
because J-PAS includes narrower and more numerous contiguous bands than the
ALHAMBRA survey, and thus we expect a higher redshift precision \citep{benitez09b}.

In Fig.~\ref{fig:AGNcomp2} we show the performance of {\small ELDAR} using
ALHAMBRA and J-PAS data as a function of the redshift and magnitude of the
source. At low-$z$, the gaps in the redshift distribution are caused by the blue
and steep continuum emission of the qso-0.2\_84 template, which makes more
difficult to detect emission lines. This is even more important for mock sources
dimmer than 21 mag., as none of them are confirmed by {\small ELDAR}.
Nonetheless, as we can see in Fig.~\ref{fig:photo-z}, at $z<1.5$ we detect
plenty of ALHAMBRA sources with F814W~$>21$. This is because the SED of real
type-{\small I} AGN is not as steep as the continuum of the qso-0.2\_84
template.

The no-detection of bright objects at $z>2$ in some redshift intervals is due to
Ly~$\alpha$ falling in between two bands. While this is an important issue for
ALHAMBRA, it gets alleviated in the case of J-PAS. This is because if we
introduce a redshift-dependent continuum or we model it using the bands which
are adjacent to the band where Ly~$\alpha$ falls, we could confirm these
sources. The smaller number of detections as we decrease the brightness of the
sources is because the SNR required to detect some of the AGN lines gathered in
Table~\ref{tab:lines} is not large enough.

Using the 3-lines mode of {\small ELDAR}, we achieve a redshift precision of
$\sigma_{\rm NMAD}=0.48\,\%$ and $\sigma_{\rm NMAD}=0.21\,\%$ for mock ALHAMBRA
and J-PAS data, respectively. As we get $\sigma_{\rm NMAD}=0.86\,\%$ for real
ALHAMBRA data (see Table~\ref{tab:accuracy}), we forecast a precision of
$\sigma_{\rm NMAD}=0.38\,\%$ for J-PAS, which is similar to the one expected for
J-PAS galaxies \citep{benitez14}.

\section{Summary and conclusions} \label{sec:conclusions}

The emergence of multi-filter surveys such as COSMOS, SHARDS, the ongoing PAUS,
and the upcoming J-PAS, open the possibility of developing new techniques to
fully exploit these data that lie at the interface between photometry and
spectroscopy. In this work we presented {\small ELDAR}, a new method that
enables the secure identification of unobscured AGN and the precise computation
of their redshifts. Using as input only the multi-band information for each
observed source, {\small ELDAR} takes advantage of the low-resolution
spectroscopic nature of the data to look for AGN emission lines, thus allowing
an unambiguous AGN identification. With this approach, {\small ELDAR} offers a
new method to confirm AGN in multi-filter surveys without the need, for example,
of spectroscopic follow-up or X-ray observations.

We started by presenting the main characteristics of {\small ELDAR}, which
consists of two main steps. In the first we apply the template fitting code
{\small LePHARE} to all the point-like objects that we want to classify,
rejecting most of the stars and producing a redshift probability distribution
function for every extragalactic object. In the second step, we confirm the AGN
candidates by looking for typical AGN emission lines in each extragalactic
object. This allows us to generate samples of AGN with a very low contamination
from galaxies.

To test the performance of {\small ELDAR} we applied it to the publicly
available data from the ALHAMBRA survey, which covered an effective area of
$2.38\,\text{deg}^2$ of the northern sky with 20 contiguous medium-bands of
FWHM~$\simeq300\,\rm{\AA}$. Given the bandwidth of the ALHAMBRA filters, we
tuned our code to detect only type-{\small I} AGN. These objects, in fact, are
characterised by emission lines that can dominate the ALHAMBRA band in which
they fall thanks to their large equivalent width. Then, we defined two different
configurations of {\small ELDAR}, where the first prioritises completeness and
requires the detection of at least 2 AGN emission lines, while the second
prioritises purity and requires the detection of 3 lines. After the
pre-selection using {\small LePHARE}, we blindly ran both configurations of
{\small ELDAR} on the ALHAMBRA data, ending up with two AGN samples of 585 and
494 sources, respectively (ALH2L and ALH3L catalogues). The ALH2L sample covers
the redshift range $1<z<5.5$ and it is limited to F814W~$<22.5$. The ALH3L
catalogue spans the range $1.5<z<5.5$, and contains objects up to F814W~$<23$.
Approximately $80\,\%$ of the sources of our catalogues were lacking a
spectroscopic identification and redshift estimation. We make publicly available
the ALH2L and ALH3L catalogues, where we provide both the ALHAMBRA photometric
data and our redshift estimate.

To characterise the properties of the ALH2L and ALH3L catalogues we ran the 2-
and 3-lines configurations of {\small ELDAR} on samples of
spectroscopically-known type-{\small I} AGN and galaxies in the ALHAMBRA fields,
estimating, for the two catalogues, a completeness of $73\,\%$ and $67\,\%$, a
redshift precision of $\sigma_{\rm NMAD}=1.01\,\%$ and $\sigma_{\rm
NMAD}=0.86\,\%$, and a galaxy contamination of $31\,\%$ and $9\,\%$,
respectively. We obtain the best results for sources at $z>2$, as the Ly-alpha
line enters the spectral coverage of ALHAMBRA. At those redshifts, the
completeness increases to $85\,\%$ and $77\,\%$ for the two-modes, and we no
longer find galaxy contamination.

Thanks to the depth of the ALHAMBRA data, we have been able to push the
detection of type-{\small I} AGN to faint sources which are typically not
accessible by spectroscopic surveys. We would like to stress that {\small
ELDAR}, when applied to multi-filter surveys such as ALHAMBRA, does not require
additional data from X-ray, radio, nor variability studies to confirm
type-{\small I} AGN.

Finally, we forecast the performance of {\small ELDAR} in surveys with narrower
bands than ALHAMBRA. We analysed the particular case of the upcoming J-PAS
survey, which will cover thousands of square degrees of the northern sky with 54
narrow-bands of FWHM~$\simeq140\,\rm{\AA}$. We generated mock J-PAS and ALHAMBRA
data using a typical AGN SED template. Then, applying {\small ELDAR} to the mock
data, we estimated that J-PAS can reach a significantly better redshift
precision than ALHAMBRA thanks to the larger number of bands.

To conclude, we point out that {\small ELDAR} can be further improved: for
example, the first obvious step will be a more detailed modelling of the AGN
continuum emission. Also, we plan to optimise the code for the detection of
narrow AGN emission lines for narrow-band surveys such as PAUS and J-PAS. With
such improvements, we expect {\small ELDAR} to perform even better in terms of
completeness and redshift precision for range of active galaxies.

\section*{Acknowledgements}

We thank the referee for the thorough review and R. Angulo and J. Varela for
productive discussions. We thank the {\small LePHARE} team for making their code
publicly available. The authors acknowledge support from FITE (Fondos de
Inversiones de Teruel), Grupos de Arag\'on E96 and E103, and the Spanish
Ministry of Economy and Competitiveness (MINECO) through projects
AYA2016-76682-C3-1-P, AYA2015-66211-C2-1, AYA2015-66211-C2-2, AYA2013-42227-P,
and AYA2012-30789. This work was supported by FCT (ref. UID/FIS/04434/2013)
through national funds and by FEDER through COMPETE2020 (ref.
POCI-01-0145-FEDER-007672). J.C. acknowledges support from the Fundaci\'on
Bancaria Ibercaja for developing this research. BA has received funding from the
European Union's Horizon 2020 research and innovation programme under the Marie
Sklodowska-Curie grant agreement No 656354. MP acknowledges financial supports
from the Ethiopian Space Science and Technology Institute (ESSTI) under the
Ethiopian Ministry of Science Science and Technology (MoST). IM acknowledges
support from an FCT post-doctoral grant (ref. SFRH/BPD/95578/2013).

\bibliographystyle{mn2e}
\bibliography{qso_alh.bib}

\begin{thebibliography}{}

\bibitem[\protect\citeauthoryear{{Allen}}{{Allen}}{1976}]{allen76}
{Allen} C.~W.,  1976, {Astrophysical Quantities}

\bibitem[\protect\citeauthoryear{{Antonucci}}{{Antonucci}}{1993}]{antonucci93}
{Antonucci} R.,  1993, \araa, 31, 473

\bibitem[\protect\citeauthoryear{{Aparicio Villegas}, {Alfaro}
  et~al.,}{{Aparicio Villegas} et~al.}{2010}]{apariciovillegas10}
{Aparicio Villegas} T.,  {Alfaro} E.~J.,    et~al., 2010, \aj, 139, 1242

\bibitem[\protect\citeauthoryear{{Arnalte-Mur} et~al.,}{{Arnalte-Mur}
  et~al.}{2014}]{arnalte14}
{Arnalte-Mur} P.,  et~al., 2014, \mnras, 441, 1783

\bibitem[\protect\citeauthoryear{{Arnouts}, {Cristiani}, {Moscardini},
  {Matarrese}, {Lucchin}, {Fontana} \& {Giallongo}}{{Arnouts}
  et~al.}{1999}]{arnouts99}
{Arnouts} S.,  {Cristiani} S.,  {Moscardini} L.,  {Matarrese} S.,  {Lucchin}
  F.,  {Fontana} A.,    {Giallongo} E.,  1999, \mnras, 310, 540

\bibitem[\protect\citeauthoryear{{Barger}, {Cowie}, {Capak}, {Alexander},
  {Bauer}, {Fernandez}, {Brandt}, {Garmire} \& {Hornschemeier}}{{Barger}
  et~al.}{2003}]{barger03}
{Barger} A.~J.,  {Cowie} L.~L.,  {Capak} P.,  {Alexander} D.~M.,  {Bauer}
  F.~E.,  {Fernandez} E.,  {Brandt} W.~N.,  {Garmire} G.~P.,    {Hornschemeier}
  A.~E.,  2003, \aj, 126, 632

\bibitem[\protect\citeauthoryear{{Ben{\'{\i}}tez} et~al.,}{{Ben{\'{\i}}tez}
  et~al.}{2009}]{benitez09b}
{Ben{\'{\i}}tez} N.,  et~al., 2009, \apjl, 692, L5

\bibitem[\protect\citeauthoryear{{Ben{\'{\i}}tez} et~al.,}{{Ben{\'{\i}}tez}
  et~al.}{2014}]{benitez14}
{Ben{\'{\i}}tez} N.,  et~al., 2014, ArXiv e-prints, arXiv:1403.5237

\bibitem[\protect\citeauthoryear{{Bertin} \& {Arnouts}}{{Bertin} \&
  {Arnouts}}{1996}]{bertin96}
{Bertin} E.,  {Arnouts} S.,  1996, \aaps, 117, 393

\bibitem[\protect\citeauthoryear{{Bixler}, {Bowyer} \& {Laget}}{{Bixler}
  et~al.}{1991}]{bixler91}
{Bixler} J.~V.,  {Bowyer} S.,    {Laget} M.,  1991, \aap, 250, 370

\bibitem[\protect\citeauthoryear{{Bohlin}, {Colina} \& {Finley}}{{Bohlin}
  et~al.}{1995}]{bohlin95}
{Bohlin} R.~C.,  {Colina} L.,    {Finley} D.~S.,  1995, \aj, 110, 1316

\bibitem[\protect\citeauthoryear{{Bolzonella}, {Miralles} \&
  {Pell{\'o}}}{{Bolzonella} et~al.}{2000}]{bolzonella00}
{Bolzonella} M.,  {Miralles} J.-M.,    {Pell{\'o}} R.,  2000, \aap, 363, 476

\bibitem[\protect\citeauthoryear{{Brandt} \& {Hasinger}}{{Brandt} \&
  {Hasinger}}{2005}]{brandt05}
{Brandt} W.~N.,  {Hasinger} G.,  2005, \araa, 43, 827

\bibitem[\protect\citeauthoryear{{Brusa}, {Comastri}, {Mignoli}, {Fiore},
  {Ciliegi}, {Vignali}, {Severgnini}, {Cocchia}, {La Franca}, {Matt}, {Perola},
  {Maiolino}, {Baldi} \& {Molendi}}{{Brusa} et~al.}{2003}]{brusa03}
{Brusa} M.,  {Comastri} A.,  {Mignoli} M.,  {Fiore} F.,  {Ciliegi} P.,
  {Vignali} C.,  {Severgnini} P.,  {Cocchia} F.,  {La Franca} F.,  {Matt} G.,
  {Perola} G.~C.,  {Maiolino} R.,  {Baldi} A.,    {Molendi} S.,  2003, \aap,
  409, 65

\bibitem[\protect\citeauthoryear{{Busca} et~al.,}{{Busca}
  et~al.}{2013}]{busca13}
{Busca} N.~G.,  et~al., 2013, \aap, 552, A96

\bibitem[\protect\citeauthoryear{{Calzetti}, {Armus}, {Bohlin}, {Kinney},
  {Koornneef} \& {Storchi-Bergmann}}{{Calzetti} et~al.}{2000}]{calzetti00}
{Calzetti} D.,  {Armus} L.,  {Bohlin} R.~C.,  {Kinney} A.~L.,  {Koornneef} J.,
    {Storchi-Bergmann} T.,  2000, \apj, 533, 682

\bibitem[\protect\citeauthoryear{{Cardamone}, {van Dokkum}, {Urry},
  {Taniguchi}, {Gawiser}, {Brammer}, {Taylor}, {Damen}, {Treister}, {Cobb},
  {Bond}, {Schawinski}, {Lira}, {Murayama}, {Saito} \& {Sumikawa}}{{Cardamone}
  et~al.}{2010}]{cardamone10}
{Cardamone} C.~N.,  {van Dokkum} P.~G.,  {Urry} C.~M.,  {Taniguchi} Y.,
  {Gawiser} E.,  {Brammer} G.,  {Taylor} E.,  {Damen} M.,  {Treister} E.,
  {Cobb} B.~E.,  {Bond} N.,  {Schawinski} K.,  {Lira} P.,  {Murayama} T.,
  {Saito} T.,    {Sumikawa} K.,  2010, \apjs, 189, 270

\bibitem[\protect\citeauthoryear{{Chabrier}, {Baraffe}, {Allard} \&
  {Hauschildt}}{{Chabrier} et~al.}{2000}]{chabrier00}
{Chabrier} G.,  {Baraffe} I.,  {Allard} F.,    {Hauschildt} P.,  2000, \apj,
  542, 464

\bibitem[\protect\citeauthoryear{{Civano}, {Marchesi}, {Comastri}
  et~al.,}{{Civano} et~al.}{2016}]{civano16}
{Civano} F.,  {Marchesi} S.,  {Comastri} A.,    et~al., 2016, \apj, 819, 62

\bibitem[\protect\citeauthoryear{{Dawson}, {Kneib}, {Percival}, {Alam},
  {Albareti}, {Anderson} et~al.,}{{Dawson} et~al.}{2016}]{dawson16}
{Dawson} K.~S.,  {Kneib} J.-P.,  {Percival} W.~J.,  {Alam} S.,  {Albareti}
  F.~D.,  {Anderson} S.~F.,    et~al., 2016, 151, 44

\bibitem[\protect\citeauthoryear{{Fitzpatrick}}{{Fitzpatrick}}{1986}]{fitzpatrick86}
{Fitzpatrick} E.~L.,  1986, \aj, 92, 1068

\bibitem[\protect\citeauthoryear{{Flesch}}{{Flesch}}{2015}]{flesch15}
{Flesch} E.~W.,  2015, \pasa, 32, e010

\bibitem[\protect\citeauthoryear{{Fotopoulou}, {Salvato}, {Hasinger},
  {Rovilos}, {Brusa}, {Egami}, {Lutz}, {Burwitz}, {Henry}, {Huang},
  {Rigopoulou} \& {Vaccari}}{{Fotopoulou} et~al.}{2012}]{fotopoulou12}
{Fotopoulou} S.,  {Salvato} M.,  {Hasinger} G.,  {Rovilos} E.,  {Brusa} M.,
  {Egami} E.,  {Lutz} D.,  {Burwitz} V.,  {Henry} J.~P.,  {Huang} J.~H.,
  {Rigopoulou} D.,    {Vaccari} M.,  2012, \apjs, 198, 1

\bibitem[\protect\citeauthoryear{{Gallerani}, {Maiolino}, {Juarez}, {Nagao},
  {Marconi}, {Bianchi}, {Schneider}, {Mannucci}, {Oliva}, {Willott}, {Jiang} \&
  {Fan}}{{Gallerani} et~al.}{2010}]{gallerani10}
{Gallerani} S.,  {Maiolino} R.,  {Juarez} Y.,  {Nagao} T.,  {Marconi} A.,
  {Bianchi} S.,  {Schneider} R.,  {Mannucci} F.,  {Oliva} T.,  {Willott} C.~J.,
   {Jiang} L.,    {Fan} X.,  2010, \aap, 523, A85

\bibitem[\protect\citeauthoryear{{Gavignaud} et~al.,}{{Gavignaud}
  et~al.}{2006}]{gavignaud06}
{Gavignaud} I.,  et~al., 2006, \aap, 457, 79

\bibitem[\protect\citeauthoryear{{Gebhardt}, {Bender}, {Bower}, {Dressler},
  {Faber}, {Filippenko}, {Green}, {Grillmair}, {Ho}, {Kormendy}, {Lauer},
  {Magorrian}, {Pinkney}, {Richstone} \& {Tremaine}}{{Gebhardt}
  et~al.}{2000}]{gebhardt00}
{Gebhardt} K.,  {Bender} R.,  {Bower} G.,  {Dressler} A.,  {Faber} S.~M.,
  {Filippenko} A.~V.,  {Green} R.,  {Grillmair} C.,  {Ho} L.~C.,  {Kormendy}
  J.,  {Lauer} T.~R.,  {Magorrian} J.,  {Pinkney} J.,  {Richstone} D.,
  {Tremaine} S.,  2000, \apjl, 539, L13

\bibitem[\protect\citeauthoryear{{Heckman} \& {Best}}{{Heckman} \&
  {Best}}{2014}]{heckman14}
{Heckman} T.~M.,  {Best} P.~N.,  2014, \araa, 52, 589

\bibitem[\protect\citeauthoryear{{Hoaglin}, {Mosteller} \& {Tukey}}{{Hoaglin}
  et~al.}{1983}]{hoaglin83}
{Hoaglin} D.~C.,  {Mosteller} F.,    {Tukey} J.~W.,  1983, {Understanding
  robust and exploratory data anlysis}

\bibitem[\protect\citeauthoryear{{Hsu} et~al.,}{{Hsu}  et~al.}{2014}]{hsu14}
{Hsu} L.-T.,  et~al., 2014, \apj, 796, 60

\bibitem[\protect\citeauthoryear{{Ilbert} et~al.,}{{Ilbert}
  et~al.}{2009}]{ilbert09}
{Ilbert} O.,  et~al., 2009, \apj, 690, 1236

\bibitem[\protect\citeauthoryear{{Jahnke} \& {Macci{\`o}}}{{Jahnke} \&
  {Macci{\`o}}}{2011}]{jahnke11}
{Jahnke} K.,  {Macci{\`o}} A.~V.,  2011, \apj, 734, 92

\bibitem[\protect\citeauthoryear{{Kormendy} \& {Richstone}}{{Kormendy} \&
  {Richstone}}{1995}]{kormendy95}
{Kormendy} J.,  {Richstone} D.,  1995, \araa, 33, 581

\bibitem[\protect\citeauthoryear{{Lacy} et~al.,}{{Lacy}  et~al.}{2004}]{lacy04}
{Lacy} M.,  et~al., 2004, \apjs, 154, 166

\bibitem[\protect\citeauthoryear{{Lehmer}, {Xue}, {Brandt}, {Alexander},
  {Bauer}, {Brusa}, {Comastri}, {Gilli}, {Hornschemeier}, {Luo}, {Paolillo},
  {Ptak}, {Shemmer}, {Schneider}, {Tozzi} \& {Vignali}}{{Lehmer}
  et~al.}{2012}]{lehmer12}
{Lehmer} B.~D.,  {Xue} Y.~Q.,  {Brandt} W.~N.,  {Alexander} D.~M.,  {Bauer}
  F.~E.,  {Brusa} M.,  {Comastri} A.,  {Gilli} R.,  {Hornschemeier} A.~E.,
  {Luo} B.,  {Paolillo} M.,  {Ptak} A.,  {Shemmer} O.,  {Schneider} D.~P.,
  {Tozzi} P.,    {Vignali} C.,  2012, \apj, 752, 46

\bibitem[\protect\citeauthoryear{{Lilly} et~al.,}{{Lilly}
  et~al.}{2009}]{lilly09}
{Lilly} S.~J.,  et~al., 2009, \apjs, 184, 218

\bibitem[\protect\citeauthoryear{{Luo}, {Brandt}, {Xue}, {Brusa}, {Alexander},
  {Bauer}, {Comastri}, {Koekemoer}, {Lehmer}, {Mainieri}, {Rafferty},
  {Schneider}, {Silverman} \& {Vignali}}{{Luo} et~al.}{2010}]{luo10}
{Luo} B.,  {Brandt} W.~N.,  {Xue} Y.~Q.,  {Brusa} M.,  {Alexander} D.~M.,
  {Bauer} F.~E.,  {Comastri} A.,  {Koekemoer} A.,  {Lehmer} B.~D.,  {Mainieri}
  V.,  {Rafferty} D.~A.,  {Schneider} D.~P.,  {Silverman} J.~D.,    {Vignali}
  C.,  2010, \apjs, 187, 560

\bibitem[\protect\citeauthoryear{{Marchesi}, {Lanzuisi}, {Civano}, {Iwasawa},
  {Suh}, {Comastri}, {Zamorani}, {Allevato}, {Griffiths}, {Miyaji}, {Ranalli},
  {Salvato}, {Schawinski}, {Silverman}, {Treister}, {Urry} \&
  {Vignali}}{{Marchesi} et~al.}{2016}]{marchesi16}
{Marchesi} S.,  {Lanzuisi} G.,  {Civano} F.,  {Iwasawa} K.,  {Suh} H.,
  {Comastri} A.,  {Zamorani} G.,  {Allevato} V.,  {Griffiths} R.,  {Miyaji} T.,
   {Ranalli} P.,  {Salvato} M.,  {Schawinski} K.,  {Silverman} J.,  {Treister}
  E.,  {Urry} C.~M.,    {Vignali} C.,  2016, \apj, 830, 100

\bibitem[\protect\citeauthoryear{{Mart{\'{\i}}}, {Miquel}, {Castander},
  {Gazta{\~n}aga}, {Eriksen} \& {S{\'a}nchez}}{{Mart{\'{\i}}}
  et~al.}{2014}]{marti14}
{Mart{\'{\i}}} P.,  {Miquel} R.,  {Castander} F.~J.,  {Gazta{\~n}aga} E.,
  {Eriksen} M.,    {S{\'a}nchez} C.,  2014, \mnras, 442, 92

\bibitem[\protect\citeauthoryear{{Matthews} \& {Sandage}}{{Matthews} \&
  {Sandage}}{1963}]{matthews63}
{Matthews} T.~A.,  {Sandage} A.~R.,  1963, \apj, 138, 30

\bibitem[\protect\citeauthoryear{{Matute} et~al.,}{{Matute}
  et~al.}{2012}]{matute12}
{Matute} I.,  et~al., 2012, \aap, 542, A20

\bibitem[\protect\citeauthoryear{{Matute} et~al.,}{{Matute}
  et~al.}{2013}]{matute13}
{Matute} I.,  et~al., 2013, \aap, 557, A78

\bibitem[\protect\citeauthoryear{{Moles} et~al.,}{{Moles}
  et~al.}{2008}]{moles08}
{Moles} M.,  et~al., 2008, \aj, 136, 1325

\bibitem[\protect\citeauthoryear{{Molino} et~al.,}{{Molino}
  et~al.}{2014}]{molino14}
{Molino} A.,  et~al., 2014, \mnras, 441, 2891

\bibitem[\protect\citeauthoryear{{Mortlock}, {Warren}, {Venemans}, {Patel},
  {Hewett}, {McMahon}, {Simpson}, {Theuns}, {Gonz{\'a}les-Solares}, {Adamson},
  {Dye}, {Hambly}, {Hirst}, {Irwin}, {Kuiper}, {Lawrence} \&
  {R{\"o}ttgering}}{{Mortlock} et~al.}{2011}]{mortlock11}
{Mortlock} D.~J.,  {Warren} S.~J.,  {Venemans} B.~P.,  {Patel} M.,  {Hewett}
  P.~C.,  {McMahon} R.~G.,  {Simpson} C.,  {Theuns} T.,  {Gonz{\'a}les-Solares}
  E.~A.,  {Adamson} A.,  {Dye} S.,  {Hambly} N.~C.,  {Hirst} P.,  {Irwin}
  M.~J.,  {Kuiper} E.,  {Lawrence} A.,    {R{\"o}ttgering} H.~J.~A.,  2011,
  \nat, 474, 616

\bibitem[\protect\citeauthoryear{{Osterbrock}}{{Osterbrock}}{1991}]{Osterbrock91}
{Osterbrock} D.~E.,  1991, Reports on Progress in Physics, 54, 579

\bibitem[\protect\citeauthoryear{{P{\^a}ris} et~al.,}{{P{\^a}ris}
  et~al.}{2017}]{paris17}
{P{\^a}ris} I.,  et~al., 2017, 597, A79

\bibitem[\protect\citeauthoryear{{Peng}}{{Peng}}{2007}]{peng07}
{Peng} C.~Y.,  2007, \apj, 671, 1098

\bibitem[\protect\citeauthoryear{{P{\'e}rez-Gonz{\'a}lez} \&
  {Cava}}{{P{\'e}rez-Gonz{\'a}lez} \& {Cava}}{2013}]{perez13}
{P{\'e}rez-Gonz{\'a}lez} P.~G.,  {Cava} A.,  2013, in Revista Mexicana de
  Astronomia y Astrofisica Conference Series Vol.~42 of Revista Mexicana de
  Astronomia y Astrofisica, vol.~27, {SHARDS: Survey for High-z Absorption Red
  \& Dead Sources}.
pp 55--57

\bibitem[\protect\citeauthoryear{{Peth}, {Ross} \& {Schneider}}{{Peth}
  et~al.}{2011}]{peth11}
{Peth} M.~A.,  {Ross} N.~P.,    {Schneider} D.~P.,  2011, \aj, 141, 105

\bibitem[\protect\citeauthoryear{{Pickles}}{{Pickles}}{1998}]{pickles98}
{Pickles} A.~J.,  1998, \pasp, 110, 863

\bibitem[\protect\citeauthoryear{{Planck Collaboration}, {Ade}, {Aghanim},
  {Arnaud}, {Ashdown}, {Aumont}, {Baccigalupi}, {Banday}, {Barreiro},
  {Bartlett} et~al.,}{{Planck Collaboration} et~al.}{2016}]{planck16a}
{Planck Collaboration} {Ade} P.~A.~R.,  {Aghanim} N.,  {Arnaud} M.,  {Ashdown}
  M.,  {Aumont} J.,  {Baccigalupi} C.,  {Banday} A.~J.,  {Barreiro} R.~B.,
  {Bartlett} J.~G.,    et~al., 2016, \aap, 594, A13

\bibitem[\protect\citeauthoryear{{Prevot}, {Lequeux}, {Prevot}, {Maurice} \&
  {Rocca-Volmerange}}{{Prevot} et~al.}{1984}]{prevot84}
{Prevot} M.~L.,  {Lequeux} J.,  {Prevot} L.,  {Maurice} E.,
  {Rocca-Volmerange} B.,  1984, \aap, 132, 389

\bibitem[\protect\citeauthoryear{{Risaliti} \& {Lusso}}{{Risaliti} \&
  {Lusso}}{2017}]{risaliti16}
{Risaliti} G.,  {Lusso} E.,  2017, Astronomische Nachrichten, 338, 329

\bibitem[\protect\citeauthoryear{{Salvato} et~al.,}{{Salvato}
  et~al.}{2009}]{salvato09}
{Salvato} M.,  et~al., 2009, \apj, 690, 1250

\bibitem[\protect\citeauthoryear{{Salvato} et~al.,}{{Salvato}
  et~al.}{2011}]{salvato11}
{Salvato} M.,  et~al., 2011, \apj, 742, 61

\bibitem[\protect\citeauthoryear{{Schmidt}, {Marshall}, {Rix}, {Jester},
  {Hennawi} \& {Dobler}}{{Schmidt} et~al.}{2010}]{schmidt10}
{Schmidt} K.~B.,  {Marshall} P.~J.,  {Rix} H.-W.,  {Jester} S.,  {Hennawi}
  J.~F.,    {Dobler} G.,  2010, \apj, 714, 1194

\bibitem[\protect\citeauthoryear{{Scoville} et~al.,}{{Scoville}
  et~al.}{2007}]{Scoville07}
{Scoville} N.,  et~al., 2007, \apjs, 172, 1

\bibitem[\protect\citeauthoryear{{Stern}, {Assef}, {Benford}, {Blain}, {Cutri},
  {Dey}, {Eisenhardt}, {Griffith}, {Jarrett}, {Lake}, {Masci}, {Petty},
  {Stanford}, {Tsai}, {Wright}, {Yan}, {Harrison} \& {Madsen}}{{Stern}
  et~al.}{2012}]{stern12}
{Stern} D.,  {Assef} R.~J.,  {Benford} D.~J.,  {Blain} A.,  {Cutri} R.,  {Dey}
  A.,  {Eisenhardt} P.,  {Griffith} R.~L.,  {Jarrett} T.~H.,  {Lake} S.,
  {Masci} F.,  {Petty} S.,  {Stanford} S.~A.,  {Tsai} C.-W.,  {Wright} E.~L.,
  {Yan} L.,  {Harrison} F.,    {Madsen} K.,  2012, \apj, 753, 30

\bibitem[\protect\citeauthoryear{{Telfer}, {Zheng}, {Kriss} \&
  {Davidsen}}{{Telfer} et~al.}{2002}]{telfer02}
{Telfer} R.~C.,  {Zheng} W.,  {Kriss} G.~A.,    {Davidsen} A.~F.,  2002, \apj,
  565, 773

\bibitem[\protect\citeauthoryear{{Urry} \& {Padovani}}{{Urry} \&
  {Padovani}}{1995}]{urry95}
{Urry} C.~M.,  {Padovani} P.,  1995, \pasp, 107, 803

\bibitem[\protect\citeauthoryear{{van der Hucht}}{{van der
  Hucht}}{2001}]{hucht01}
{van der Hucht} K.~A.,  2001, \nar, 45, 135

\bibitem[\protect\citeauthoryear{{Vanden Berk} et~al.,}{{Vanden Berk}
  et~al.}{2001}]{vandenberk01}
{Vanden Berk} D.~E.,  et~al., 2001, \aj, 122, 549

\bibitem[\protect\citeauthoryear{{Wang}, {Du}, {Hu}, {Netzer}, {Bai}, {Lu},
  {Kaspi}, {Qiu}, {Li}, {Wang} \& {SEAMBH Collaboration}}{{Wang}
  et~al.}{2014}]{wang14}
{Wang} J.-M.,  {Du} P.,  {Hu} C.,  {Netzer} H.,  {Bai} J.-M.,  {Lu} K.-X.,
  {Kaspi} S.,  {Qiu} J.,  {Li} Y.-R.,  {Wang} F.,    {SEAMBH Collaboration}
  2014, \apj, 793, 108

\bibitem[\protect\citeauthoryear{{Watson}, {Denney}, {Vestergaard} \&
  {Davis}}{{Watson} et~al.}{2011}]{watson11}
{Watson} D.,  {Denney} K.~D.,  {Vestergaard} M.,    {Davis} T.~M.,  2011,
  \apjl, 740, L49

\bibitem[\protect\citeauthoryear{{White}, {Becker}, {Gregg},
  {Laurent-Muehleisen}, {Brotherton}, {Impey}, {Petry}, {Foltz}, {Chaffee},
  {Richards}, {Oegerle}, {Helfand}, {McMahon} \& {Cabanela}}{{White}
  et~al.}{2000}]{white00}
{White} R.~L.,  {Becker} R.~H.,  {Gregg} M.~D.,  {Laurent-Muehleisen} S.~A.,
  {Brotherton} M.~S.,  {Impey} C.~D.,  {Petry} C.~E.,  {Foltz} C.~B.,
  {Chaffee} F.~H.,  {Richards} G.~T.,  {Oegerle} W.~R.,  {Helfand} D.~J.,
  {McMahon} R.~G.,    {Cabanela} J.~E.,  2000, \apjs, 126, 133

\bibitem[\protect\citeauthoryear{{Wolf}, {Hildebrandt}, {Taylor} \&
  {Meisenheimer}}{{Wolf} et~al.}{2008}]{wolf08}
{Wolf} C.,  {Hildebrandt} H.,  {Taylor} E.~N.,    {Meisenheimer} K.,  2008,
  \aap, 492, 933

\bibitem[\protect\citeauthoryear{{Wolf}, {Meisenheimer}, {Kleinheinrich},
  {Borch}, {Dye}, {Gray}, {Wisotzki}, {Bell}, {Rix}, {Cimatti}, {Hasinger} \&
  {Szokoly}}{{Wolf} et~al.}{2004}]{wolf04}
{Wolf} C.,  {Meisenheimer} K.,  {Kleinheinrich} M.,  {Borch} A.,  {Dye} S.,
  {Gray} M.,  {Wisotzki} L.,  {Bell} E.~F.,  {Rix} H.-W.,  {Cimatti} A.,
  {Hasinger} G.,    {Szokoly} G.,  2004, \aap, 421, 913

\bibitem[\protect\citeauthoryear{{Zhao}, {Wang}, {Ross}, {Shandera}, {Percival}
  et~al.,}{{Zhao} et~al.}{2016}]{zhao16}
{Zhao} G.-B.,  {Wang} Y.,  {Ross} A.~J.,  {Shandera} S.,  {Percival} W.~J.,
  et~al., 2016, 457, 2377

\end{thebibliography}

\noindent\makebox[\linewidth]{\rule{0.45\textwidth}{1pt}}

\parbox[h]{0.45\textwidth}{
\hspace*{3pt}$^1$ Centro de Estudios de F\'isica del Cosmos de Arag\'on, Plaza
San Juan 1, Planta-3, 44001, Teruel, Spain.
\\\hspace*{3pt}$^2$ Max-Planck Institut f\"ur extraterrestrische Physik,
Postfach 1312, 85741 Garching bei M\"unchen, Germany.
\\\hspace*{3pt}$^3$ Instituto de F\'isica de Cantabria (CSIC-UC), 39005
Santander, Spain.
\\\hspace*{3pt}$^4$ Unidad Asociada Observatorio Astron\'omico (IFCA-UV), 46980
Patema, Spain.
\\\hspace*{3pt}$^5$ Ethiopian Space Science and
Technology Institute (ESSTI), Entoto Observatory and Research Center (EORC),
Astronomy and Astrophysics Research Division, P.O. Box 33679, Addis Ababa,
Ethiopia.
\\\hspace*{3pt}$^6$ Instituto de Astrofísica de Andalucía (IAA-CSIC), Glorieta
de la Astronom\'ia s/n, 18008 Granada, Spain.
\\\hspace*{3pt}$^{7}$ APC, AstroParticule et Cosmologie,
Universit\`e Paris Diderot, CNRS/IN2P3, CEA/Irfu, Observatoire de Paris,
Sorbonne Paris Cit\`e, 10, rue Alice Domon et L\`eonie Duquet, 75205 Paris Cedez
13, France.
\\\hspace*{3pt}$^8$ Observatori Astron\`omic, Universitat de Val\`encia, c/
Catedr\`atic Jos\'e Beltra\'an 2, 46980 Patema, Spain.
\\\hspace*{3pt}$^9$ Departament d'Astronomia i Astrof\'isica, Universitat de
Val\`encia, 46100 Burjassot, Spain.
\\$^{10}$ Instituto de Astrof\'isica e Ci\^encias do Espa\c o,
Universidade de Lisboa, OAL, Tapada da Ajuda, PT1349-018 Lisbon, Portugal.
\\$^{11}$ Departamento de F\'isica Matem\'atica, Instituto de F\'isica, 
Universidade de S\~ao Paulo, CP 66318, CEP 05314-970 S\~ao Paulo, Brazil.
\\$^{12}$ Instituto de Astrof\'isica de Andaluc\'ia, CSIC, Apdo 3004, 18080
Granada, Spain.
}


\begin{figure*}
\begin{center}
\includegraphics[width=0.475\textwidth]{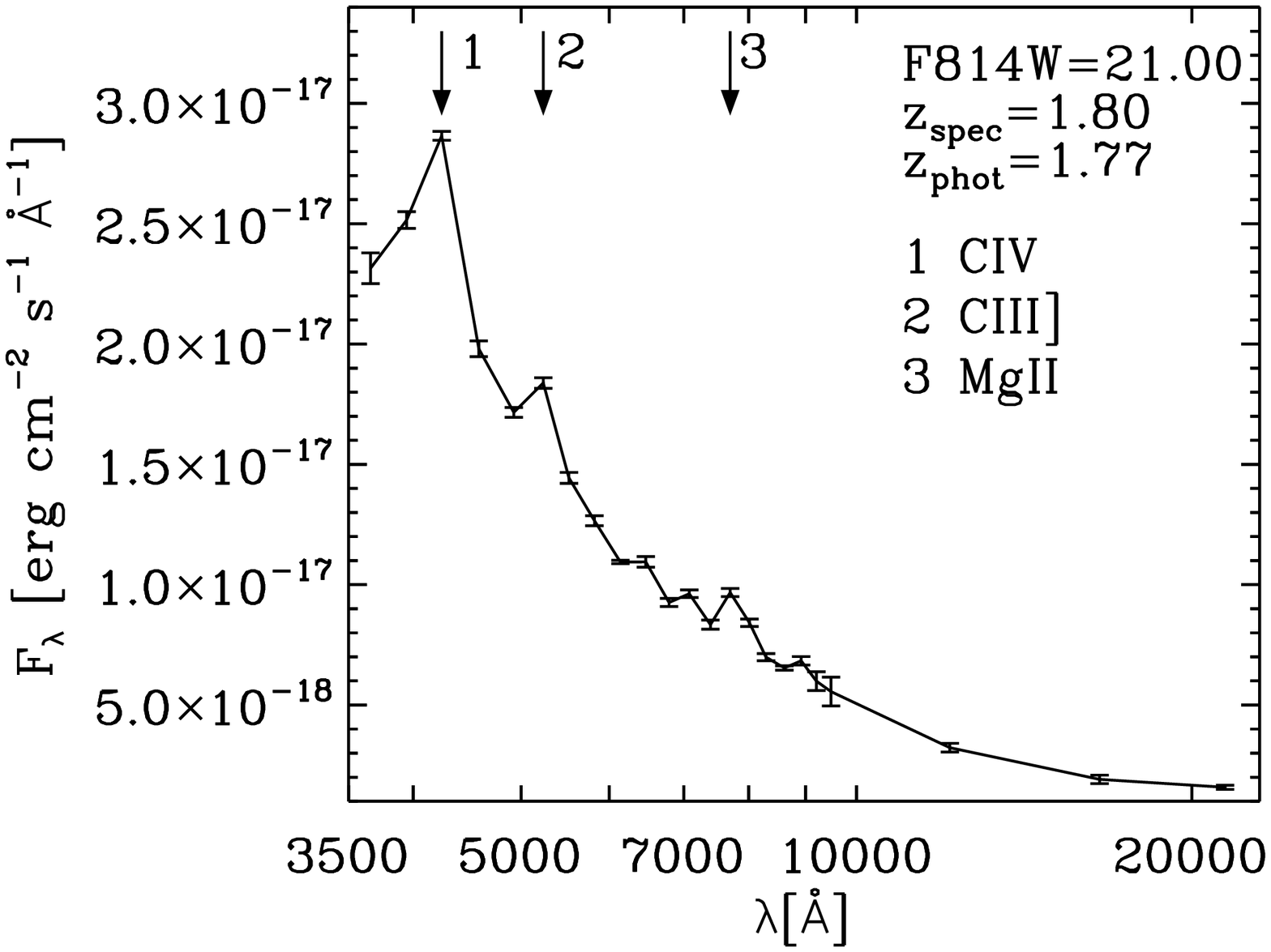}\includegraphics[width=0.475\textwidth]{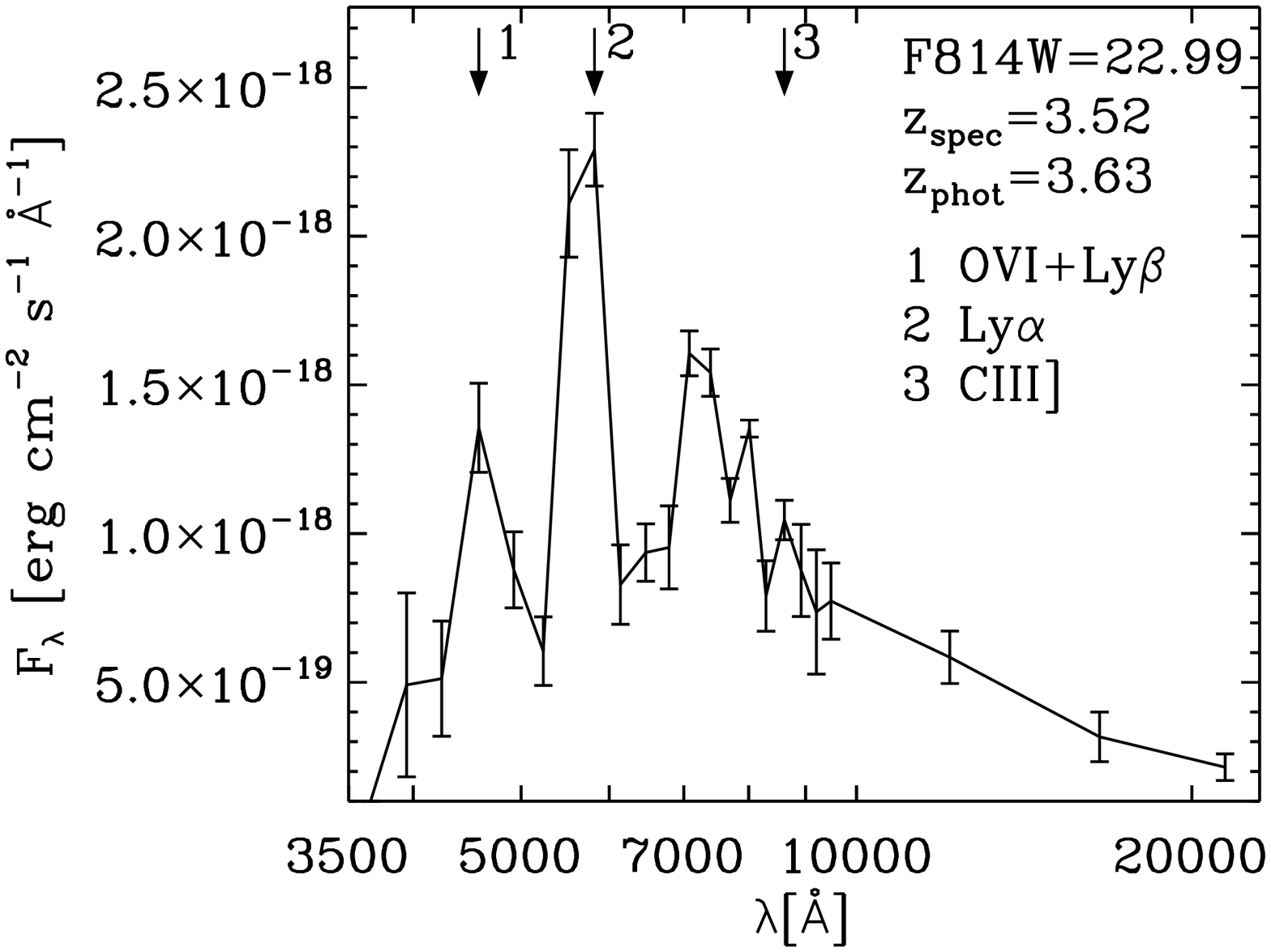}
\includegraphics[width=0.475\textwidth]{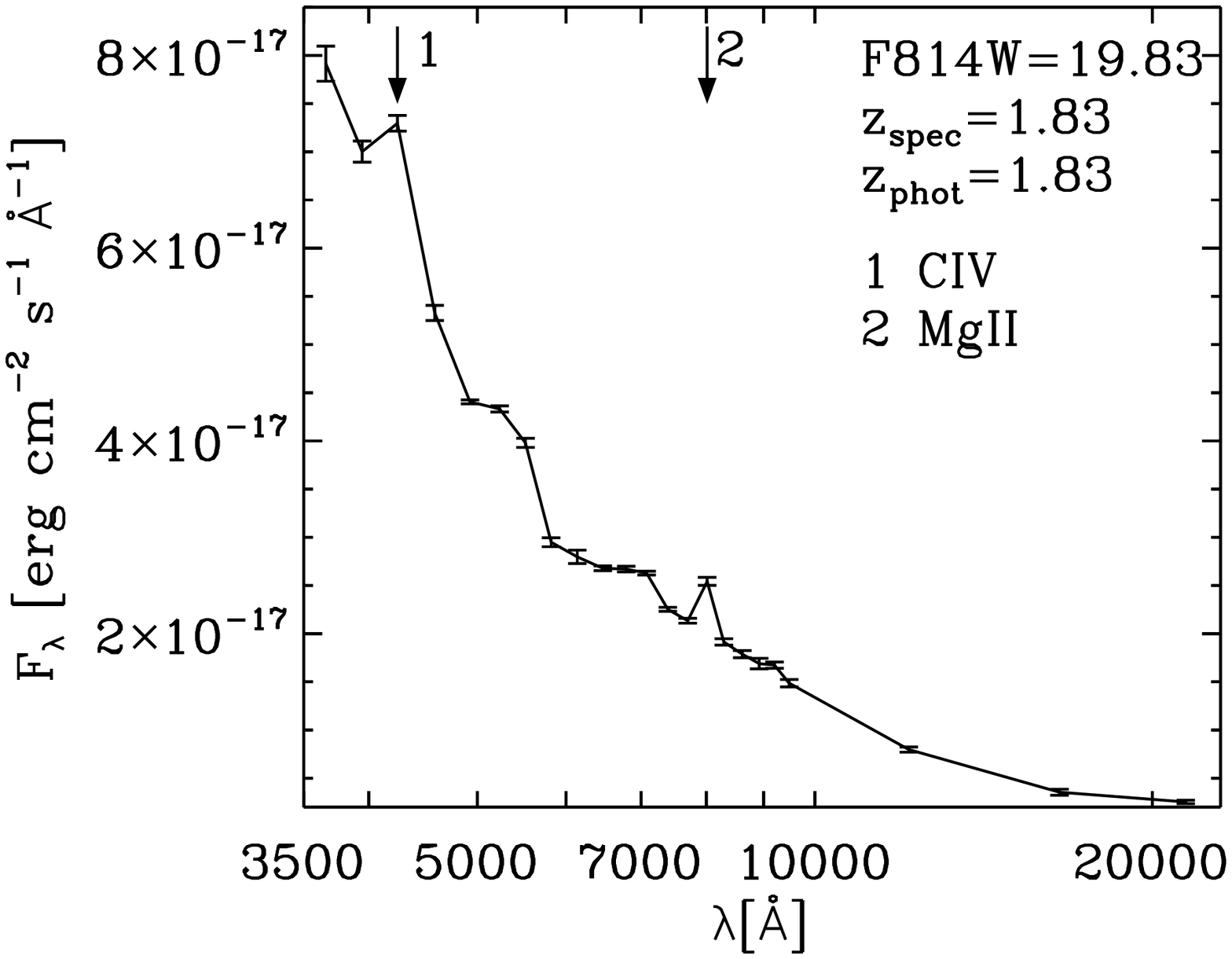}\includegraphics[width=0.475\textwidth]{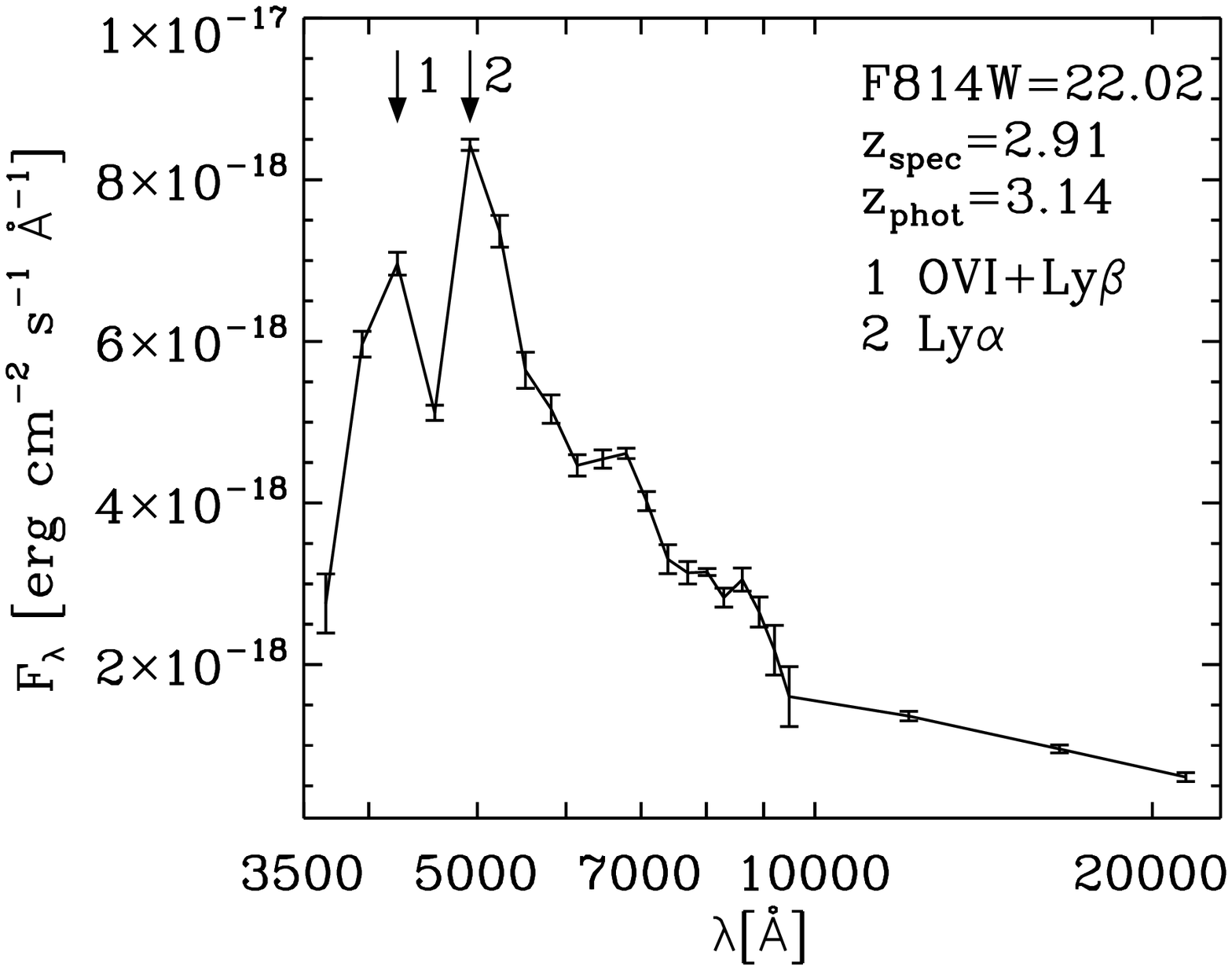}
\includegraphics[width=0.475\textwidth]{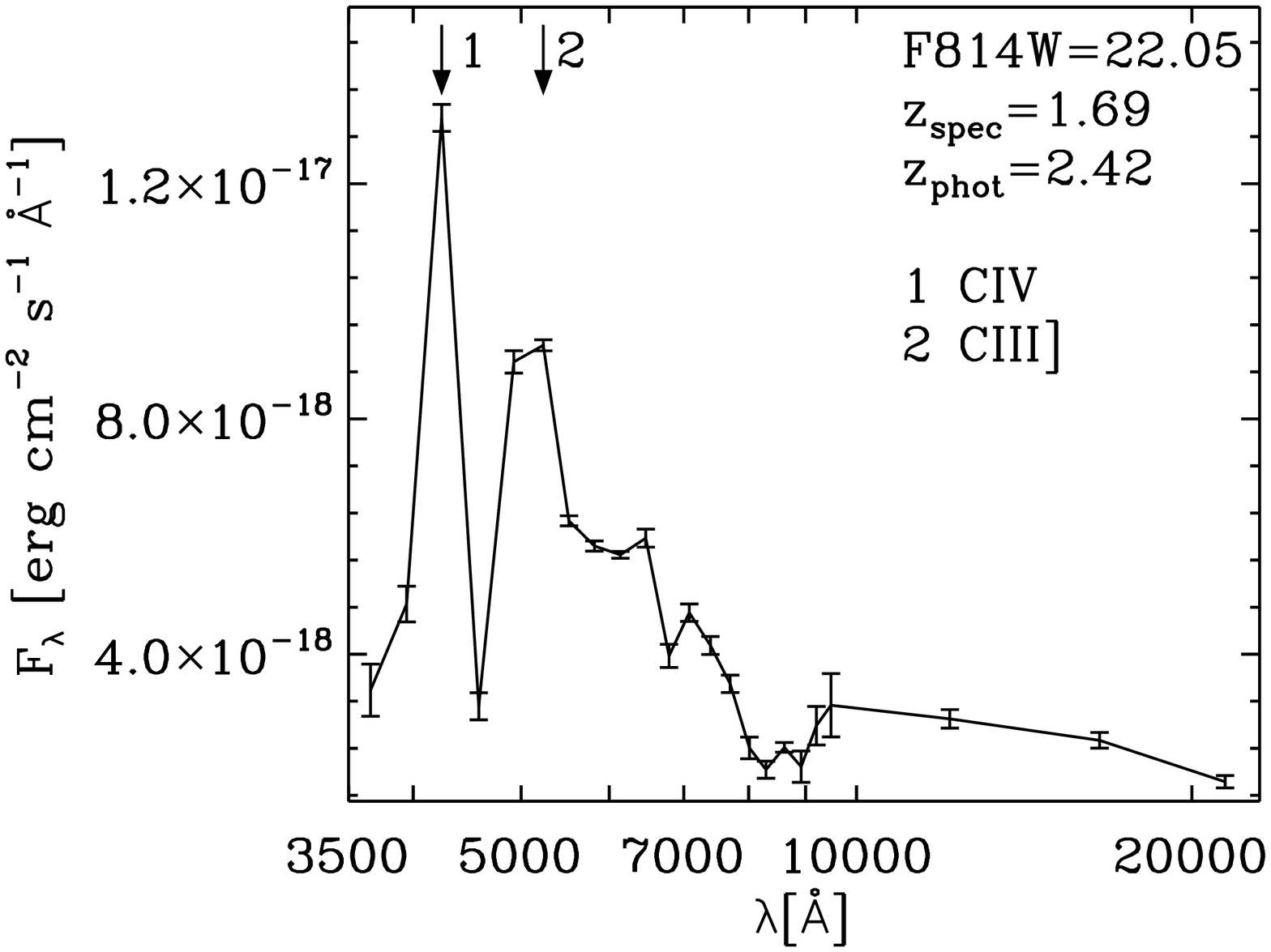}\includegraphics[width=0.475\textwidth]{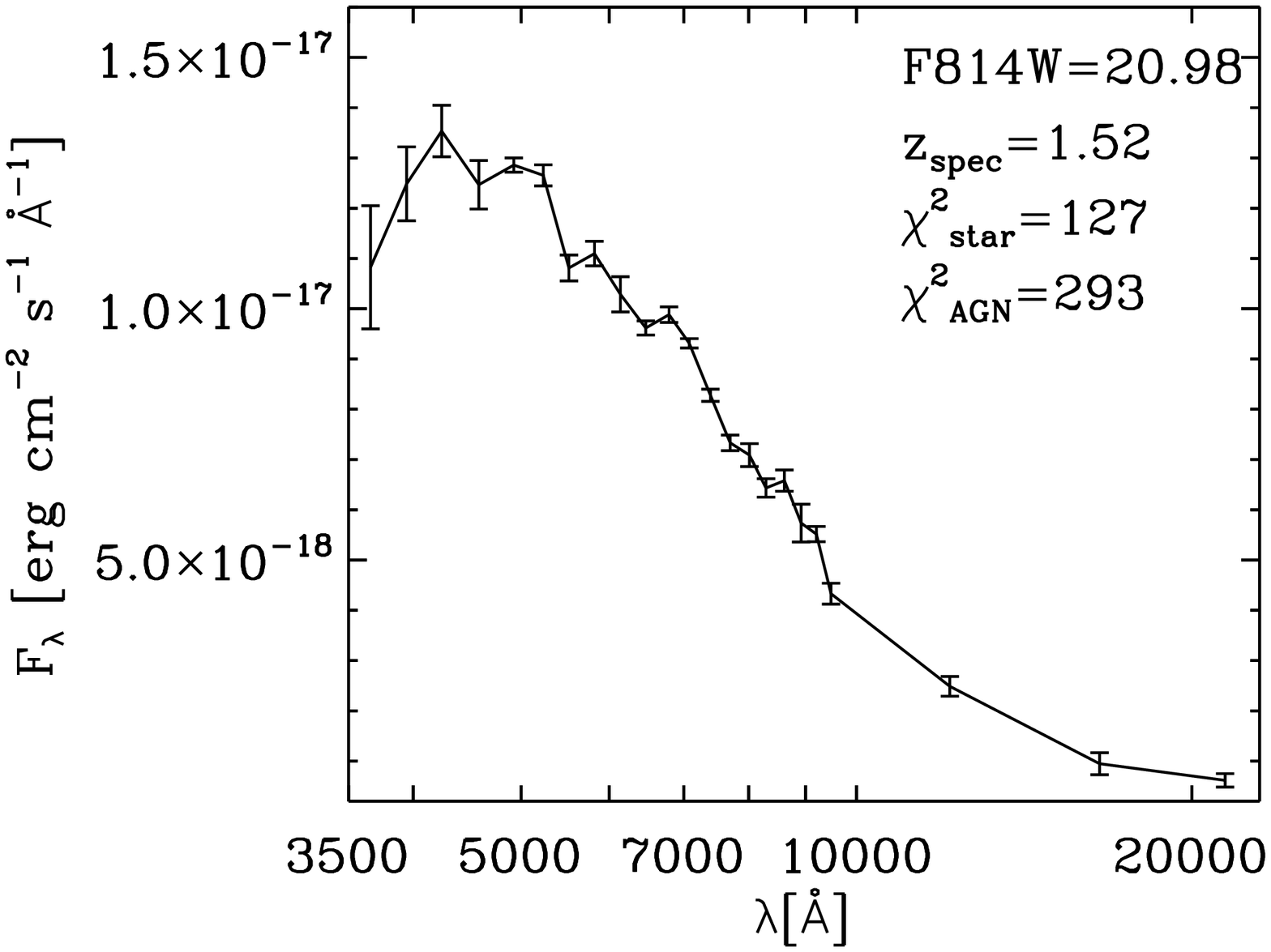}
\end{center}
\caption{\label{fig:ex_sed1} {\small ELDAR} results for spectroscopically-known
type-{\small I} AGN. We show illustrative examples of objects correctly
classifies as type-{\small I} AGN, redshift outliers, objects for which {\small
ELDAR} does not detect all the available lines, and the only source of the AGN-S
sample misclassified as a star. In all panels, arrows point to the bands where
AGN emission lines are confirmed. {\bf Top-left panel:} low-$z$ AGN ($z_{\rm
spec}=1.80$) correctly classified as type-{\small I} AGN by the 2- and 3-lines
modes of {\small ELDAR}. {\bf Top-right panel:} high-$z$ AGN ($z_{\rm
spec}=3.52$) correctly classified as type-{\small I} AGN by the 3-lines mode of
{\small ELDAR} (it is not classified as AGN by the 2-lines mode because it is
fainter that the magnitude limit for this mode, F814W~$22.5$). {\bf Medium-left
panel:} low-$z$ AGN ($z_{\rm spec}=1.83$) classified as type-{\small I} AGN just
by the 2-lines mode of {\small ELDAR}. This is because the 3-lines mode does not
detect C~{\small III}] at $5404\rm{\AA}$. {\bf Medium-right panel:} high-$z$ AGN
($z_{\rm spec}=2.91$) only classified as type-{\small I} AGN by the 2-lines mode
of {\small ELDAR}. The 3-lines mode does not detect C~{\small IV} at
$6407\rm{\AA}$ nor C~{\small III}] at $7895\rm{\AA}$. {\bf Bottom-left panel:}
low-$z$ at $z_{\rm spec}=1.69$ classified as AGN by the 2- and 3-lines modes of
{\small ELDAR} at $z_{\rm phot}=2.42$. This is because the the pair
\{Ly~$\alpha$, C~{\small IV}\} at $z=2.42$ is degenerated with the position
of the pair \{C~{\small IV}, C~{\small III}]\} at $z=1.69$. {\bf Bottom-right
panel:} only source of the AGN-S sample that it is best-fitted by a stellar
template. We cannot see any clear emission line in the ALHAMBRA data.}
\end{figure*}

\section*{Appendix A: AGN examples}
\addcontentsline{toc}{section}{Appendix A: AGN examples}
\sectionmark{Appendix A: AGN examples}
\label{app:A}

To illustrate the objects of the AGN-S sample that {\small ELDAR} confirms and
rejects and why it does so, in Fig.~\ref{fig:ex_sed1} we display the SED of
multiple sources and we discuss whether they fulfil all {\small ELDAR}'s
criteria or not. In the panels we use arrows to point to the bands where {\small
ELDAR} detects AGN emission lines. We indicate the name of the lines according
to $z_{\rm spec}$. In the top-left panel we show an object of the AGN-S sample
at $z_{\rm spec}=1.80$. After applying the 2- and 3-lines modes of {\small
ELDAR}, we find that it is correctly classified as AGN by both. This is because
our method detects C~{\small IV} in the 3rd band, C~{\small III}] in the 5th
band, and Mg~{\small II} in the 14th band. In addition, the redshift that
{\small ELDAR} assigns to this object, $z_{\rm phot}=1.77$, is compatible with
its spectroscopic redshift, $z_{\rm phot}=1.78$. In top-right panel we display
an object of the AGN-S sample at $z_{\rm spec}=3.52$ that is classified as
type-{\small I} AGN by the 3-lines mode of {\small ELDAR}. This is because our
method detects the complex O~{\small VI}+Ly~$\beta$ in the 4th band, Ly~$\alpha$
in the 8th band, and C~{\small III}] in the 17th band. On the other hand, the
line C~{\small IV} is not detected because it falls between the 12th and 13th
bands. According to $z_{\rm spec}$, the central wavelength of the lines
Ly~$\alpha$ and C~{\small IV} should be very close to the central wavelength of
the 7th and 12th band, respectively. However, the first falls between the 7th
and 8th band and the second between the 12th and 13th band. This is because AGN
emission lines may be shifted with respect to their rest-frame wavelength and/or
have anisotropic profiles \citep[see ][]{vandenberk01}, where these effects can
modify the band where they fall. As a consequence, the photometric redshift
computed for this source, $z_{\rm phot}=3.63$, is $\simeq 3\,\%$ greater than
its spectroscopic redshift.

In the medium-left panel we plot a source of the AGN-S sample at $z_{\rm
spec}=1.83$. This object is classified as AGN just by the 2-lines mode of
{\small ELDAR}. This is because our method detects C~{\small IV} in the 3rd
band, Mg~{\small II} in the 15th band, but not C~{\small III}] because it falls
between the 6th and 7th bands. The photometric redshift computed by {\small
ELDAR}, $z_{\rm phot}=1.83$, is the same as its spectroscopic redshift, $z_{\rm
spec}=1.83$. In the medium-right panel we show the only object of the AGN-S
sub-sample with $z_{\rm spec}>2.75$ not confirmed as type-{\small I} AGN by the
3-lines mode. However, it is classified as type-{\small I} AGN by the 2-lines
mode. This is because our code does not detect C~{\small IV}, which should fall
in the 10th band, nor C~{\small III}], which should fall in the 15th band. It is
the consequence of the ALHAMBRA bands not been narrow enough for detecting these
lines. The lack of these lines causes the computed photometric redshift, $z_{\rm
phot}=3.14$, to be $\simeq 8\,\%$ greater than the spectroscopic redshift for
this object, $z_{\rm spec}=2.91$. In these two panels we have shown a low-$z$
and a high-$z$ spectroscopically-known object that are not classified as
type-{\small I} AGN by the 3-lines mode of {\small ELDAR}. Objects like these
ones explain why the 3-lines mode has a lower completeness than the 2-lines
mode. In the following, we will show some examples of spectroscopically-known
objects for which {\small ELDAR} produces catastrophic redshift solutions.

In the bottom-left panel we display a source of the AGN-S sample at $z_{\rm
spec}=1.69$ classified as type-{\small I} AGN by the 2- and 3-lines modes at
$z_{\rm phot}=2.42$. Thus, this object is an outlier according to our definition
(see \S\ref{sec:precision}). This is because i) PDZ$(z_{\rm spec})<0.5$ and ii)
there is a degeneracy between the pair \{Ly~$\alpha$, C~{\small IV}\} at
$z=2.42$ and the pair \{C~{\small IV}, C~{\small III}]\} at $z=1.69$. This
source is also confirmed by the 3-lines mode because C~{\small III}] is confused
with a spurious line detected in the 9th band. In the bottom-right panel we
display the only object of the AGN-S sample best-fitted by a stellar template.
This object is at $z_{\rm spec}=1.52$ and it does not show any clear emission
lines. The best-fitting AGN templates has a $\chi^2$ more than twice the
$\chi^2$ of the best-fitting stellar templates. Even if this object is not
best-fitted by an AGN template, it will not be confirmed as type-{\small I} AGN
because {\small ELDAR} does not detect any AGN emission lines. No objects from
the AGN-X sample are best-fitted by stellar templates.


\section*{Appendix B: Dependence of the results on the criteria adopted in ELDAR}
\addcontentsline{toc}{section}{Appendix B: Dependence of the results on the 
criteria adopted in ELDAR}
\sectionmark{Appendix B: Different ELDAR configurations}
\label{app:C}

In \S\ref{sec:methodology} and \S\ref{sec:ELDARALH} we introduced multiple
parameters in the configuration of {\small ELDAR}. In this section we show the
dependence of the results for the objects of the AGN-S and GAL-S samples on
these criteria. In all the tables we underline the results for the fiducial
configuration of {\small ELDAR}.

\begin{table}
\begin{center}
\caption{\label{tab:ap1} Results for the AGN-S and GAL-S samples as a function
of the PDZ cut-off.}
\begin{tabular}{cccccc}
\hline
PDZ&Mode&Compl.$(\%)$&$\sigma_{\rm NMAD}(\%)$&$\eta(\%)$&Galaxies\\ \hline
0.90&2-lines&71.1&1.00&7.3&4\\
    &3-lines&64.4&0.86&5.9&1\\
0.50&2-lines&{\bf71.7}&{\bf1.01}&{\bf8.1}&{\bf4}\\
    &3-lines&{\bf65.2}&{\bf0.86}&{\bf5.8}&{\bf1}\\
0.01&2-lines&72.2&1.02&8.8&4\\
    &3-lines&66.7&0.92&6.8&1\\ \hline
\end{tabular}
\end{center}
{\bf Notes.} Bold numbers denote fiducial values for the 2- and 3-lines
modes of {\small ELDAR}.
\end{table}

We introduced a PDZ cut-off of 0.5 in {\small ELDAR} to reject redshift
solutions for which the $\chi^2$ is very low. In Table~\ref{tab:ap1} we gather
the results for the AGN-S and GAL-S samples using different values of the PDZ
cut-off. The quality of the ALH2L and ALH3L catalogues is largely independent of
the value of this parameter. This is because most of the objects with
F814W~$<22.5$ have only one peak in their PDZ with PDZ~$>0.5$.

\begin{table}
\begin{center}
\caption{\label{tab:ap2} Results for the AGN-S and GAL-S samples as a function
of the Ly~$\alpha$ criterion.}
\begin{tabular}{cccccc}
\hline
Ly~$\alpha$&Mode&Compl.$(\%)$&$\sigma_{\rm NMAD}(\%)$&$\eta(\%)$&Galaxies\\
\hline
1.25&2-lines&67.3&1.00&6.8&4\\
    &3-lines&60.6&0.86&3.8&1\\
0.75&2-lines&{\bf71.7}&{\bf1.01}&{\bf8.1}&{\bf4}\\
    &3-lines&{\bf65.2}&{\bf0.86}&{\bf5.8}&{\bf1}\\
0.25&2-lines&72.2&1.00&7.2&4\\
    &3-lines&65.9&0.86&5.7&1\\ \hline
\end{tabular}
\end{center}
\end{table}

Another criterion that we included in {\small ELDAR} is that the flux in the
band where the Ly~$\alpha$ line falls has to be $75\,\%$ greater than the flux
in the rest of the bands. In Table~\ref{tab:ap2} we present the results for the
AGN-S and GAL-S using different percentages. We find that increasing this
percentage the completeness is reduced.

\begin{table}
\begin{center}
\caption{\label{tab:ap3} Results for the AGN-S and GAL-S samples as a function
of the line acceptance criterion.}
\begin{tabular}{cccccc}
\hline
$\sigma_{\rm line}$&Mode&Compl.$(\%)$&$\sigma_{\rm NMAD}(\%)$&$\eta(\%)$&Galaxies\\
\hline
0.50&2-lines&78.6&1.10&9.6&13\\
0.75& 			&77.4&1.07&8.2&11\\
1.00& 			&75.1&1.06&7.7& 7\\
1.25& 			&75.1&1.05&8.5& 4\\
1.50& 			&{\bf71.7}&{\bf1.01}&{\bf8.1}&{\bf4}\\
1.75& 			&69.4&0.98&8.3& 1\\
0.50&3-lines&67.4&0.96&5.6& 1\\
0.75& 			&{\bf65.2}&{\bf0.86}&{\bf5.8}&{\bf1}\\
1.00& 			&61.4&0.96&6.2& 1\\
1.25& 			&56.8&0.86&5.7& 1\\
1.50& 			&49.2&0.86&4.9& 0\\
1.75& 			&44.7&0.86&5.1& 0\\ \hline
\end{tabular}
\end{center}
{\bf Notes.} $\sigma_{\rm line}$ indicates the minimum number of $\sigma$s that
we require to confirm an emission lines.
\end{table}

We set different requirements to confirm emission lines for the 2- and 3-lines
modes. For the 2-lines mode we established a stricter acceptance criterion than
for the 3-lines mode to reduce possible galaxy contamination. In
Table~\ref{tab:ap3} we display the results for the AGN-S and GAL-S samples using
different acceptance criteria. We find that the smaller is the value of
$\sigma_{\rm line}$, the higher is the completeness and the galaxy
contamination. Moreover, the galaxy contamination strongly grows by reducing
$\mathcal{N}$, and thus $\sigma_{\rm line}$ has to be carefully chosen depending
on $\mathcal{N}$.

\begin{table}
\begin{center}
\caption{\label{tab:ap4} Results for the AGN-S and GAL-S samples as a function
of $z_{\rm min}$.}
\begin{tabular}{cccccc}
\hline
$z_{\rm min}$&Mode&Compl.$(\%)$&$\sigma_{\rm NMAD}(\%)$&$\eta(\%)$&Galaxies\\
\hline
1.0&2-lines&{\bf71.7}&{\bf1.01}&{\bf8.1}&{\bf4}\\
   &3-lines&   -&   -&   -&-\\
1.5&2-lines&79.8&0.87& 6.3&3\\
   &3-lines&{\bf65.2}&{\bf0.86}&{\bf5.8}&{\bf1}\\
2.0&2-lines&84.8&0.92& 7.1&0\\
   &3-lines&77.0&0.97& 8.8&0\\
2.5&2-lines&80.0&0.91& 0.0&0\\
   &3-lines&75.7&0.86& 4.0&0\\ \hline
\end{tabular}
\end{center}
{\bf Notes.} The 3-lines mode is not defined at $z=1$ because there are less
than 3 AGN emission lines that {\small ELDAR} looks for within the ALHAMBRA
wavelength coverage.
\end{table}

The condition of detecting at least 2 or 3 AGN emission lines to confirm
objects sets a minimum redshift, $z_{\rm min}$, for the sources. In order to
check whether the {\small ELDAR}'s performance depends on the redshift of the
sources, we apply the 2- and 3-lines modes to the AGN-S and GAL-S samples using
different values of $z_{\rm min}$. In Table~\ref{tab:ap4} we gather the
results. We find that the completeness increases as a function of the redshift,
and the galaxy contamination decreases. Moreover, the redshift precision is
largely independent of $z_{\rm min}$.

\begin{table}
\begin{center}
\caption{\label{tab:ap5} Results for the AGN-S and GAL-S samples as a function
of the magnitude limit in the detection band, F814W.}
\begin{tabular}{cccccc}
\hline
F814W&Mode&Compl.$(\%)$&$\sigma_{\rm NMAD}(\%)$&$\eta(\%)$&Galaxies\\ \hline
21.5&2-lines&73.8&0.98&5.6&2\\
    &3-lines&65.8&0.77&1.8&1\\
22.0&2-lines&70.3&0.97&5.9&3\\
    &3-lines&62.9&0.78&1.6&1\\
22.5&2-lines&{\bf71.7}&{\bf1.01}&{\bf8.1}&{\bf4}\\
    &3-lines&64.7&0.86&3.9&1\\
23.0&2-lines&70.0&1.06&9.8&4\\
    &3-lines&{\bf65.2}&{\bf0.86}&{\bf5.8}&{\bf1}\\\hline
\end{tabular}
\end{center}
\end{table}

Finally, we address the dependence of the results on the magnitude limit. In
Table~\ref{tab:ap5} we summarize the results for the AGN-S and GAL-S samples
using the 2- and 3-lines modes. For both modes, the completeness does not depend
strongly on the magnitude limit; however, the redshift precision grows for 
brighter objects and the galaxy contamination decreases.

\section*{Appendix C: Description of the ALH2L and ALH3L catalogues}
\addcontentsline{toc}{section}{Appendix C: Description of the ALH2L and ALH3L
catalogues}
\sectionmark{Appendix C: Description of the ELDAR catalogues}
\label{app:E}

The catalogues ALH2L and ALH3L are available as binary ASCII tables. They are
documented in an accessory README file (column, bytes, format, units, label,
description) and it is also shown in \S\ref{sec:results}.

Notes on the catalogue columns:

\begin{center}
\begin{tabular}{rp{7cm}}
1 & The identification number of each object. The format is ALHXLYYY, where the
value of X is 2 and 3 for the ALH2L and ALH3L catalogues, respectively, and YYY
is the number of the object. The IDs are ranked according to $z_{\rm phot}$.\\
2 - 4 & J2000 coordinates (right ascension, sign of the declination, and
declination). The astrometry is from ALHAMBRA.\\
5 		& {\small ELDAR} redshift solution.\\
6 & Flag that indicates whether an object is inside the ALHAMBRA mask (1) or not
(0).\\
7 		& Index of the AGN template that best-fit the data.\\
8 - 9 & Extinction law and colour excess of the extragalactic template that
best-fit the data. The extinction is 0 for templates without extinction and 1 for
the \citet{calzetti00} extinction law, 2 for the \citet{allen76} extinction law,
3 for the \citet{prevot84} extinction law, and 4 for the \citet{fitzpatrick86}
extinction law.\\
10 - 11 & PSF-magnitude and uncertainty in the F814W band.\\
12 & Stellarity parameter of SExtractor. In ALHAMBRA it does not provide
accurate results for objects with F814W~$>23$.\\
13 - 50 & PSF-magnitude and uncertainty in ALHAMBRA medium-bands.\\
51 - 56 & PSF-magnitude and uncertainty in ALHAMBRA infrared broad-bands.\\
57 - 74 & ALHAMBRA band where the AGN emission lines of Table~\ref{tab:lines}
fall. We set this value to 99 for no detections and to 0 for lines outside the
wavelength range. For detected lines we also include the SNR in the band where
they fall, and the significance with which they are detected, ${\rm S}_{\rm
lin}$, defined as:\\
&\begin{equation}
{\rm S}_{\rm lin} = {\rm min} \left\{
\begin{aligned}
&\frac{F_{\rm cen}-F_{\rm blue}}{S_{\rm cen}}-\sigma_{\rm line},\\
&\frac{F_{\rm cen}-F_{\rm red}} {S_{\rm cen}}-\sigma_{\rm line},\\
&\frac{F_{\rm cen}-F_{\rm blue}}{S_{\rm cen}}-\sigma_{\rm line}
\frac{S_{\rm blue}}{S_{\rm cen}},\\
&\frac{F_{\rm cen}-F_{\rm red}} {S_{\rm cen}}-\sigma_{\rm line}
\frac{S_{\rm red}}{S_{\rm cen}}.\\
\end{aligned}
\right.
\end{equation}
\end{tabular}
\end{center}

\onecolumn
\captionsetup[longtable]{labelfont=bf,singlelinecheck=off,font=small,labelsep=period}
\begin{longtable}[!p]{rrccll}
\caption{Byte-by-byte description of the ALH2L and ALH3L catalogues.
\label{tab:cat}}\\
\hline \hline
Column&Bytes&Format&Units&Label&Description\\ \hline
\endfirsthead
\caption*{\small {\bf Table E1.} Continued.}\\
\hline \hline
Column&Bytes&Format&Units&Label&Description\\ \hline
\endhead
\hline
\endfoot
\hline
\endlastfoot
 1&     1-8  &A8      &--      &ID        	&Identification number\\
 2&   10-17  &F8.4    &deg     &RA       		&Right Ascension J2000 [0, 360]\\
 3&      19  &A1      &--      &DE-       	&Declination J2000 (sign)\\
 4&   20-26  &F7.4    &deg     &DEC      		&Declination J2000 [-90, 90]\\
 5&   28-32  &F5.3    &--      &Z       		&Photometric redshift\\
 6&      34  &I1      &--      &MASK      	&Mask [0 outside, 1 inside]\\
 7&   36-37  &I2      &--      &TEMP      	&Best-fit extragalactic template\\
 8&   39-42	 &F4.2    &--      &EXTB      	&Best-fit colour excess\\
 9&   44-49  &F6.3    &mag     &F814W     	&F814W magnitude\\
10&   51-55  &F5.3    &mag     &eF814W  		&F814W uncertainty\\
11&   57-60  &F4.2    &--      &STELL       &SExtractor Stellarity parameter\\
&&&&&[1 point-like, 0 extended]\\
12&   62-68	 &F7.3    &mag     &F365W    		&F365W magnitude\\
13&   70-76	 &F7.3    &mag     &eF365W  		&F365W uncertainty\\
14&   78-84	 &F7.3    &mag     &F396W    		&F396W magnitude\\
15&   86-92	 &F7.3    &mag     &eF396W  		&F396W uncertainty\\
16&  94-100  &F7.3    &mag     &F427W     	&F427W magnitude\\
17& 102-108  &F7.3    &mag     &eF427W  		&F427W uncertainty\\
18& 110-116  &F7.3    &mag     &F458W     	&F458W magnitude\\
19& 118-124  &F7.3    &mag     &eF458W  		&F458W uncertainty\\
20& 126-132  &F7.3    &mag     &F489W     	&F489W magnitude\\
21& 134-140  &F7.3    &mag     &eF489W  		&F489W uncertainty\\
22& 142-148  &F7.3    &mag     &F520W     	&F520W magnitude\\
23& 150-156  &F7.3    &mag     &eF520W  		&F520W uncertainty\\
24& 158-164  &F7.3    &mag     &F551W     	&F551W magnitude\\
25& 166-172  &F7.3    &mag     &eF551W  		&F551W uncertainty\\
26& 174-180  &F7.3    &mag     &F582W     	&F582W magnitude\\
27& 182-188  &F7.3    &mag     &eF582W  		&F582W uncertainty\\
28& 190-196  &F7.3    &mag     &F613W     	&F613W magnitude\\
29& 198-204  &F7.3    &mag     &eF613W  		&F613W uncertainty\\
30& 206-212  &F7.3    &mag     &F644W     	&F644W magnitude\\
31& 214-220  &F7.3    &mag     &eF644W  		&F644W uncertainty\\
32& 222-228  &F7.3    &mag     &F675W     	&F675W magnitude\\
33& 230-236  &F7.3    &mag     &eF675W  		&F675W uncertainty\\
34& 238-244  &F7.3    &mag     &F706W     	&F706W magnitude\\
35& 246-252  &F7.3    &mag     &eF706W  		&F706W uncertainty\\
36& 254-260  &F7.3    &mag     &F737W     	&F737W magnitude\\
37& 262-268  &F7.3    &mag     &eF737W  		&F737W uncertainty\\
38& 270-276  &F7.3    &mag     &F768W     	&F768W magnitude\\
39& 278-284  &F7.3    &mag     &eF768W  		&F768W uncertainty\\
40& 286-292  &F7.3    &mag     &F799W   	 	&F799W magnitude\\
41& 294-300  &F7.3    &mag     &eF799W  		&F799W uncertainty\\
42& 302-308  &F7.3    &mag     &F830W     	&F830W magnitude\\
43& 310-316  &F7.3    &mag     &eF830W  		&F830W uncertainty\\
44& 318-324  &F7.3    &mag     &F861W     	&F861W magnitude\\
45& 326-332  &F7.3    &mag     &eF861W  		&F861W uncertainty\\
46& 334-340  &F7.3    &mag     &F892W     	&F892W magnitude\\
47& 342-348  &F7.3    &mag     &eF892W  		&F892W uncertainty\\
48& 350-356  &F7.3    &mag     &F923W     	&F923W magnitude\\
49& 358-364  &F7.3    &mag     &eF923W  		&F923W uncertainty\\
50& 366-372  &F7.3    &mag     &F954W     	&F954W magnitude\\
51& 374-380  &F7.3    &mag     &eF954W  		&F954W uncertainty\\
52& 382-388  &F7.3    &mag     &J						&{\it J} magnitude\\
53& 390-396  &F7.3    &mag     &eJ					&{\it J} uncertainty\\
54& 398-404  &F7.3    &mag     &H						&{\it H} magnitude\\
55& 406-412  &F7.3    &mag     &eH					&{\it H} uncertainty\\
56& 414-420  &F7.3    &mag     &Ks					&$K_s$ magnitude\\
57& 422-428  &F7.3    &mag     &eKs					&$K_s$ uncertainty\\
58& 430-431  &I2      &--      &LINE1       &Band where the O~{\small VI}+Ly~$\beta$\\
&&&&&complex is detected [2,19]\\
59& 433-438  &F6.3    &--      &SNLINE1     &$\log_{10}({\rm SNR})$ in the band where\\
&&&&&the O~{\small VI}+Ly~$\beta$ complex is detected\\
60& 440-445  &F6.3    &--      &SLINE1      &$\log_{10}({\rm S}_{\rm lin})$ in the band where\\
&&&&&the O~{\small VI}+Ly~$\beta$ complex is detected\\
61& 447-448  &I2      &--      &LINE2       &Band where the Ly~$\alpha$ line\\
&&&&&is detected [2,19]\\
62& 450-455  &F6.3    &--      &SNLINE2     &$\log_{10}({\rm SNR})$ in the band where\\
&&&&&the Ly~$\alpha$ line is detected\\
63& 457-462  &F6.3    &--      &SLINE2      &$\log_{10}({\rm S}_{\rm lin})$ in the band where\\
&&&&&the Ly~$\alpha$ line is detected\\
64& 464-469  &I2      &--      &LINE3       &Band where the Si~{\small IV}+O~{\small IV}]\\
&&&&&complex is detected [2,19]\\
65& 471-476  &F6.3    &--      &SNLINE3     &$\log_{10}({\rm SNR})$ in the band where\\
&&&&&the Si~{\small IV}+O~{\small IV}] complex is detected\\
66& 478-479  &F6.3    &--      &SLINE3      &$\log_{10}({\rm S}_{\rm lin})$ in the band where\\
&&&&&the Si~{\small IV}+O~{\small IV}] complex is detected\\
67& 481-482  &I2      &--      &LINE4       &Band where the C~{\small IV} line\\
&&&&&is detected [2,19]\\
68& 484-489  &F6.3    &--      &SNLINE4     &$\log_{10}({\rm SNR})$ in the band where\\
&&&&&the C~{\small IV} line is detected\\
69& 491-496  &F6.3    &--      &SLINE4      &$\log_{10}({\rm S}_{\rm lin})$ in the band where\\ 
&&&&&the C~{\small IV} line is detected\\
70& 498-499  &I2      &--      &LINE5       &Band where the C~{\small III}] line\\
&&&&&is detected [2,19]\\
71& 501-506  &F6.3    &--      &SNLINE5     &$\log_{10}({\rm SNR})$ in the band where\\
&&&&&the C~{\small III}] line is detected\\
72& 508-513  &F6.3    &--      &SLINE5      &$\log_{10}({\rm S}_{\rm lin})$ in the band where\\ 
&&&&&the C~{\small III}] line is detected\\
73& 515-516  &I2      &--      &LINE6       &Band where the Mg~{\small II} line\\
&&&&&is detected [2,19]\\
74& 518-523  &F6.3    &--      &SNLINE6     &$\log_{10}({\rm SNR})$ in the band where\\
&&&&&the Mg~{\small II} line is detected\\
75& 525-530  &F6.3    &--      &SLINE6      &$\log_{10}({\rm S}_{\rm lin})$ in the band where\\ 
&&&&&the Mg~{\small II} line is detected\\
\end{longtable}
\twocolumn

\label{lastpage} \end{document}